\documentclass{appolb}
\usepackage{graphicx}

\usepackage{a4p}
\usepackage{rotating}
\usepackage{subfigure}
\usepackage{longtable}
\usepackage{multirow}
\usepackage{textcomp}
\usepackage{authblk,colortbl}
\usepackage{units}
\usepackage{amsmath,amssymb,amsfonts} 
\usepackage{mathrsfs}
\usepackage{color} 
\usepackage{authblk}
\usepackage{hyperref} 

\def\beq{\begin{equation}} 
\def\eeq{\end{equation}}
\def\bea{\begin{eqnarray}}
\def\eea{\end{eqnarray}}
\def\eqref#1{eq.~(\ref{eq:#1})}

\def\nn{\nonumber}   

\begin{document}



\title{TauSpinner algorithms for including spin and New Physics effects in $\gamma \gamma \rightarrow \tau \tau$ process} 
\author{A~Yu.~Korchin
\address{NSC Kharkiv Institute of Physics and Technology, 61108 Kharkiv, Ukraine } 
\address{V.~N.~Karazin Kharkiv National University, 61022 Kharkiv, Ukraine } 
\address{Institute of Physics, Jagellonian University, ul. Lojasiewicza 11, 30-348 Krakow, Poland} \\ 
E.~Richter-Was
\address{Institute of Physics, Jagellonian University, ul. Lojasiewicza 11, 30-348 Krakow, Poland} \\
\vspace{0.25cm}
Z.~Was
\address{Institute of Nuclear Physics Polish Academy of Sciences, PL-31342 Krakow, Poland}
}

\maketitle

\begin{abstract}
The possible anomalous New Physics contributions to electric and magnetic dipole moments of the $\tau$ lepton 
has brought renewed interest in development of new charge-parity violating signatures
in the $\tau$-pair production at Belle II energies, and also at higher energies of the LHC and the FCC.
In this paper, we discuss effects of anomalous contributions to cross-section and spin correlations
in the  $\gamma\gamma \to \tau^-\tau^+$ production processes, with $\tau$ decays included.
Such processes have been observed in the $pp$ and PbPb collisions at CERN LHC experiments.
Because of complex nature of the resulting distributions, Monte Carlo techniques are useful,
in particular of event reweighting with studied New Physics phenomena.
For the $\gamma\gamma$ processes, extensions of the Standard Model amplitudes  are implemented
in the {\tt TauSpinner} program. This is mainly with electric and magnetic dipole moments in mind,
but algorithm can be easily extended to other New Physics interactions,
provided they can be encapsulated into similar form-factors in the Standard Model structure of matrix elements.
Basic formulas and algorithm principles are presented, numerical examples are provided as illustration. 
Information on how to use the program is given in Appendix of the paper. 
\end{abstract}


%

\section{Introduction}
\label{sec:intro}
Electric and magnetic dipole moments of the $\tau$ lepton are sensitive
to violation of fundamental symmetries, such as charge-parity (CP) violation~\cite{Ramsey:1982pq, Cheng:1983mh, Bernreuther:1989kc}.
Recent measurements of dipole moments of the $\tau$ lepton at the Belle experiment~\cite{Belle:2021ybo},
as well as observation of $\gamma\gamma \to \tau^-\tau^+$ production at the hadron colliders~\cite{ATLAS:2022ryk,CMS:2022arf}
had brought renewed interest in electric and magnetic dipole moments of the $\tau$ lepton.
Deviation from predicted and measured values of the magnetic moment of the muon~\cite{Muong-2:2021ojo},
and possibly enhanced contributions from New Physics (NP) models to the magnetic moment of the $\tau$ lepton,
proportional to the square of its mass, makes these studies important and of contemporary interest.
Several beyond the Standard Model (SM) scenarios introduce dark weakly-interacting scalars or vector states
accompanying production of heavy fermions, e.g. $\tau$ leptons, or through the loop corrections with new virtual particles, which can serve as a source of anomalous contributions to electric or magnetic dipole moments of 
the $\tau$ lepton, as mentioned in~\cite{Bernreuther:1996dr, Eidelman:2007sb}  and references therein. 

In this paper, we discuss effects on the cross-section and  spin correlations in the $\tau$-pair production and decay.
First, in the SM, because they determine nature of interfering background for
the impact of anomalous dipole moments and later of potential NP augmentation of dipole moments themselves.
These studies include calculation of the analytical formulas, implementation and validation in the tool
and presentation of impact on typical kinematical distributions.
The impact of NP contributions to dipole moments can be introduced on top of simulations
of $\gamma\gamma \to \tau^-\tau^+$ processes assuming the SM couplings,
involving multi-body final states. The Monte Carlo (MC) solutions are convenient for this purpose
and presented below calculations have been implemented in the {\tt TauSpinner} program for reweighting
events with  $\tau$ pair produced in the $pp$ or PbPb collisions.

The {\tt TauSpinner} program~\cite{Czyczula:2012ny, Przedzinski:2018ett} is a convenient tool to study observables
sensitive to the NP effects in hadron colliders.
It allows to include NP and spin effects in case they are absent in event samples generated with general purpose MC generators like {\tt Pythia} \cite{Bierlich:2022pfr} or {\tt Sherpa} \cite{Sherpa:2019gpd}. Program development 
has a long history driven by expanding scope of its initially designed applications~\cite{Czyczula:2012ny}. 
First,  the longitudinal spin effects
in the case of  Drell-Yan ($Z, W$) and Higgs decay processes of $\tau$ leptons at the LHC \cite{Czyczula:2012ny}
were implemented. Later, it was extended to applications for the NP interactions in the hard processes,
in which lepton pairs are accompanied with one or two hard jets \cite{Kalinowski:2016qcd},
and to the transverse spin effects~\cite{Przedzinski:2014pla}.
The implementations were then further extended to allow study of  the electroweak effects 
in Refs.~\cite{Richter-Was:2018lld,Richter-Was:2020jlt} and anomalous dipole moments of the $\tau$-lepton couplings
to vector bosons in Ref.~\cite{Banerjee:2023qjc}. In the present paper we extend calculations of 
Ref.~\cite{Banerjee:2023qjc} including the higher order terms of  dipole moments in case of 
$\gamma \gamma \to \tau \tau$ process and present more numerical results.

Let us introduce terminology, which is used throughout the paper. 
In the reaction $\gamma \gamma \to \tau^- \tau^+$ with real 
photons, we include the anomalous magnetic and electric dipole moments of the $\tau$ lepton
as form-factors. At the real-photon point, $q^2=0$, the electromagnetic form-factors reduce to the 
corresponding dipole moments. These form-factors are connected to the chirality flipping operators. 
The terms proportional to the electric form-factor are also $CP$ violating. 
In general, the magnetic form-factor has a contribution from radiative corrections in the SM,
and we separate this contribution from the NP term. As for electric dipole form-factor --  
it is highly suppressed in the SM, and one can assume that this form-factor comes exclusively from NP.       
           
Our paper is organized as follows.  
In Section \ref{sec:theory} we present analytic results for the spin-correlation matrix including effects 
of the $\tau$-lepton anomalous magnetic and electric dipole form-factors, the SM and NP ones.
In Section \ref{sec:algorithm} we recall main points of the {\tt TauSpinner} reweighting algorithms
for inclusion of the dipole moments. Discussions on some numerical results are collected in
Section~\ref{sec:numerical}. First, elements of the spin-correlation matrix are shown in 
Subsection~\ref{sec:Rij} and we discuss main features of the 
$\gamma \gamma \to \tau^- \tau^+$ process spin correlations. Then, in Subsection~\ref{sec:pipi} shown is impact
from spin correlations in the SM and NP extensions on typical kinematical distributions in the case of 
both $\tau^\pm \to \pi^\pm \nu_\tau$ decays. It is followed by similar discussions of the case with both 
$\tau^\pm \to \pi^\pm \pi^0 \nu_\tau$
in Subsection~\ref{sec:rhorho}, and in the case of 
one $\tau$ decaying $\tau^\pm \to \mu^\pm \nu_\tau \nu_\mu$ and another $\tau$ decaying 
$\tau^\mp \to \pi^\mp \nu_\tau$ in Subsection~\ref{sec:mupi}.
Summary and outlook, Section \ref{sec:Outlook}, closes the paper. Technical description about {\tt TauSpinner}
initialisation and available weights is given in Appendix~\ref{app:TauSpinner}, 
and expressions for the elements of the spin-correlation matrix used in the code are listed in 
Appendix~\ref{app:Rij_elements}. 


\section{Amplitudes and spin correlations}
\label{sec:theory}
In this Section the formulas  for the two-photon production of the polarised $\tau$ leptons, 
$\gamma \gamma  \to \tau^-\tau^+$, are considered.  
In~\cite{Banerjee:2022sgf, Banerjee:2023qjc} we discussed formulas for including magnetic and electric dipole form-factors in elementary $2 \to 2$ parton processes of $\tau$-pair production. Here, we extend the formulas for the
 $\gamma \gamma  \to \tau^-\tau^+$ process to include also higher order terms in the dipole moments.

Let us  recall first introduction presented in~\cite{Banerjee:2023qjc} for the sake of more smooth reading.
The spin amplitude for  $AB \to \tau^+\tau^-$ process  with $AB$ $=$  $\gamma\gamma$
can be written as follows
\begin{equation}
A (k_1) + B (k_2) \to \tau^- (p_-) + \tau^+ (p_+)
\label{eq:001}
\end{equation}
with the four-momenta satisfying $k_1 +k_2 = p_- + p_+ $.

In the center-of-mass (CM) frame the components of the momenta are  
\begin{eqnarray}
&& p_- =(E, \vec{p}), \qquad \;  p_+ =(E, \, -\vec{p}), \qquad  \vec{p} = (0, \, 0, \, p), 
\nonumber \\ 
&& k_1 = (E, \, \vec{k}) , \qquad k_2  = (E, \, -\vec{k}),  \qquad 
\vec{k} =  (E \, \sin \theta, \, 0, \, E \, \cos \theta ), 
\label{eq:002}
\end{eqnarray}
so that $\hat{3}$ axis is along the momentum $\vec{p}$, the reaction plane is spanned on $\hat{1}$ and 
$\hat{3}$ axes defined 
by the momenta $\vec{p}$ and $\vec{k}$, and $\hat{2}$ axis is along $\vec{p} \times \vec{k}$. 
Here, $E=\tfrac{1}{2} \sqrt{s}$ is the photon ($\tau$-lepton) energy,  
$p = \beta E$ is the $\tau$-lepton three-momentum, where $\beta = (1-4 \tfrac{m_\tau^2}{s})^{1/2}$ is the velocity and $m_\tau$ is the mass of the $\tau$ lepton. 
The quantization frames of $\tau^-$ and  $\tau^+$ are connected to this reaction frame by 
the appropriate boosts along  $\hat{3}$ direction.  Note that $\hat{3}$ axis  is parallel to 
momentum of $\tau^-$ but anti-parallel to momentum of $\tau^+$. 
Only the reaction frame, the $\tau^-$ and the $\tau^+$ rest 
frames are used for calculations throughout the paper.

We assume that the $\gamma \tau \tau$ electromagnetic vertex has the following structure  
\begin{equation}
\Gamma^\mu_\gamma (q)  = -i e Q_\tau  \Big\{ \gamma^\mu F_1 (q^2)+ \frac{\sigma^{\mu \nu} q_\nu}{2 m_\tau} \,  
\big[ i  A(q^2)   + B(q^2)  \gamma_5  \big] \Big\},
\label{eq:003}
\end{equation}
where $q$ is the photon four-momentum, $e$ is the positron charge, $Q_\tau = -1$, 
$\sigma^{\mu \nu}= \frac{i}{2} [\gamma^\mu, \, \gamma^\nu]$,  
$F_1 (q^2)$ is the Dirac form-factor, 
$A(q^2) =F_2(q^2)$ is the Pauli form-factor and $B(q^2) = F_3(q^2)$ is the electric dipole form-factor.   
At the real-photon point, $F_1 (0) =1$, $A(0)$ is the anomalous magnetic 
dipole moment $a_\tau$, and $B(0)$ is related to the $CP$ violating electric dipole moment $d_\tau$  
\begin{equation}
A(0) = a_\tau,             
\qquad \quad B(0)  = \frac{2m_\tau}{e Q_\tau} d_\tau.
\label{eq:004}
\end{equation}

In the description of the reaction $\gamma(k_1) + \gamma(k_2) \to \tau^- (p_-) + \tau^+ (p_+)$ with the 
real photons ($k_1^2=k_2^2=0$),  we separate contribution from NP including   
the total SM contribution $A(0)_{SM}= 1.17721(5) \times 10^{-3}$~\cite{Eidelman:2007sb}, and define  
\begin{equation}
A(0) = A(0)_{SM} + A(0)_{NP}, \qquad \quad B(0) = B(0)_{NP}
\label{eq:006}
\end{equation}  
neglecting minor contribution to $B(0)$ in the SM. We assume that for the real photons 
the dipole moments are real-valued.   
  
After squaring the matrix element in the order $e^2$ and averaging result over the polarisations 
of the photons we obtain
\begin{equation}
|{\cal M}|^2 =    \sum_{i, j=1}^4 \, R_{i j} \, s^-_i  s^+_j  
=R_{44} + \sum_{i, j=1}^3 \, R_{i j} \, s^-_i  s^+_j ,
\label{eq:010}
\end{equation}
where $s^\mp_i \equiv (\vec{s}^{\, \mp}, \, 1)$,  and elements $R_{4 i}=R_{i 4} =0$ for $i=1,2,3$.

The $ s^-_i , s^+_j $ represent spin density states of the outgoing $\tau^-, \tau^+$, 
respectively\footnote{
In earlier version of the code we kept in $R_{i j}$ only terms
linear in the dipole moments, as published in \cite{Banerjee:2023qjc}. Here we include higher
order terms as well, that allows one to study impact on the total cross-section from NP models for $B(0)$.}.
The elements of the spin-correlation matrix $R_{ij}$ depend on the invariant mass of 
the $\tau$ pair $m_{{\tau^+} {\tau^-}}=\sqrt{s}$ and the scattering angle $\theta$.
They are written below in terms of the velocity $\beta$, Lorentz factor 
$\gamma = {\sqrt{s}}/(2m_\tau)$ and $\theta$. 

Elements $R_{ij}$, in form as used in the {\tt TauSpinner} code, are explicitly given in 
Appendix~\ref{app:Rij_elements}. A more convenient representation of $R_{ij}$ is expansion in 
powers of $A$ and $B$ (here $A \equiv A(0)$ and $B \equiv B(0)$): 
\begin{equation}
R_{ij} = \frac{e^4}{(1- \beta^2 \cos^2 \theta)^2}  \sum_{m, \, n =0 \, (m+n \le 4)}^{4}
A^m \, B^n \, c_{m n}^{(ij)} \qquad  \; (i, j = 1,2,3,4) .
\label{eq:R_{ij}}
\end{equation}
The nonzero coefficients $c_{m n}^{(i j)}$ are presented below.  

For the transverse-transverse spin-diagonal elements $R_{11}$, we obtain
\bea
\label{eq:R11}
 c_{00}^{(11)} &=& \frac{1}{8} [22 \beta^2 -11 \beta^4 -\beta^2 (\beta^2 -2) 
(-4 \cos 2 \theta + \cos 4 \theta) -8 ] ,     \\
c_{10}^{(11)} &=&   \frac{1}{2} (7 \beta^2+\beta^2 \cos 4 \theta -8),      \nn \\ 
c_{20}^{(11)} &=&  - \frac{1}{4} \gamma^2 
\{20-35 \beta^2+12 \beta^4+\beta^2 
[ (6 \beta^2 -2) \cos 2 \theta \nn \\ 
&+& (2 \beta^2 -3) \cos 4 \theta ] \},       \nn \\ 
 c_{30}^{(11)} &=& c_{12}^{(11)} = -\frac{1}{4} \gamma^2 [8-14 \beta^2
+3 \beta^4+4 \beta^4 \cos 2 \theta +\beta^2  (\beta^2 -2) \cos 4 \theta],
       \nn \\
 c_{40}^{(11)} &=& c_{04}^{(11)} = \frac{1}{2} c_{22}^{(11)}  \nn\\ 
&=& 
 \frac{1}{128} \gamma^4 \{6 \beta^2 (\beta^2 -8) (\beta^2 -2) +
\beta^2 [ (17 \beta^4 -16 \beta^2 -16) \cos 2 \theta   \nn \\ 
&+& 2 (8+5 \beta^2 (\beta^2 -2)) \cos 4 \theta  - \beta^4 \cos 6 \theta ] -32 \} , \nn \\ 
c_{02}^{(11)} &=&  \frac{1}{2} [ 6 (\beta^2 -1) +\beta^2 ( \cos 4 \theta -\cos 2 \theta  ) ] ,       \nn  
\eea
and for the transverse-transverse spin-diagonal element $R_{22}$, we have 
\bea
\label{eq:R22}   
c_{00}^{(22)} &=& 2 \beta^2 - 2 \beta^4 - \beta^4 \cos^2 \theta  (\cos^2 \theta -2) -1, \\
 c_{10}^{(22)} &=& -4 (1- \beta^2 \cos^2 \theta), \nn \\
c_{20}^{(22)}  &=& -5  (1- \beta^2 \cos^2 \theta),   \nn \\
c_{30}^{(22)}  &=& c_{12}^{(22)} =   -2  (1- \beta^2 \cos^2 \theta),   \nn \\
c_{40}^{(22)} &=& c_{04}^{(11)} = \frac{1}{2} c_{22}^{(11)}   \nn\\ 
&=& -\frac{1}{4} \gamma^4 (1- \beta^2 \cos^2 \theta) [1-2 \beta^2+2 \beta^4  
+ \beta^4 \cos^2 \theta (\cos^2 \theta -2)] , \nn  \\
c_{02}^{(22)} &=&
 -\gamma^2 [3-5 \beta^2+4 \beta^4-(\beta^2+3 \beta^4) \cos^2 \theta+2 \beta^4 \cos^4 \theta] . \nn
\eea

For the transverse-transverse spin-nondiagonal element $R_{12}$, the  coefficients are
\bea
\label{eq:R12}
c_{01}^{(12)} &=& \frac{1}{4} \beta (15 \beta^2+4 \cos 2 \theta + \beta^2 \cos 4 \theta -20),  \\
 c_{03}^{(12)} &=&  c_{21}^{(12)}  =  - \frac{1}{2} \beta  \gamma^2 (2-3 \beta^2+\beta^2 \cos 2 \theta ) 
\sin^2 \theta, \nn \\ 
c_{11}^{(12)} &=&  - \frac{1}{4} \beta \gamma^2 
[12-19 \beta^2+4 \beta^4+4 (\beta^4 +\beta^2 -1) \cos 2 \theta - \beta^2 \cos 4 \theta ]. \nn
\eea 

Next, we present coefficients for the longitudinal-longitudinal element $R_{33}$ as 
\bea
\label{eq:R33}
c_{00}^{(33)}& =&  -1+2 \beta^4+\beta^2 \cos^2 \theta 
[ 2-2 \beta^2+(\beta^2 -2) \cos^2 \theta ]  ,      \\
 c_{10}^{(33)} &=& \frac{1}{2} (-8+9 \beta^2 -\beta^2 \cos 4 \theta) ,        \nn \\ 
c_{20}^{(33)} &=&  -\gamma^2 \{5-5 \beta^2+2 \beta^4+\beta^2 \cos^2 \theta 
[-8+5 \beta^2+(3-2 \beta^2) \cos 2 \theta] \} ,       \nn \\ 
 c_{30}^{(33)} &=& c_{12}^{(33)} =  
-\gamma^2 [2 -\beta^2 (\beta^2 -2) \cos^2 \theta (\cos 2 \theta -3) ]   ,     \nn \\
 c_{40}^{(33)} &=& c_{04}^{(33)} = \frac{1}{2} c_{22}^{(33)}   \nn \\
&=& 
\frac{1}{4} \gamma^4 \{-1+2 \beta^4+\beta^2 \cos^2 \theta [7+2 \beta^2 (\beta^2 -6) 
+ (7 \beta^2 -4) \cos^2 \theta - \beta^4 \cos^4 \theta] \},             \nn \\
c_{02}^{(33)} &=&  -\gamma^2 \{3-\beta^2+\beta^2 \cos^2 \theta [-9+5 \beta^2
-2 (\beta^2 -2) \cos^2 \theta] \} .     \nn  
\eea

There are nonzero longitudinal-transverse elements $R_{13}$ and $R_{23}$ with the coefficients
\bea
\label{eq:R13}
 c_{00}^{(13)} &=&    \frac{\beta^2}{\gamma}  \sin^2 \theta \sin 2 \theta,      \\
 c_{10}^{(13)} &=&  \beta^2 \gamma  [1+(\beta^2 -2) \cos^2 \theta] \sin 2 \theta,        \nn \\ 
c_{20}^{(13)} &=& \frac{1}{2} \beta^2 \gamma  [ 5+(\beta^2 -6) \cos^2 \theta ] \sin 2 \theta,       \nn \\ 
 c_{30}^{(13)} &=& c_{12}^{(13)} = 2 \beta^2 \gamma  \sin^2 \theta \sin 2\theta,     \nn \\
 c_{40}^{(13)} &=& c_{04}^{(13)} = \frac{1}{2} c_{22}^{(13)} = 
 \frac{1}{2} \beta^2 \gamma  \sin^2 \theta \sin 2 \theta, \nn \\
c_{02}^{(13)} &=& \frac{1}{2} \beta^2 \gamma  [5-2 \beta^2 +(\beta^2 -4) \cos^2 \theta ] \sin 2 \theta,   \nn  
\eea
and 
\bea
\label{eq:R23}
c_{01}^{(23)} &=&  c_{11}^{(23)} = 2 c_{03}^{(23)} = 2c_{21}^{(23)}   \\ 
& =& \frac{1}{2} \beta \gamma (2 -3 \beta^2+ \beta^2 \cos 2 \theta) \sin 2 \theta .\nn
\eea 

For the spin-independent element $R_{44}$, the nonzero coefficients $c_{m n}^{(44)}$ read
\bea
\label{eq:R44}
c_{00}^{(44)} &=& 1+ 2 \beta^2 -2 \beta^4 -2 {\beta^2} (1 - \beta^2) \cos^2 \theta  
-\beta^4 \cos^4 \theta,          \\
 c_{10}^{(44)} &=& 4 (1- \beta^2 \cos^2 \theta),        \nn \\ 
c_{20}^{(44)} &=&  \frac{1}{2} 
\gamma^2 (\beta^2+\beta^2 \cos 2 \theta -2) (3 \beta^2+2 \beta^2 \cos 2 \theta -5) ,       \nn \\ 
 c_{30}^{(44)} &=& c_{12}^{(44)} = 
\gamma^2 (\beta^2 \cos 2 \theta -1) (\beta^2+\beta^2 \cos 2 \theta -2)   ,     \nn \\
  c_{40}^{(44)} &=& c_{04}^{(44)} = \frac{1}{2} c_{22}^{(44)} \nn \\ 
	&=&  \frac{1}{4} \gamma^4 (1 -\beta^2 \cos^2 \theta)   
[1+2 \beta^2 - 2 \beta^4+2 \beta^2 (\beta^2 -2)  \cos^2 \theta+ \beta^4 \cos^4 \theta],
         \nn \\
c_{02}^{(44)} &=&  - \gamma^2 (1-\beta^2 \cos^2 \theta) (\beta^2+2 \beta^2 \cos 2 \theta -3) .     \nn  
\eea
The following symmetry  relations are fulfilled: $R_{21} =- R_{12}$, $R_{31} = R_{13}$ 
and $R_{32} = - R_{23}$.  

Note that the the formulas published in~\cite{Banerjee:2023qjc} correspond to keeping only coefficients 
$c^{(i j)}_{00}, \, c^{(i j)}_{10}$ and $c^{(i j)}_{01}$ in Eqs.~(\ref{eq:R11})-(\ref{eq:R44}).
The expressions in the present paper include the higher order terms in dipole moments $A$ and $B$ up to the power 
of 4. In particular,  $R_{44}$ depends not only on $A$ but also on $B^2$ and $B^4$, allowing in experimental analysis to study impact of the electric dipole moment on the             
cross-section summed over all spin density configurations.
Also, the contribution from the electric dipole moment is no longer different from the rest of the terms, 
as it was the case of~\cite{Banerjee:2023qjc}, where only
$R_{23}$ and $R_{12}$ were dependent on $B$ while all other elements $R_{ij}$ were dependent 
on $A$ only.
                                                                                                                                 
Finally, the cross-section of the process $\gamma \gamma \to \tau^- \tau^+$ is expressed as
\begin{equation}
\frac{d \sigma }{d \Omega} (\gamma \gamma  \to \tau^- \tau^+) = \frac{\beta}{64 \pi^2   s } 
\Bigl( R_{44} +\sum_{i, j=1}^3 \, R_{i j}  \, s^-_i  s^+_j  \Bigr).
\label{eq:014}
\end{equation}
The element $R_{44}$ determines the cross-section in which the sum is taken over the spins of the $\tau$
leptons\footnote{Note that in the SM, the one-loop electroweak corrections     
to the cross-section of the processes $\gamma \gamma \to \mu^- \mu^+$ and $\gamma \gamma \to \tau^- \tau^+$ with unpolarized leptons have been calculated in Ref.~\cite{Demirci:2021zwz}.}
\begin{equation} 
\frac{d \sigma}{d \Omega} (\gamma \gamma  \to \tau^- \tau^+)  = \frac{\beta}{16 \pi^2   s } 
\, R_{44}.
\label{eq:015}
\end{equation}

One can further rearrange Eq.~(\ref{eq:014}) and introduce the normalized elements 
$r_{ij} = R_{ij}/R_{44}$, factorizing out explicitly spin-correlation components of the cross-section 
\begin{equation} 
\frac{d \sigma}{d \Omega} (\gamma \gamma  \to \tau^- \tau^+) = \frac{\beta}{64 \pi^2   s } 
\, R_{44} \; \Bigl( r_{44} +\sum_{i, j=1}^3 \, r_{i j}  \, s^-_i  s^+_j  \Bigr), \qquad  r_{44}=1.
\label{eq:016}
\end{equation}

Let us stress now, that the frame and sign convention of $R_{ij}$ presented in Eqs.~(\ref{eq:R11})-(\ref{eq:R44}) differ from the ones used later in the paper, and in the {\tt TauSpinner} program,
when the matrix is contracted with $\tau$-lepton polarimetric vectors.
The change in convention reads as follows:  
\begin{eqnarray}
\label{eq:framesR}
 R_{tt}\;\leftarrow \;\; R_{44}, & R_{tx}\leftarrow-R_{42}, & R_{ty}\leftarrow -R_{41},\;  R_{tz}\leftarrow-R_{43}, \nonumber \\
 R_{xt}\leftarrow -R_{24}, & R_{xx}\leftarrow  \;\;R_{22}, & R_{xy}\leftarrow  \;\;  R_{21},\;  R_{xz}\leftarrow \;\; R_{23}, \\
 R_{yt}\leftarrow -R_{14}, & R_{yx}\leftarrow \;\;  R_{12}, & R_{yy}\leftarrow  \;\;  R_{11},\;  R_{yz}\leftarrow \;\; R_{13},  \nonumber \\
 R_{zt}\leftarrow -R_{34}, & R_{zx}\leftarrow \;\; R_{32}, & R_{zy}\leftarrow  \;\; R_{31}, \;  R_{zz}\leftarrow \;\; R_{33}.\nonumber 
  \end{eqnarray}

Let us briefly explain that there are several reasons for that frame orientation differences 
used in {\tt TauSpinner} and {\tt Tauola} decay library~\cite{Jadach:1993hs}. Historical
one; in the past, reactions were organized having $z$ axis along $e^+$ or antiquark direction, whereas
now, many authors prefer to use  $z$ axis along $e^-$ or quark directions. This was the convention used
for {\tt KKMC}~\cite{Arbuzov:2020coe} and {\tt Tauola}~\cite{Jadach:1993hs} programs implementations. To adjust, this require
$\pi$ angle rotation around axis usually perpendicular to the reaction plane. There is also an overall
sign which is affecting $\tau^+$ spin indices of the spin-correlation matrices. This is moved in
all {\tt Tauola/TauSpinner} interfaces~\cite{Davidson:2010rw}  from the $\tau$ decay to spin-correlation matrices.
Also for many past calculations which we rely on as reference, the rest
frames of $\tau^\pm$ were chosen to have the common $z$ axis direction, parallel to $z$ axis of
reaction frame (where incoming partons are not defining $z$ direction). For calculations involving
NP, we have found that for present day authors it is most convenient to allow for distinct frame orientations
than that. The easiest way to solve this different conventions was to provide for {\tt TauSpinner} implementation
an adjustment internal routine.

There is also another adjustment, this time  for the $\tau^+$ polarimetric vector orientation, also of
its overall sign. At present,
this adjustment is shifted into $R_{ij} $ matrix redefinition too\footnote{It is done separately,
in different place of the code, just before the event weight calculation.}, even though it does not correspond to change of
its orientation, but is for the $\tau^+$ polarimetric vector. Finally
\begin{eqnarray}
\label{eq:frames}
 R_{tt}= \;\; R_{44}, & R_{tx}=     -R_{42}, & R_{ty}=     -R_{41},\;  R_{tz}=     -R_{43}, \nonumber \\
 R_{xt}=     -R_{24}, & R_{xx}= \;\; R_{22}, & R_{xy}= \;\; R_{21},\;  R_{xz}= \;\; R_{23},           \\
 R_{yt}= \;\; R_{14}, & R_{yx}=     -R_{12}, & R_{yy}=     -R_{11},\;  R_{yz}=     -R_{13}, \nonumber \\
 R_{zt}=     -R_{34}, & R_{zx}= \;\; R_{32}, & R_{zy}= \;\; R_{31},\;  R_{zz}= \;\; R_{33}. \nonumber 
  \end{eqnarray}

With this transformation, expressions of the $R_{ij} $ matrices of Eq.~(\ref{eq:frames}) with $i, j=t,x,y,z$, 
are used  in {\tt TauSpinner} event  reweighting algorithm discussed in Section~\ref{sec:algorithm}
for calculating weight implemented in {\tt TauSpinner} code and in Section~\ref{sec:numerical} for presenting numerical results.


\section{The reweighting algorithm for {\tt TauSpinner} }
\label{sec:algorithm}
The basis formalism  of {\tt TauSpinner} is documented in Ref.~\cite{Przedzinski:2018ett}, 
Section 2.2,  Eqs.~(7) to (12). We do not repeat details of this formalism here, nor details how kinematics
of the hard process is deciphered from kinematics of the $\tau$-decay products. We recall however
a few basic equations for calculating final weights, which allow one to take into account changes in the cross-section
and spin correlations in the SM and SM+NP models.

The  basic  equation in the calculation of the cross-section is
 \begin{eqnarray}
&d \sigma = \sum_{flavors} \int dx_1 \, dx_2 \, f(x_1,...)\, f(x_2,...) \, d\Omega^{parton\; level}_{prod} \; d\Omega_{\tau^+} \; d\Omega_{\tau^-} \nonumber \\
& \times \Bigl(\sum_{\lambda_1,  \lambda_2 }|{\cal M}^{prod}_{parton\; level}|^2 \Bigr)
 \Bigl(\sum_{\lambda_1 }|{\cal M}^{\tau^+}|^2 \Bigr)
 \Bigl(\sum_{\lambda_2 }|{\cal M}^{\tau^-}|^2 \Bigr) \, wt_{spin}, 
\label{eq:parton-level}
\end{eqnarray}
where $x_1$, $x_2$ denote fractions of the beam momenta carried by the partons, $f(x_1,...)$, $f(x_2,...)$
are parton distribution functions (PDF)s of the beams\footnote{The {\tt TauSpinner} algorithm does not use information of the flavour of incoming partons from the event record, allowing for application of its weight also on experimental data.}
and $d\Omega$ denote phase-space integration elements. 
Eq.~(\ref{eq:parton-level}) represents product of distribution for the $\tau^\pm$ production and decay, $ \Bigl(\sum_{\lambda_i }|{\cal M}^{\tau^\pm}|^2 \Bigr)$ stands for the decay matrix element squared, and 
$ \Bigl(\sum_{\lambda_1,  \lambda_2 }|{\cal M}^{prod}_{parton\; level}|^2 \Bigr)$ -- 
 for the production matrix element squared.
 Only the spin weight $wt_{spin}$ needs input both from $\tau^\pm$ production and decay.

The $R_{i j}$ used in calculation of components of Eq.~(\ref{eq:parton-level}) are taken as weighted average
(with PDFs and production matrix elements squared) 
over all flavour configurations, as in the following equation:
{\small
 \begin{eqnarray}
R_{i j} \to \frac{ \sum_{flav.} f(x_1,...)f(x_2,...) 
\Bigl(\sum_{\lambda_1,  \lambda_2 }|{\cal M}^{prod}_{parton\ level}|^2 \Bigr)  R_{i j} }
{\sum_{flav.}  f(x_1,...)f(x_2,...) 
\Bigl(\sum_{\lambda_1,  \lambda_2 }|{\cal M}^{prod}_{parton\ level}|^2 \Bigr)\;\;\;\;\;\;\;} . \label{eq:Rij-ave}
 \end{eqnarray}
}
No  approximation is introduced in this  way, the denominator of Eq.~(\ref{eq:Rij-ave}) cancels 
explicitly the corresponding factor of Eq.~(\ref{eq:parton-level}).

For the $wt_{spin}$ calculation, the normalised  elements $ r_{i j} = R_{i j}/R_{tt}$ are used, 
following Eq.~(\ref{eq:016})
\begin{equation}
wt_{spin} = \sum_{i ,j=t,x,y,z} r_{i j} h^i_{\tau^+} h^j_{\tau^-}. \label{eq:wtspin}
\end{equation}
Here $ h^i_{\tau^+}, \,  h^j_{\tau^-}$
stand for decay mode dependent $\tau^+, \, \tau^-$ polarimetric vectors.
This weight is the only term which needs input from both $\tau^\pm$ production and decay.
Please note, that the $wt_{spin}$ is independent of the PDF's, except through
already averaged over partons contribution  $ r_{i j}$ elements of the spin-correlations matrix which are used.
 
To introduce the corrections due to different spin effects and modified production
process in the generated sample (i.e. without re-generation of events), one can define the weight $wt$, representing
the ratio of the new to old cross-sections at each point in the phase space.

Eq.~(\ref{eq:parton-level}) for the modified cross-section takes then the form
 \begin{equation}
   d \sigma_{new} = d\sigma_{old} \; wt_{prod}^{new/old} \; wt_{spin}^{new/old} , 
	\label{eq:sigma_new}
 \end{equation}
 where $ \sigma_{old}$ is calculated with Eq.~(\ref{eq:015}),  $wt_{spin}$ using Eq.~(\ref{eq:wtspin}) and 
$ wt_{prod}^{new/old}$ using Eq.~(\ref{eq:wtprod}) below
 {\small
\begin{equation}
wt_{prod}^{new/old}=  \frac{\sum_{flav.}f(x_1,...)f(x_2,...)\Bigl(\sum_{spin }|{\cal M}^{prod}_{part. lev.}|^2\Bigr) \Big|_{new}}
{\sum_{flav.}f(x_1,...)f(x_2,...)\Bigl(\sum_{spin }|{\cal M}^{prod}_{part. lev.}|^2\Bigr) \Big|_{old}}\\
= \frac{R_{tt}|_{new}}{R_{tt}|_{old}}.
\label{eq:wtprod}
\end{equation}
The present implementation  assumes, that the generated sample has no spin correlations included, however it can
easily be extended\footnote{Such special case is available for $\bar q q \to \tau \tau$ processes, where, e.g.
polarisation but not spin correlations, was included in the generated sample.}
to provide weight calculated as
\begin{equation}
wt_{spin}^{new/old} =\frac{ \sum_{i ,j=t,x,y,z} r_{i j} h^i_{\tau^+} h^j_{\tau^-}\Big|_{new}}{ \sum_{i ,j=t,x,y,z} r_{i j} h^i_{\tau^+} h^j_{\tau^-}\Big|_{old}} . \label{eq:wtspinr}
\end{equation}

The {\tt TauSpinner} program provides both weights, $ wt_{spin}$ and $wt_{prod}$,  which allow one to modify
per-event distributions of sample generated according to $ d\sigma_{old}$ model and should be used as
multiplicative components.

The  combined weight should be used as multiplicative product as in Eq.~(\ref{eq:combi})
\begin{equation}
wt = {wt_{prod}^{new/old}} \; \times \; {wt_{spin}^{new/old}}, \label{eq:combi}
\end{equation}
where the first term of the weight represents modification of the matrix elements for production,
the second term -- of the spin correlations. It is nothing else than ratios of spin averaged amplitudes squared 
for the whole process; new to old. If the analysis is sensitive to changes in the PDFs parametrisations
used for sample generation and {\tt TauSpinner} weights calculations, it should be taken
into account in calculation of $wt_{prod}^{new/old}$ and  $wt_{spin}^{new/old}$ in Eq.~(\ref{eq:combi}).
In case the production process is not modified, $wt_{prod}^{new/old}$ is equal to 1.
In case of originally generated sample without spin correlations, $wt_{spin}^{new/old}$ alone allows to introduce
the desired spin effects. 

In Eq.~(\ref{eq:wtprod}), indirectly through~(\ref{eq:Rij-ave}) also in~(\ref{eq:wtspinr}), 
$\sum_{flav.}$ stands for including all components of the beam
which lead to the $\tau$ pair in the final state being produced.
For calculating weights  of  the $\gamma \gamma \to \tau^{-} \tau^{+}$ events, the sum over flavours
includes now also photons.
This requires  on one side,  structure functions available
for the quasi-real photon as a parton in the proton or heavy ion, and
on the other side, the $t$- and $u$-channels matrix element for the hard process $\gamma \gamma \to \tau^- \tau^+$,  with the spin correlations included. 

Let us now give more technical details on the implementation for the $\gamma \gamma \to \tau \tau$ process.
The  $\gamma \gamma \to \tau^-\tau^+$ hard process, its spin amplitudes, cross-section and spin-correlation
matrix were described in Section~\ref{sec:theory}.
The non-normalized spin-correlation matrix $R_{ij}$ of Eq.~(\ref{eq:frames}) contains all the necessary information
for calculating the cross-sections and introducing spin-correlation effects in the 
$\gamma \gamma \to \tau^{-} \tau^{+}$ events.
Each parton process contributes incoherently to the final
state. That is why, introduction of nearly real photon as an extra parton, 
and a corresponding hard process, were possible with straightforward extension  
of the sums in Eq.~(\ref{eq:parton-level}).
As for the set of the PDFs, in case of $pp$ collision, one could take the ones described, e.g. in Refs.~\cite{Klein:2016yzr, Xie:2021equ}. These structure functions include the photon PDF as well. Alternatively,
and that is our choice, recommended  for the
PbPb collision, where parametrisation of the photon flux is not easily available, we can simply request that 
$\gamma\gamma$ contributes in proportion to all other processes. 

In the present implementation, the $\tau\tau$ event are  analyzed as if produced
through combination of parton level $ \bar{q} q $ and  $\gamma \gamma$ processes, and  to contribute 
in the proportion to quark processes set by the user\footnote{%
Parameters {\tt GAMfraci, GAMfrac2i} set at the initialisation step.}.
This can be changed and the corresponding photon PDFs can be installed. At present,
the $R_{ij}$ elements of  $\bar q q$ part are averaged over all parton flavours  according to the density obtained from PDFs library. Then the  contribution from  $\gamma \gamma$ process  to  $R_{ij}$  is added  with a fixed proportion to the ones of quarks.
This may look as over-simplification, but we firstly avoid evaluation of the $\gamma$ PDFs variants, as the  $\gamma$
PDFs may be strongly experimental conditions dependent. Secondly, it gives the flexibility to use {\tt TauSpinner} weights for  the $\gamma \gamma \to  \tau\tau $ produced in PbPb collisions, where the issue of incoming photon fluxes
are modeled with specialized MC generators and are sensitive to experimental conditions.
In fact, we assume that the case, in which only the $\gamma\gamma$ process contributes, is unphysical, as there 
are always accompanying processes from the $\bar q q$ interactions, also in PbPb beam case.
User interested nonetheless in such a case, can always set parameters {\tt GAMfraci, GAMfrac2i} to very large values,
so the quark process contribution is way smaller than measurements ambiguity threshold. 
 


\section{ Numerical results} 
\label{sec:numerical}

Numerical results presented below are based on events generated with {\tt Pythia 8.3}~\cite{Bierlich:2022pfr}
using $pp$ scattering at 13 TeV, hard process {\tt PhotonCollision:gmgm2tautau}, internal parametrisation of structure functions
{\tt PDF:pSet = 13} and  restricted to low mass range of the $\tau\tau$ pair, 
$m_{\tau\tau} =$ 5-50 GeV and $p_T^{\tau\tau} >$ 5 GeV.
This choice of phase-space corresponds roughly to the range
covered by the $\gamma \gamma \rightarrow \tau \tau$ processes in PbPb collisions at the LHC. 
Then, the $\tau$ decays were modeled with {\tt Tauola} decay library~\cite{Jadach:1993hs} with no 
spin correlations between decaying $\tau$ leptons assumed. 
In total, we have available about $0.8 \times 10^6$ events for each decay mode combination generated.
Then, the spin correlations were added using weight calculated with {\tt TauSpinner} program discussed
in Section~\ref{sec:algorithm}, both for the spin-correlation effects in the SM and SM+NP models, and the cross-section
normalisation in SM+NP models.

In Fig.~\ref{Fig:mtautau_costheta} distributions of invariant mass $m_{\tau\tau}$ and $\cos\theta$ of the scattering
angle for the generated sample are shown. Using the weight calculated with {\tt TauSpinner}, the distributions
are shown for the SM and two SM+NP models: $A=0.02$, $B=0.0$ and $A=0.0$, $B=0.02$.
Effect from the NP models on these distributions is small and will be quantified with Table~\ref{Tab:WTsec}
discussed later.
The $\tau$ leptons are decayed in following decay mode configurations: both $\tau \to \pi \nu_\tau$,
both $\tau \to \rho \nu_\tau$  and one $\tau \to \mu \nu_\tau \nu_\mu$ with another one $\tau \to \pi \nu_\tau$.

\begin{figure} 
  \begin{center}                               
{
  \includegraphics[width=7.5cm,angle=0]{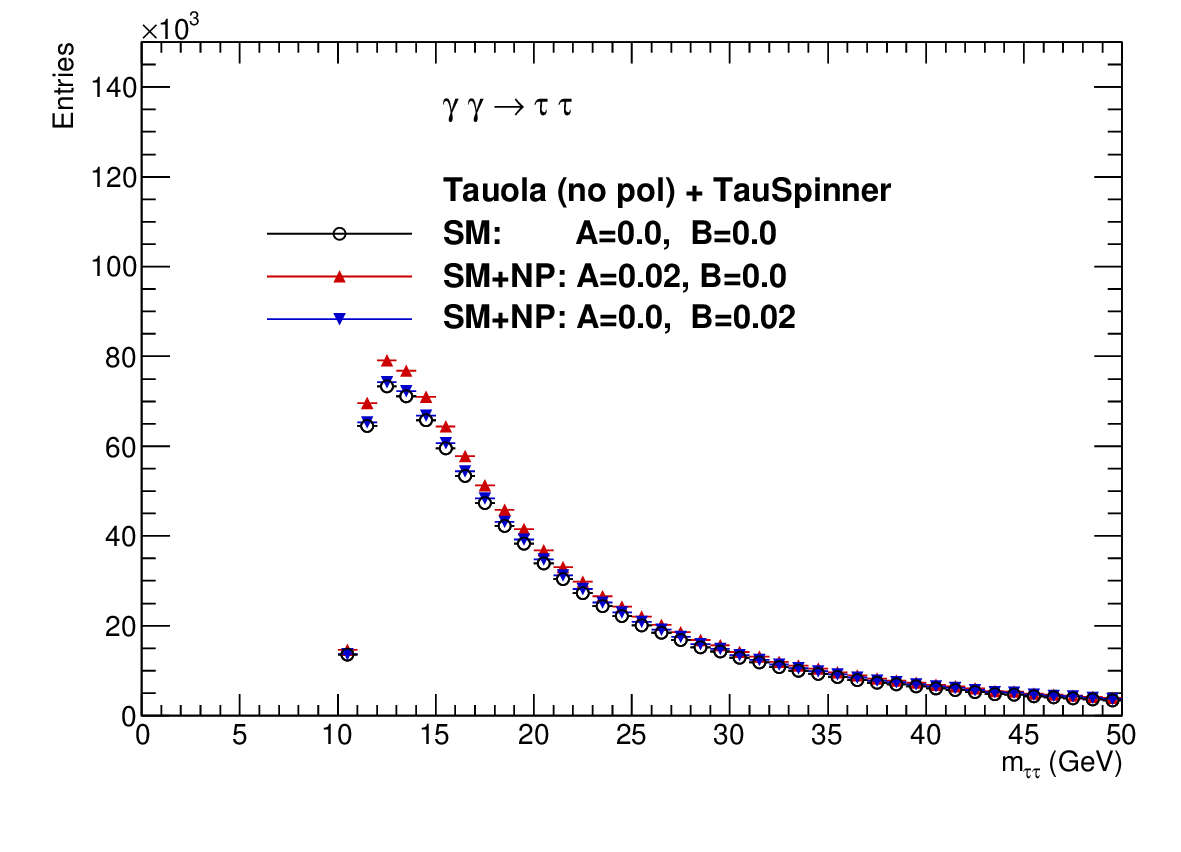}
  \includegraphics[width=7.5cm,angle=0]{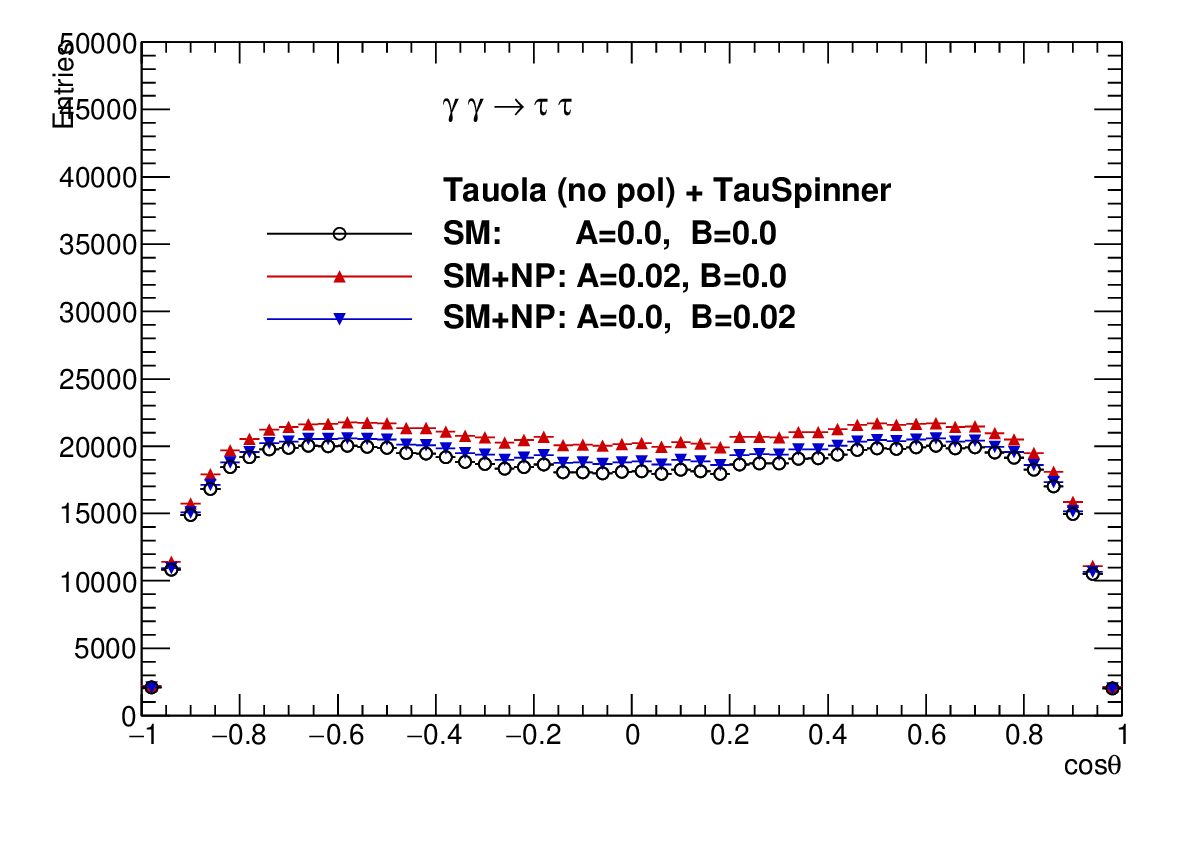}
}
\end{center}
  \caption{ Distribution of invariant mass of the $\tau\tau$ system and $\cos\theta$ of the scattering angle
    for the analyzed sample of $\gamma \gamma \to \tau \tau$ events.
 \label{Fig:mtautau_costheta} }
\end{figure}

In the following we discuss numerical results for elements of spin-correlation matrix  $R_{ij}$
and identify, which component may be the most sensitive to dipole moments and provide
some evidence of NP.
Then we move to discussing impact on a few kinematical variables, typically studied in the
experimental analysis. The aim is to quantify effect from spin correlations as present in the SM and then
impact from NP extensions due to the anomalous dipole moments.


\subsection{\texorpdfstring{Spin-correlation elements $R_{ij}$}{}}
\label{sec:Rij}
To simplify discussion we use as a reference the SM with $A=B=0$. It is assumed that
for the application, dipole moments due to the SM loop corrections are small and will be dropped out
when discussing the NP effects.  Therefore for numerical results, we compare the SM predictions
with six different settings  for the dipole
moments: (i) $A=0.002$, $0005$, $0.02$ with $B=0$, and (ii) $A=0$ with $B=0.002$, $0.005$, $0.02$.
This choice is somewhat arbitrary, covering the range which is plausible and
behaving numerically stable with higher powers of $A$ and $B$ terms included.
The smallest considered value $A=0.002$ roughly corresponds to $2 \times A_{SM}$.

Impact on the cross-section is quantified in Table~\ref{Tab:WTsec}. For the same values of anomalous
$A$ and $B$, the impact on the cross-section is few times bigger (of the order of 5) from $A$ than from $B$,
being about 0.6\% for $A = 0.002$ corresponding to $\sim 2 \times A_{SM}$,  and about 9\% for $A = 0.02$ corresponding to $\sim 20 \times A_{SM}$.
Correspondingly the impact of $B$ is below 0.1\% for  $B=0.002$, and 3\% for  $B=0.02$.

\begin{table}
 \vspace{2mm}
 \caption{The $\sigma^{SM+NP}/\sigma^{SM}$  for scan of different SM+NP models.}
 \label{Tab:WTsec}
 \begin{center}
     \begin{tabular}{|l||r|}
     \hline\hline
     NP model  & $\sigma^{SM+NP}/\sigma^{SM}$ \\
     \hline
      $A=0.002$; $B=0.0$  & 1.006   $\pm$ 0.001     \\ 
      $A=0.005$; $B=0.0$  & 1.017   $\pm$ 0.001     \\ 
      $A=0.020$; $B=0.0$  & 1.091   $\pm$ 0.001     \\ 
     \hline
      $A=0.0$; $B=0.002$  & 1.000   $\pm$  0.001     \\ 
      $A=0.0$; $B=0.005$  & 1.002   $\pm$  0.001     \\ 
      $A=0.0$; $B=0.020$  & 1.030   $\pm$  0.001     \\ 
     \hline
    \end{tabular}
 \end{center}
 \vspace {1cm}
\end{table}

Let us remind that we expect the effect from the NP extension of the SM 
to be very small, given the plausible range of the parameters $A$ and $B$.
But the spin correlations in the SM should not be neglected in the first place when extracting
limits from experimental data analysis.

Figure~\ref{Fig:Rtt} shows distribution of $R_{tt}$ as a function of $m_{\tau \tau}$ (top plots),
restricted to range  $\theta =  \pi/3 \times [0.8-1.2]$  (middle plots) and restricted
to  $\theta =  2\pi/3 \times [0.8-1.2]$ (bottom plots). By construction, for the SM $R_{tt}$ is chosen 1.0, and
it increases with the NP dipole moments.
The element $R_{tt}$ is more sensitive to changes
in the magnetic dipole moment $A$ than in the electric one $B$, and for both the sensitivity increases
with increasing invariant mass of the $\tau\tau$ system.
We observe that effects are larger and of the same sign for the middle and bottom plots, than for the top one,
which is integrated over full range of $\theta$. This indicates  that regions of $\theta$ close to $\pi/2$ are of lesser
sensitivity to $A$ and $B$. 
At the highest mass point studied,
$m_{\tau \tau}$ = 50~GeV, and integrated over the full phase space, $R_{tt}$ reaches about 20\% for
$A=0.02$ and 10\% for $B=0.02$. When restricting scattering angle to the range of
$\theta =  \pi/3 \times [0.8-1.2]$, the effect is magnified, reaching 25\% and 15\%, respectively.
In the opposite hemisphere, $\theta =  2\pi/3 \times [0.8-1.2]$, the effect is even larger, about 35\% and 25\%
respectively, indicating some asymmetry which can be explored further with experimental analysis.
Note, however, that such significant effects are observed for non-realistically large values of $A$ and $B$.

Fig.~\ref{Fig:Rij_a} shows elements $r_{ij} = R_{ij}/R_{tt}$ in the SM 
and discussed above SM+NP models. The diagonal elements of $r_{ij}$ are sizable (top and middle lines), 
in bottom line shown is $r_{xy}$ element in which the SM contributions are close to zero but NP ones are not.
In the left column the effect from varying $A$ is shown, on the right column -- from varying $B$.
For $A=0.02$, some shift is observed on $r_{yy}$, largely independent on $m_{\tau\tau}$.
For $B=0.02$, largely independent on $m_{\tau\tau}$ effect is visible in $r_{xy}$
(which has zero contribution from the SM) and
some effect on $r_{zz}$ at the higher end of the studied $m_{\tau\tau}$ range.
The other non-diagonal elements $r_{ij}$ have nearly zero SM contributions, then the dipole moment contributions are tiny as well.  That is why we have dropped out $r_{zx}$ and $r_{zy}$ plots.

In Figs.~\ref{Fig:Rij_a_costheta},~\ref{Fig:Rij_b_costheta} we show effect on $r_{ij}$
in function of $\cos\theta$. These plots help to identify details of anomalous coupling
dependence, which were observed in Fig.~\ref{Fig:Rtt}. 

We can conclude that the SM spin correlations are in many cases sizable and dominate over impact from the
dipole moments. But it is not always the case, for example, $r_{xy}$ and $r_{zx}$ at certain angles are zero in the SM,
but attain contribution from NP. That may give some hints on how to optimize choice of observables
and minimize background, at the same time underlying importance of spin correlations as possible bias
for the cross-section $R_{tt}$ based signatures.

\begin{figure} 
  \begin{center}                               
{
  \includegraphics[width=7.5cm,angle=0]{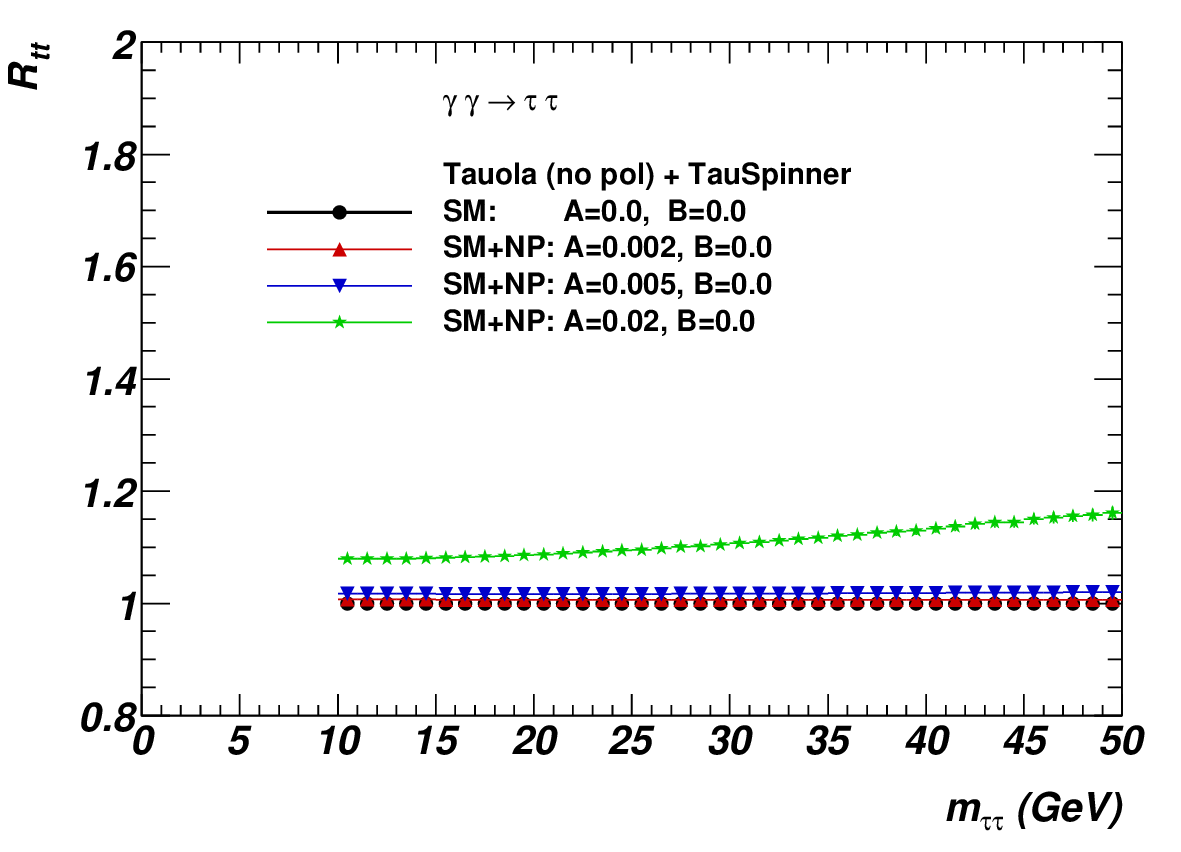}
  \includegraphics[width=7.5cm,angle=0]{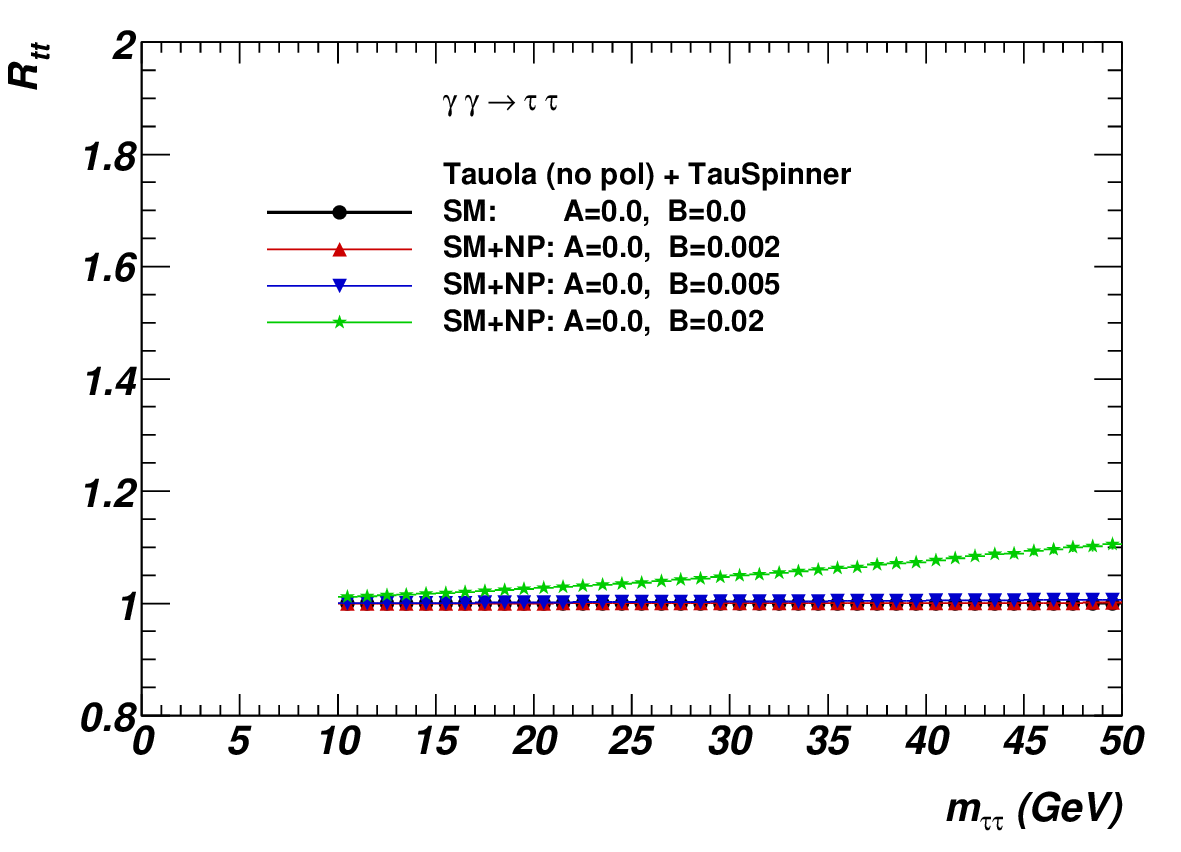}
  \includegraphics[width=7.5cm,angle=0]{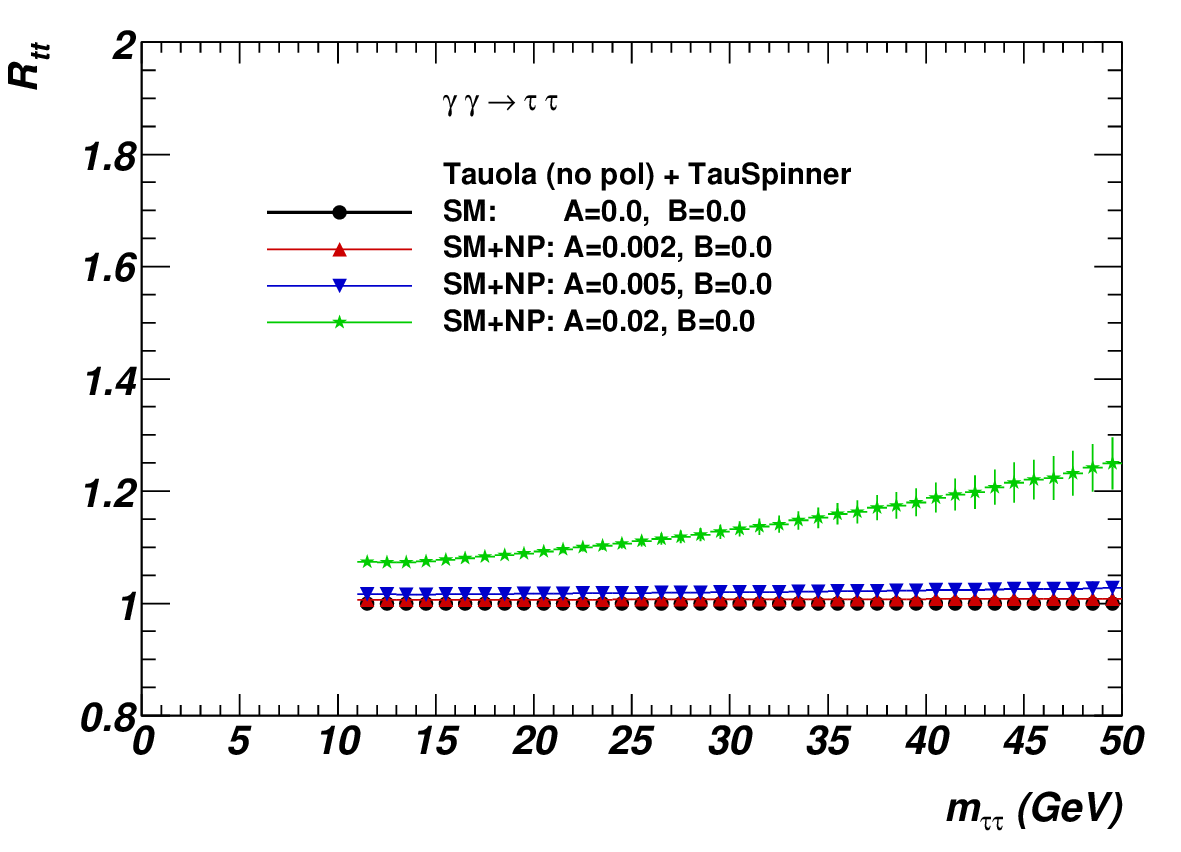}
  \includegraphics[width=7.5cm,angle=0]{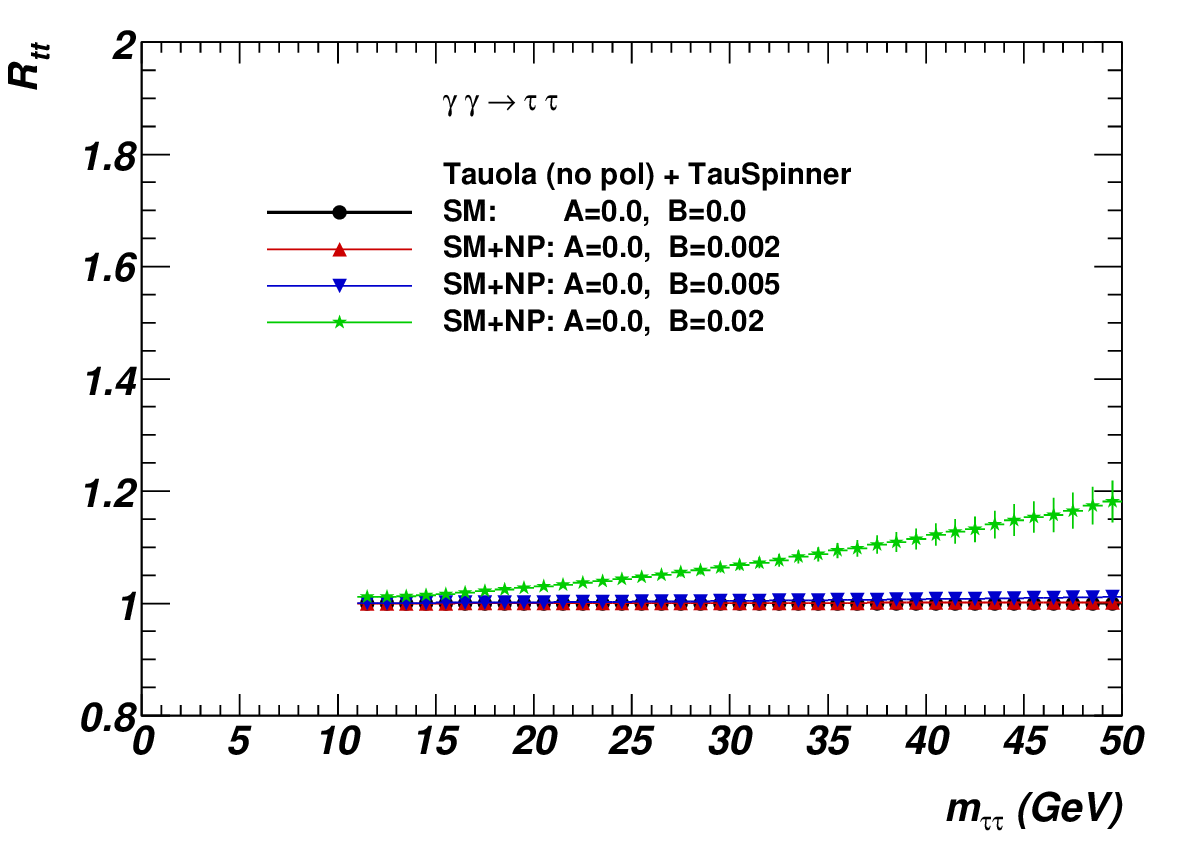}
  \includegraphics[width=7.5cm,angle=0]{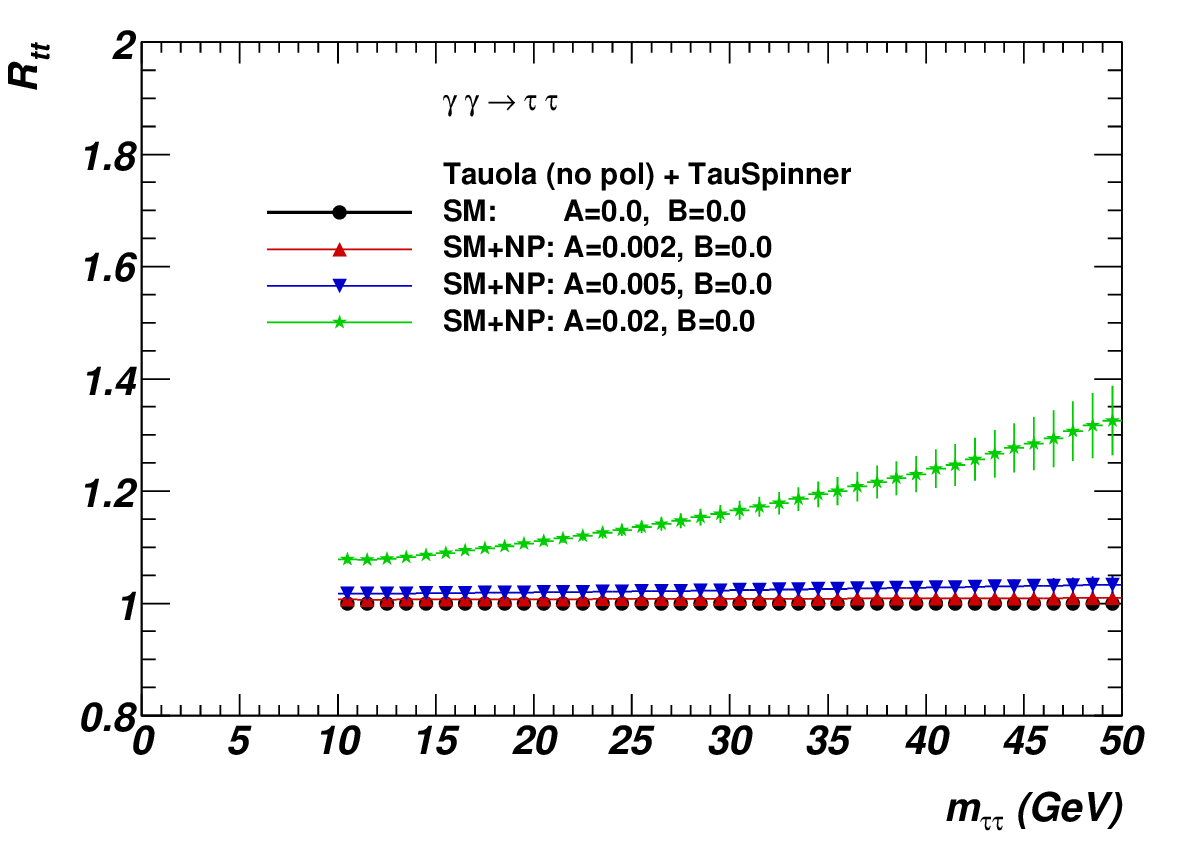}
  \includegraphics[width=7.5cm,angle=0]{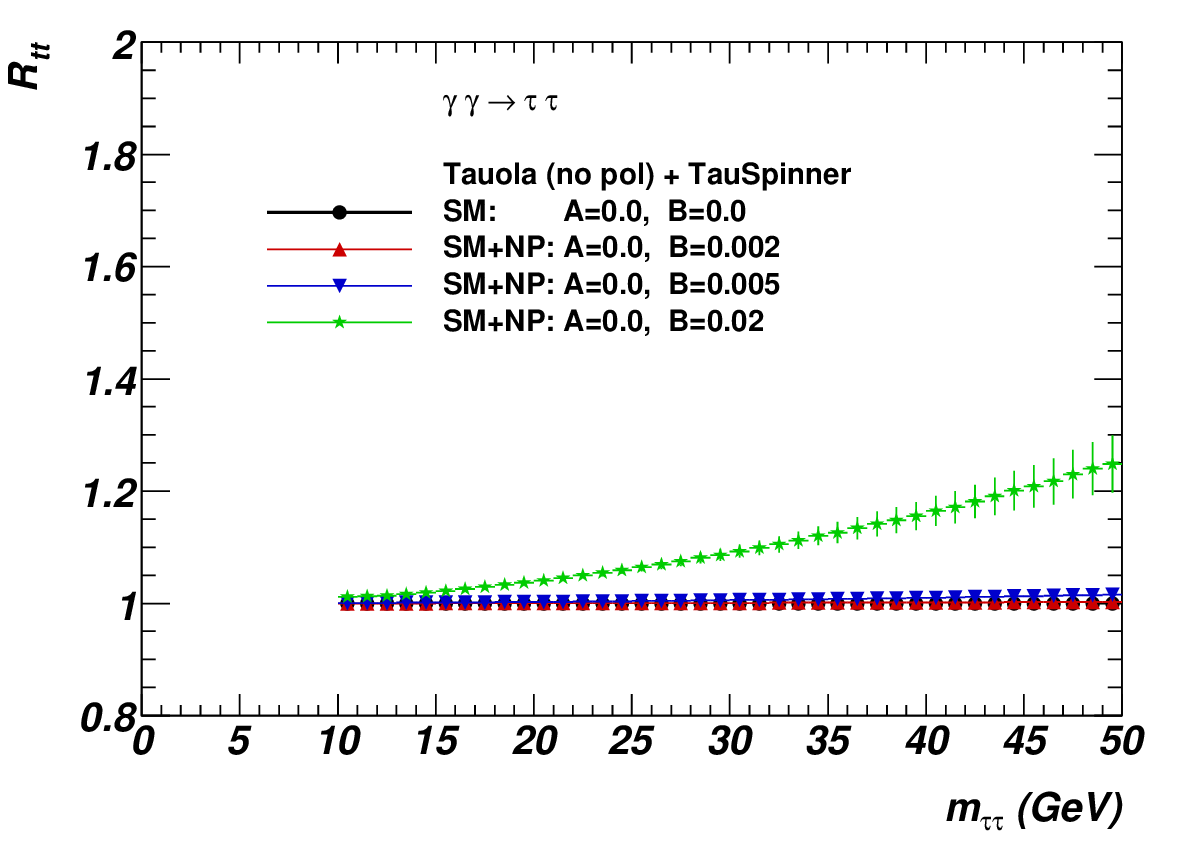}
}
\end{center}
  \caption{Spin averaged element $R_{tt}$ as a function of $m_{\tau \tau}$: integrated over 
	the full phase space (top plots),    restricted to $\theta =  \pi/3 \times [0.8-1.2]$  (middle plots) and restricted 
	to $\theta =  2\pi/3 \times [0.8-1.2]$ (bottom plots). Compared SM and SM+NP with six models:
   $A=0.002$, $0.005$, $0.02$ and $B=0$ (left column) and   $A=0.0$, $B=0.002$, $0.005$, $0.02$ (right column).
   Curves marked with $\star$ (shown in green) always denote the largest
anomalous moment: $A=0.02$ (left column), or $B=0.02$ (right column).
 \label{Fig:Rtt} }
\end{figure}

\begin{figure} 
\centering                               
{
  \includegraphics[width=7.5cm,angle=0]{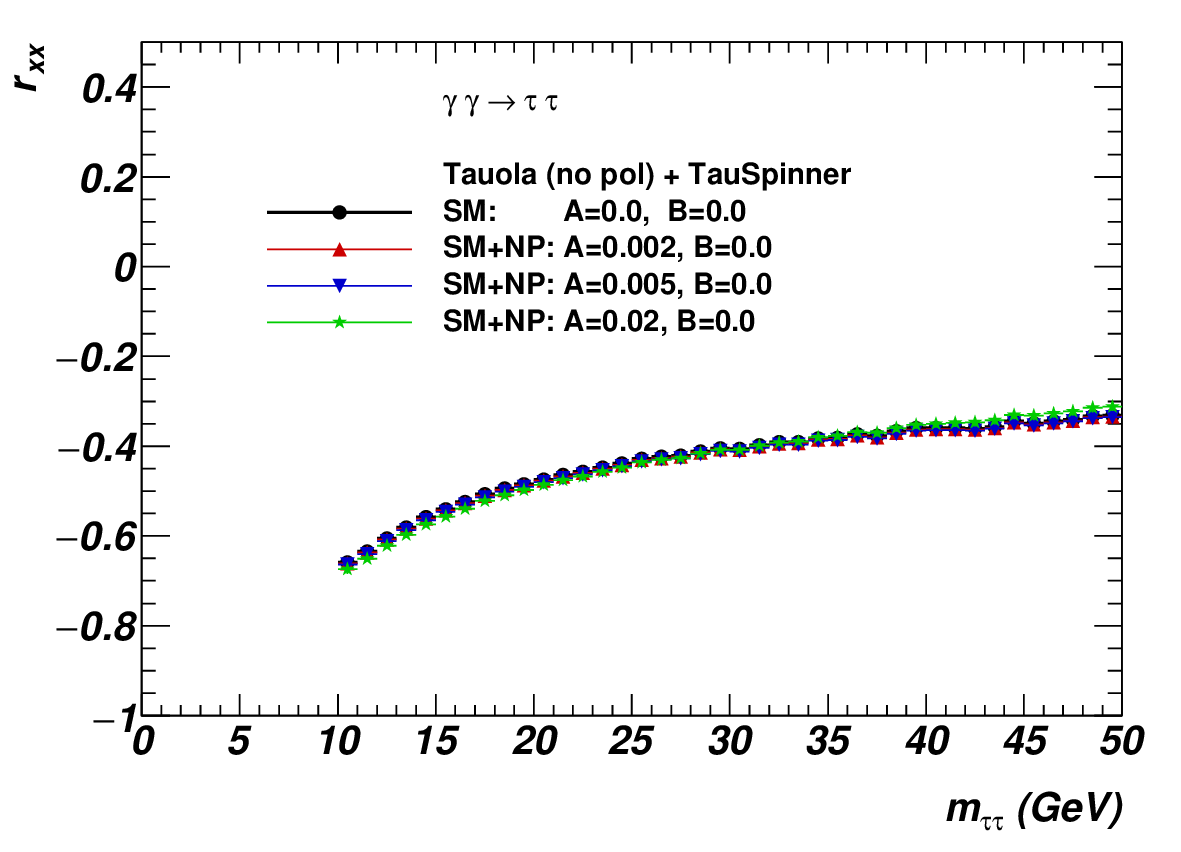}
  \includegraphics[width=7.5cm,angle=0]{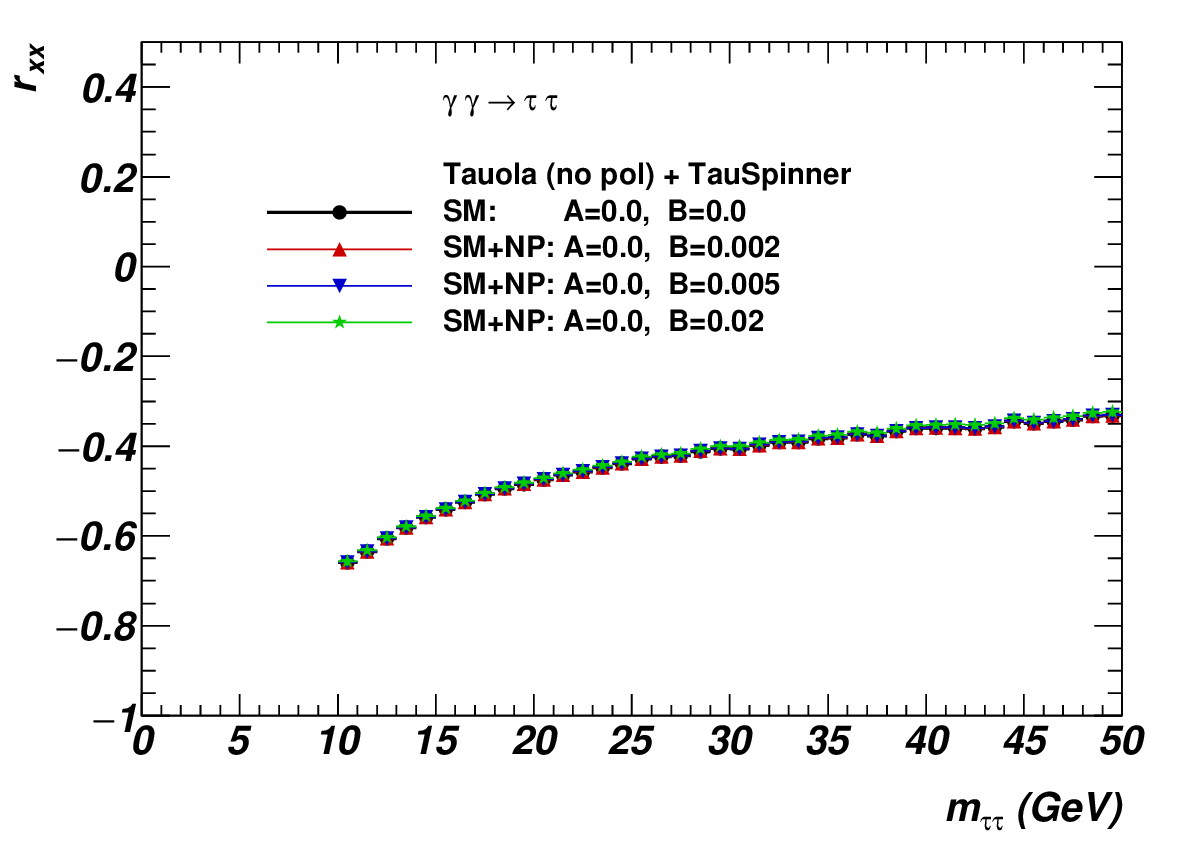}
  \includegraphics[width=7.5cm,angle=0]{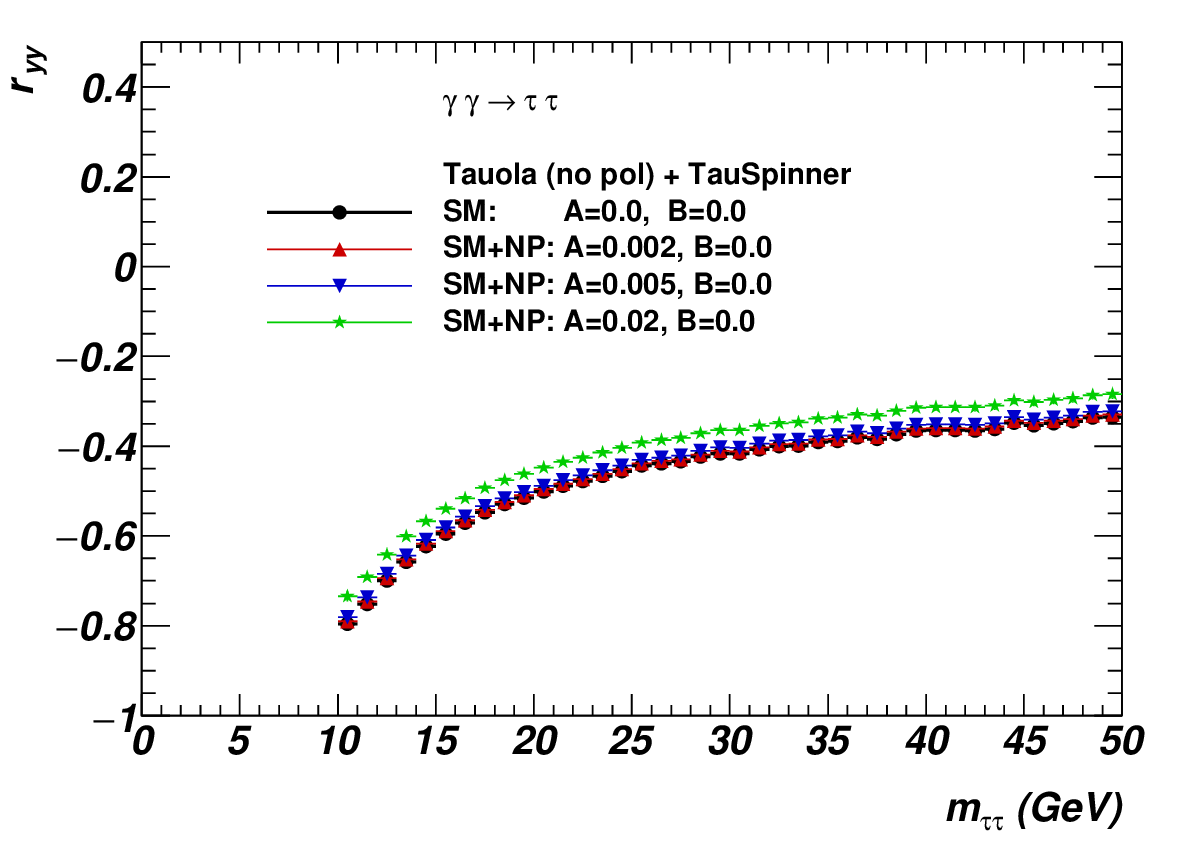}
  \includegraphics[width=7.5cm,angle=0]{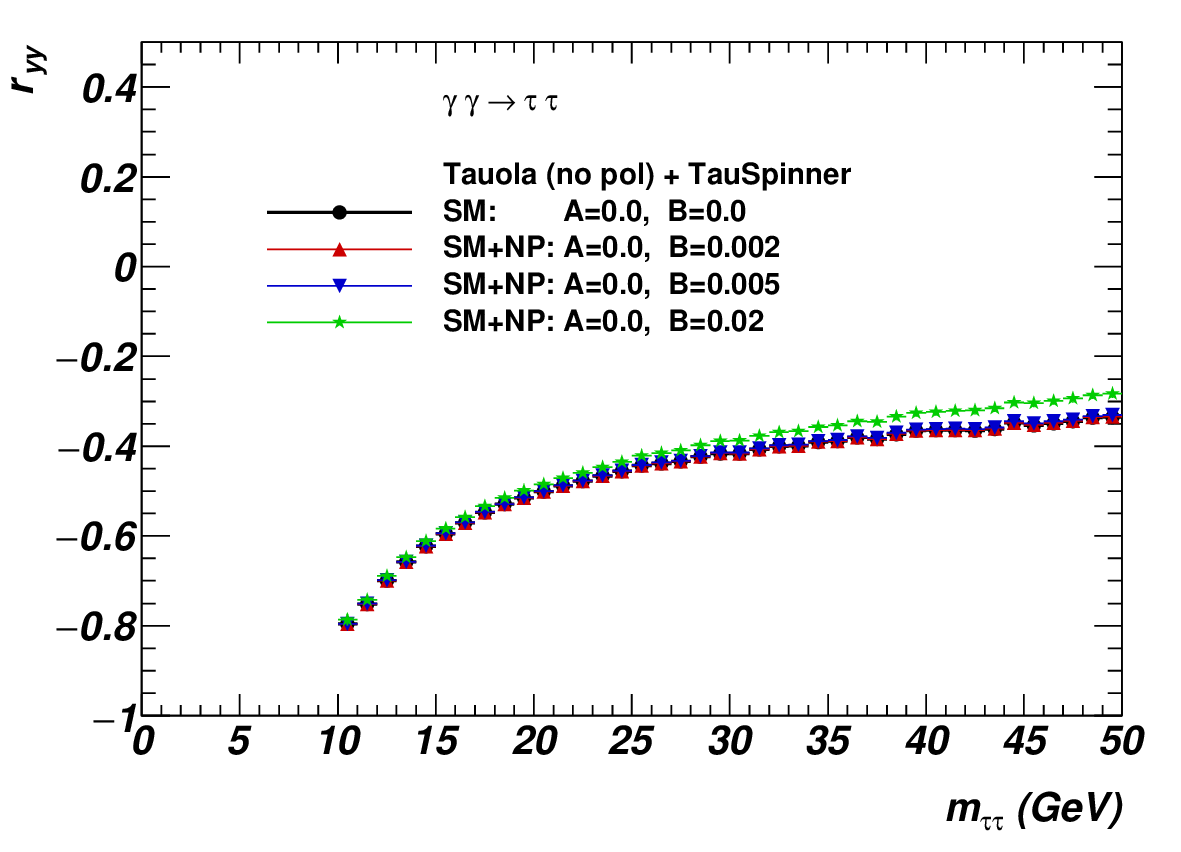}
  \includegraphics[width=7.5cm,angle=0]{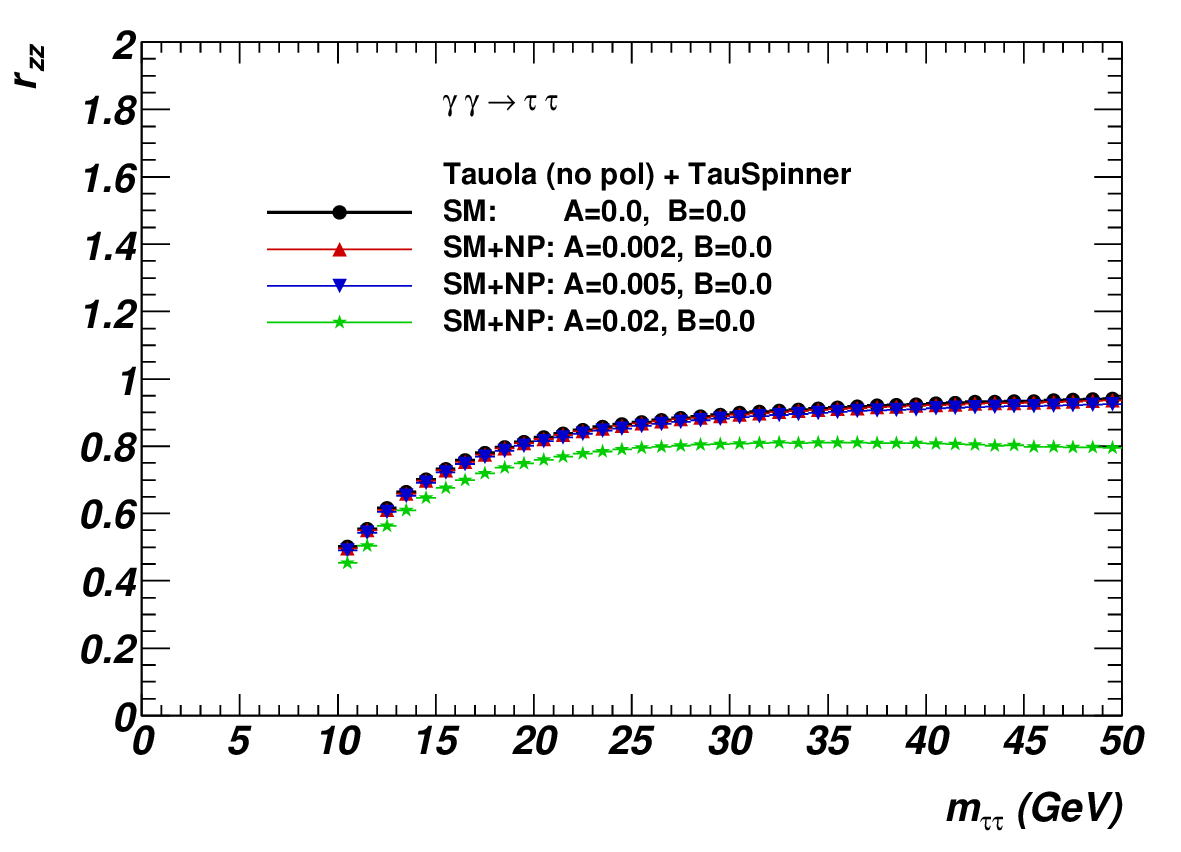}
  \includegraphics[width=7.5cm,angle=0]{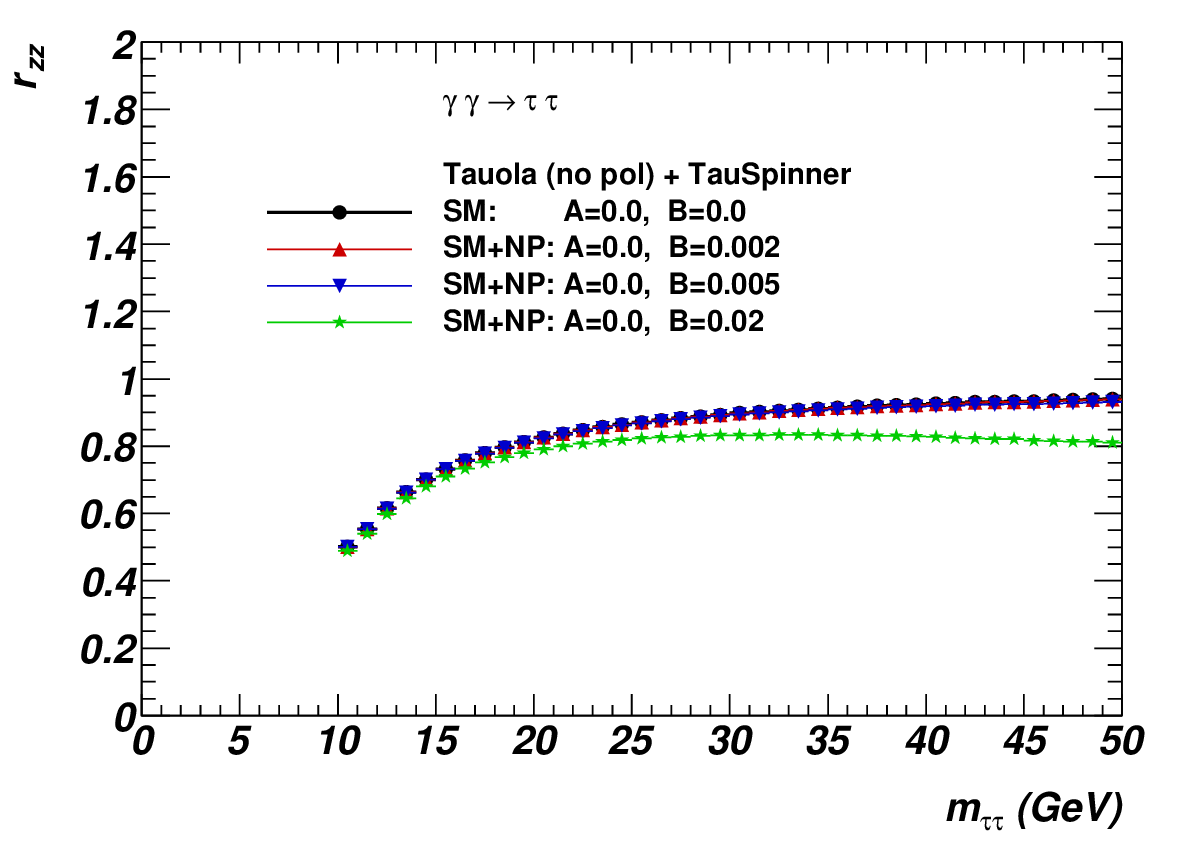}
  \includegraphics[width=7.5cm,angle=0]{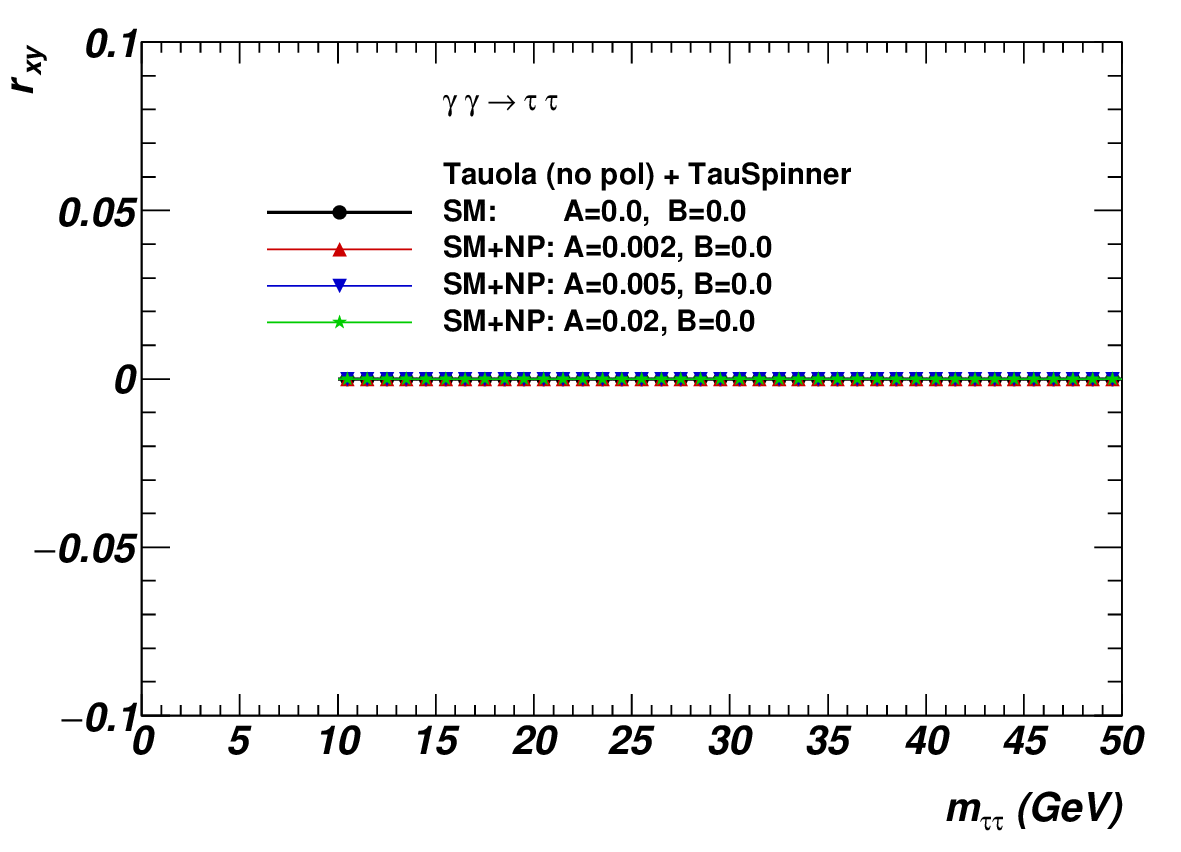}
  \includegraphics[width=7.5cm,angle=0]{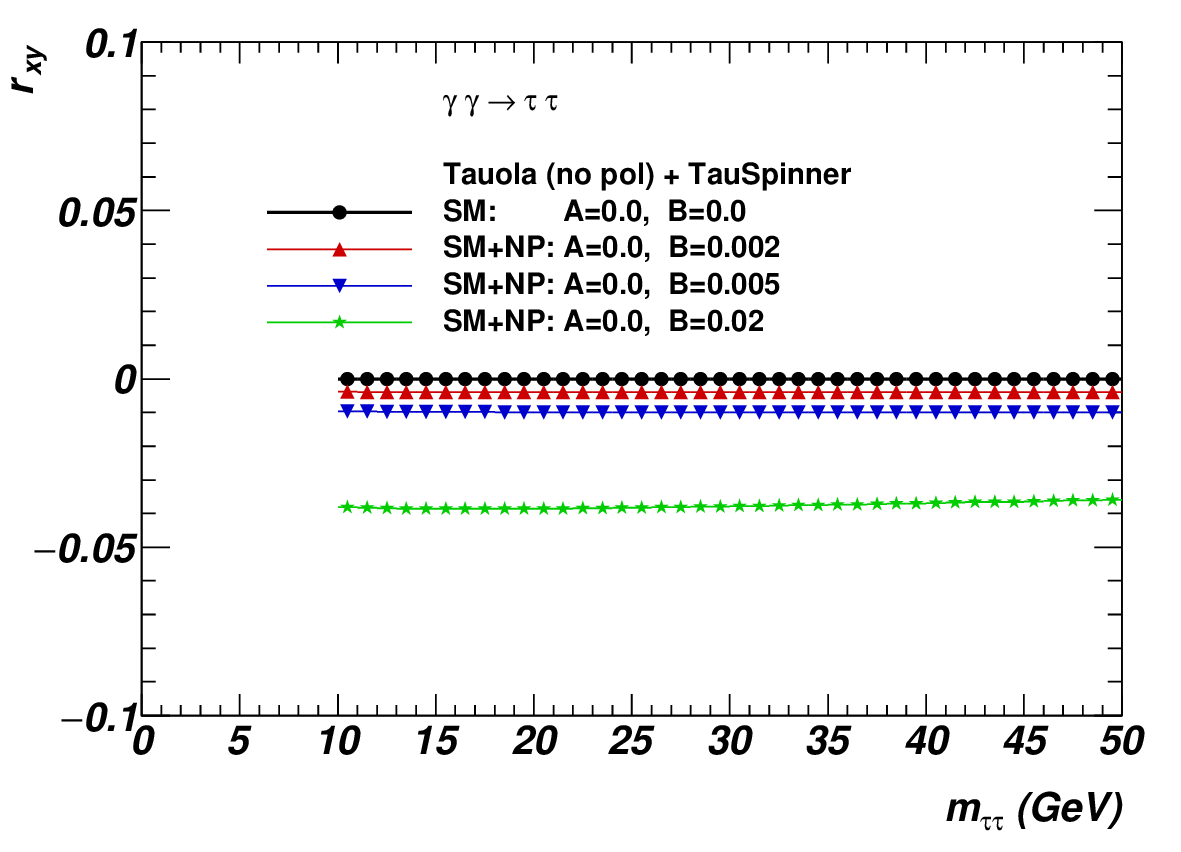}
}
  \caption{Spin-correlation matrix elements $r_{xx}$, $r_{yy}$, $r_{zz}$ and $r_{xy}$ as functions of $m_{\tau \tau}$.  Notation is the same as in Fig.~\ref{Fig:Rtt}. Except $r_{xy}$, these are elements with sizable SM contributions.
 \label{Fig:Rij_a} }
\end{figure}

\begin{figure} 
  \begin{center}                                
{
   \includegraphics[width=7.5cm,angle=0]{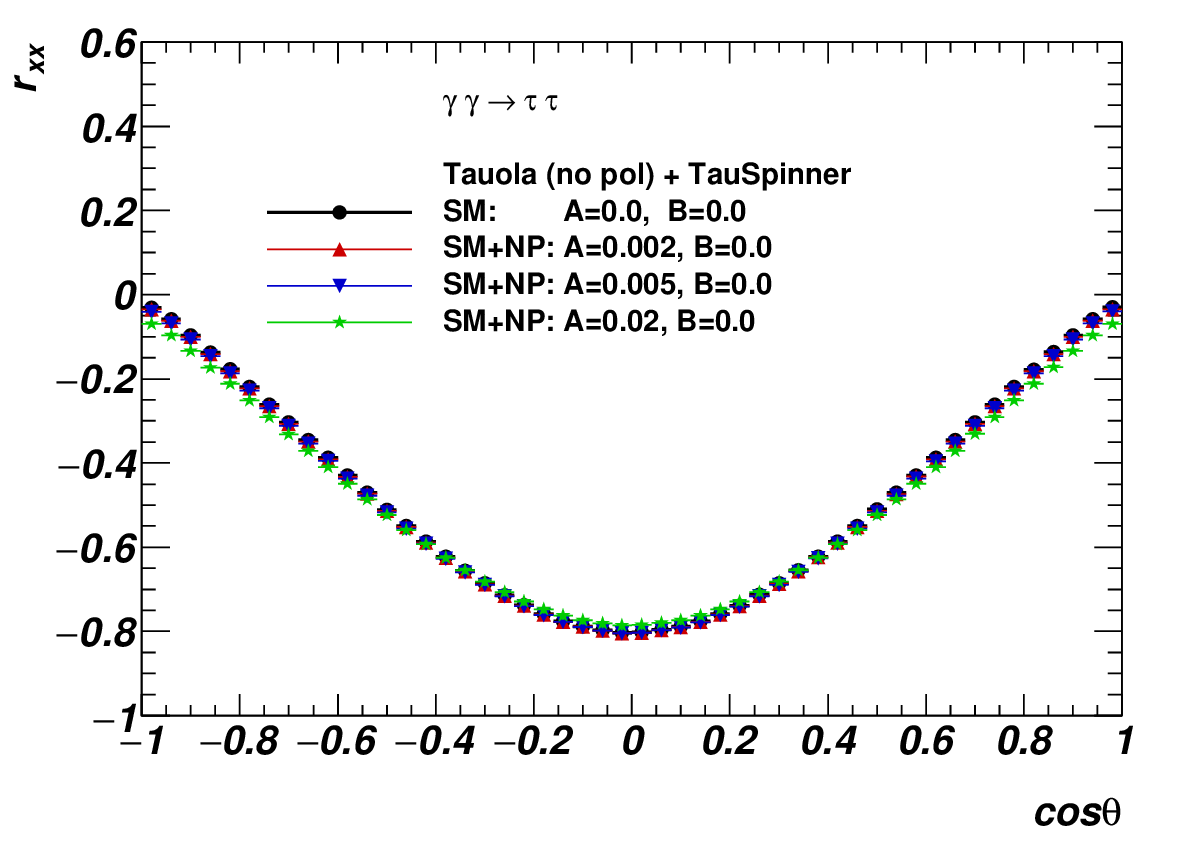}
   \includegraphics[width=7.5cm,angle=0]{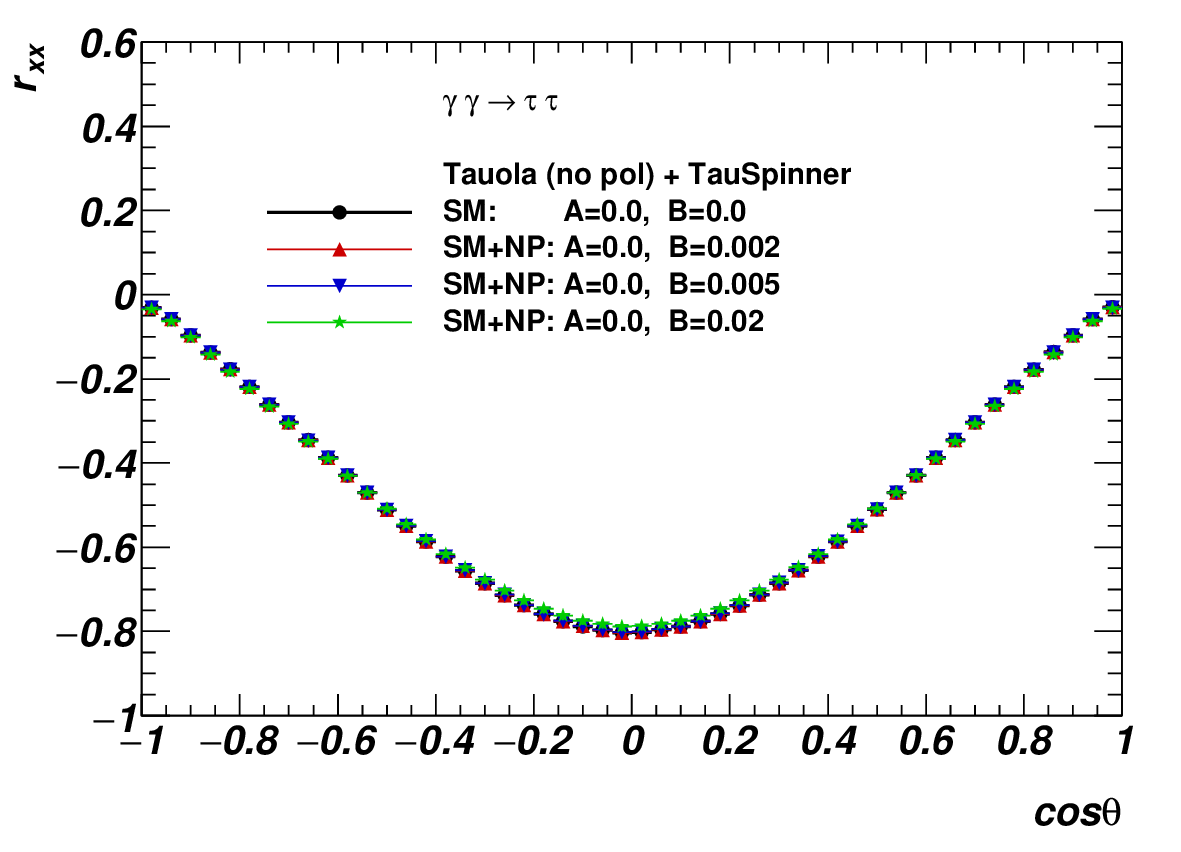}
   \includegraphics[width=7.5cm,angle=0]{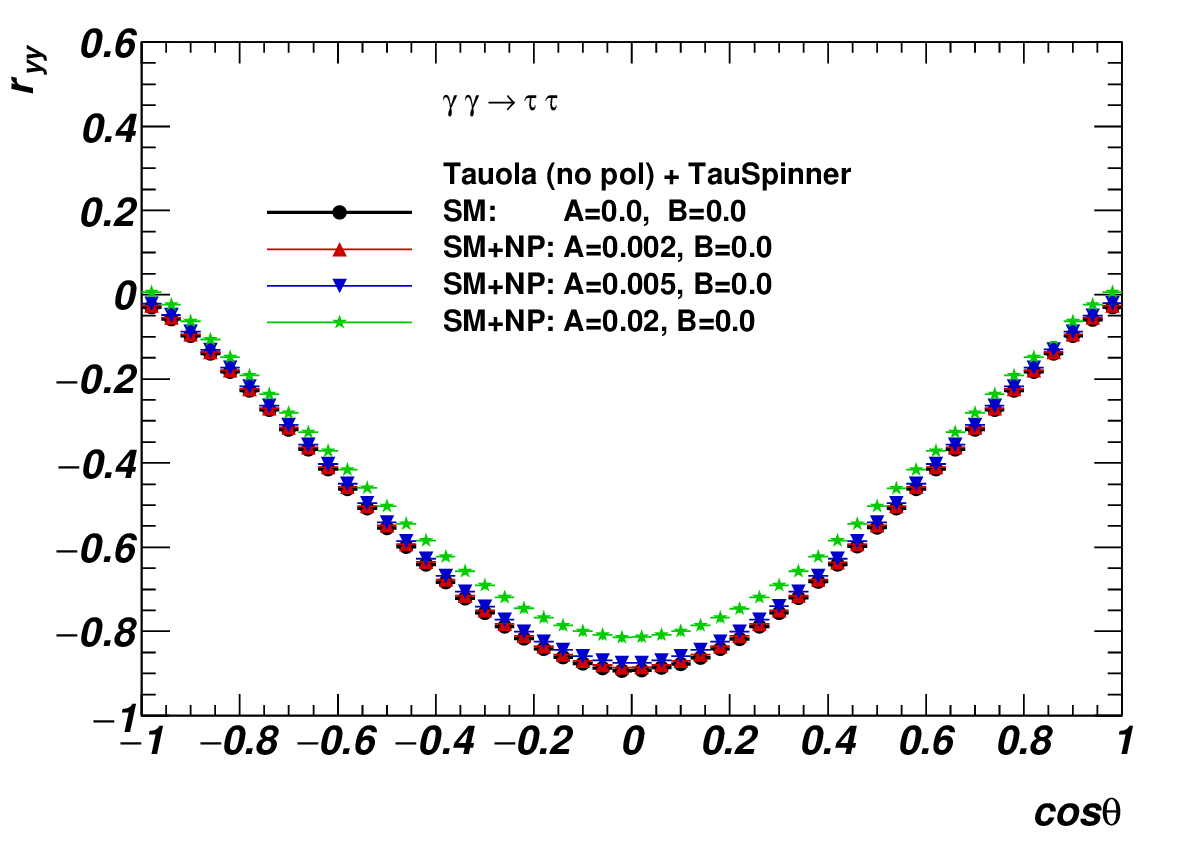}
   \includegraphics[width=7.5cm,angle=0]{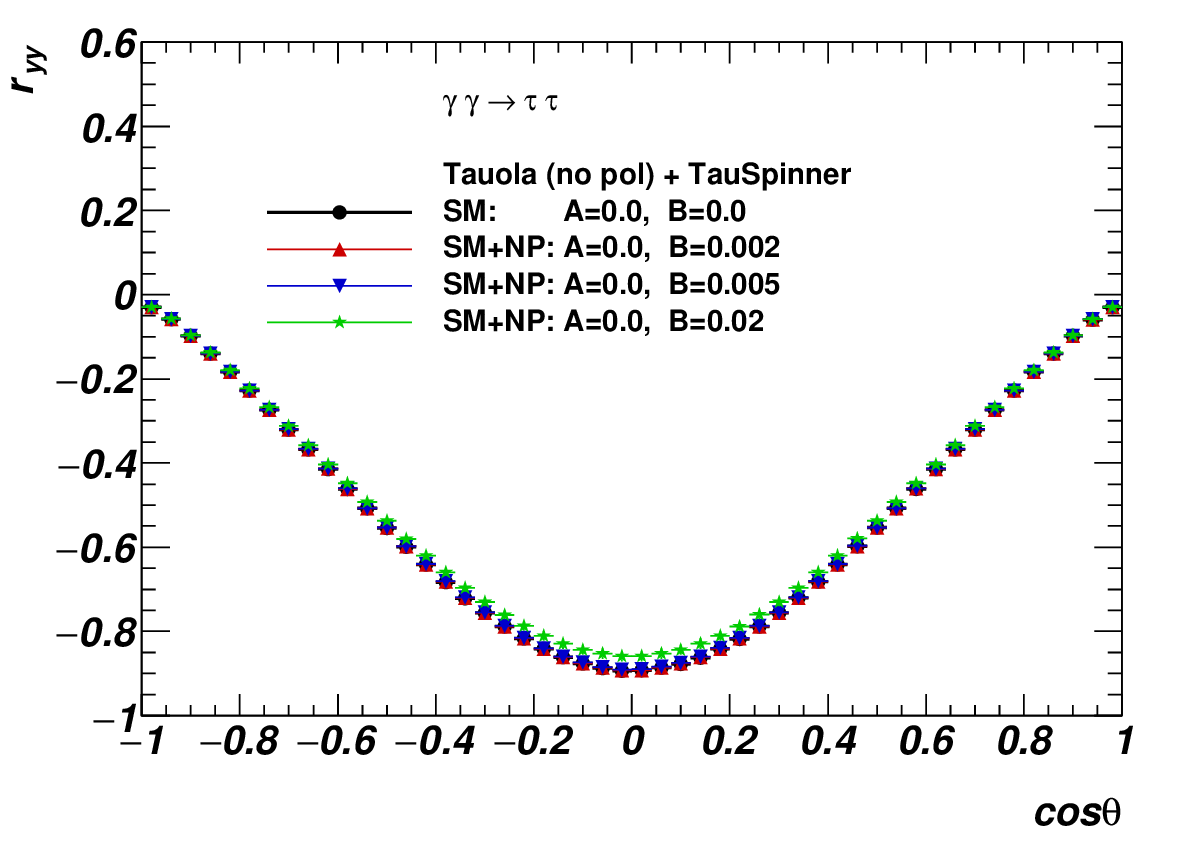}
   \includegraphics[width=7.5cm,angle=0]{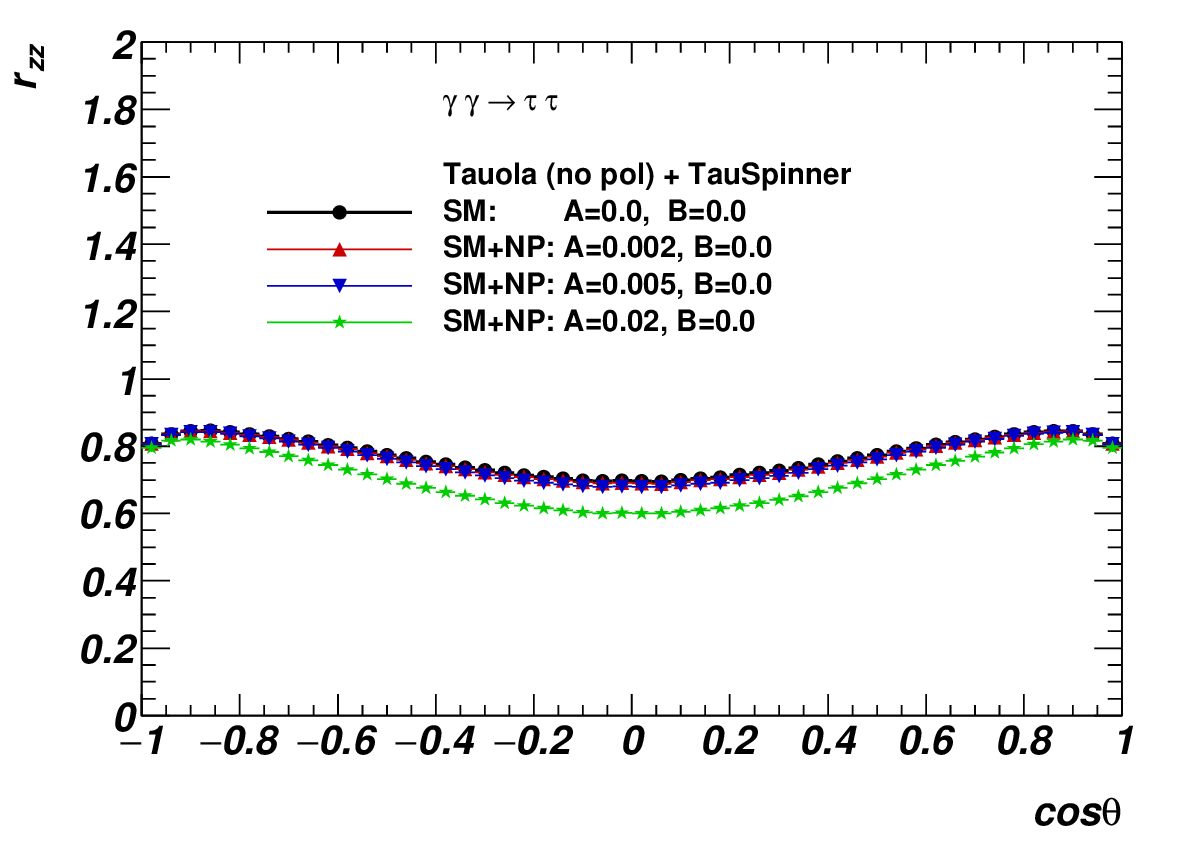}
   \includegraphics[width=7.5cm,angle=0]{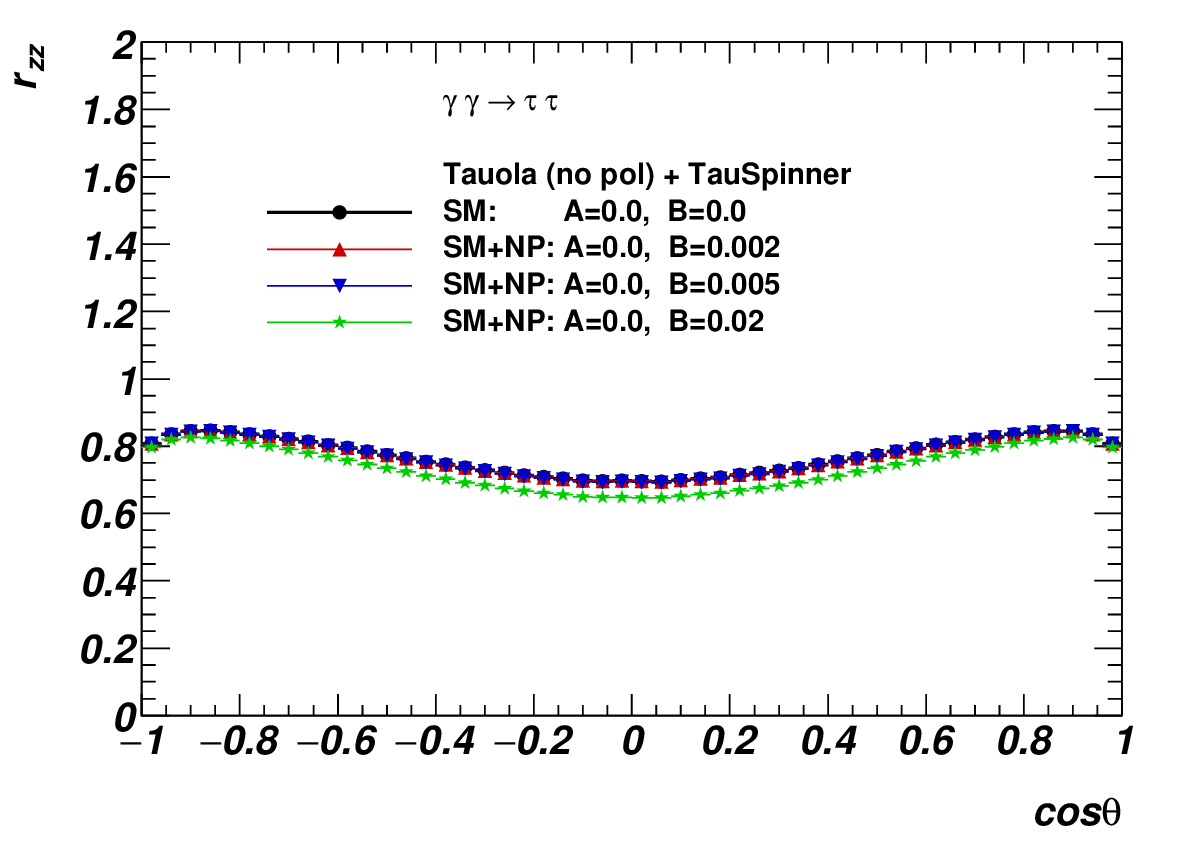}
}
\end{center}
  \caption{Spin-correlation matrix elements $r_{xx}$, $r_{yy}$ and $r_{zz}$ as functions of $\cos\theta$.
Notation is the same as in Fig.~\ref{Fig:Rtt}. These elements have sizable contributions from the SM.
 \label{Fig:Rij_a_costheta} }
\end{figure}

\begin{figure} 
  \begin{center}                               
{
  \includegraphics[width=7.5cm,angle=0]{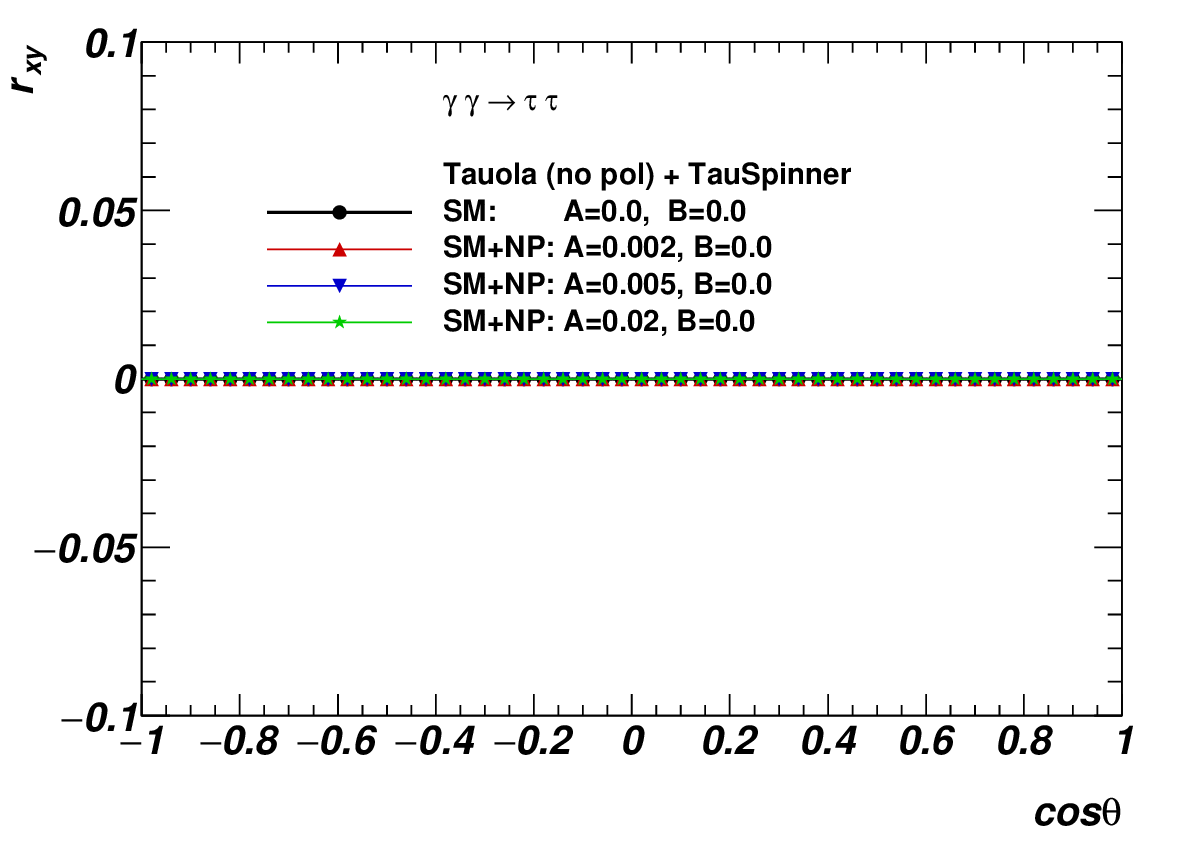}
  \includegraphics[width=7.5cm,angle=0]{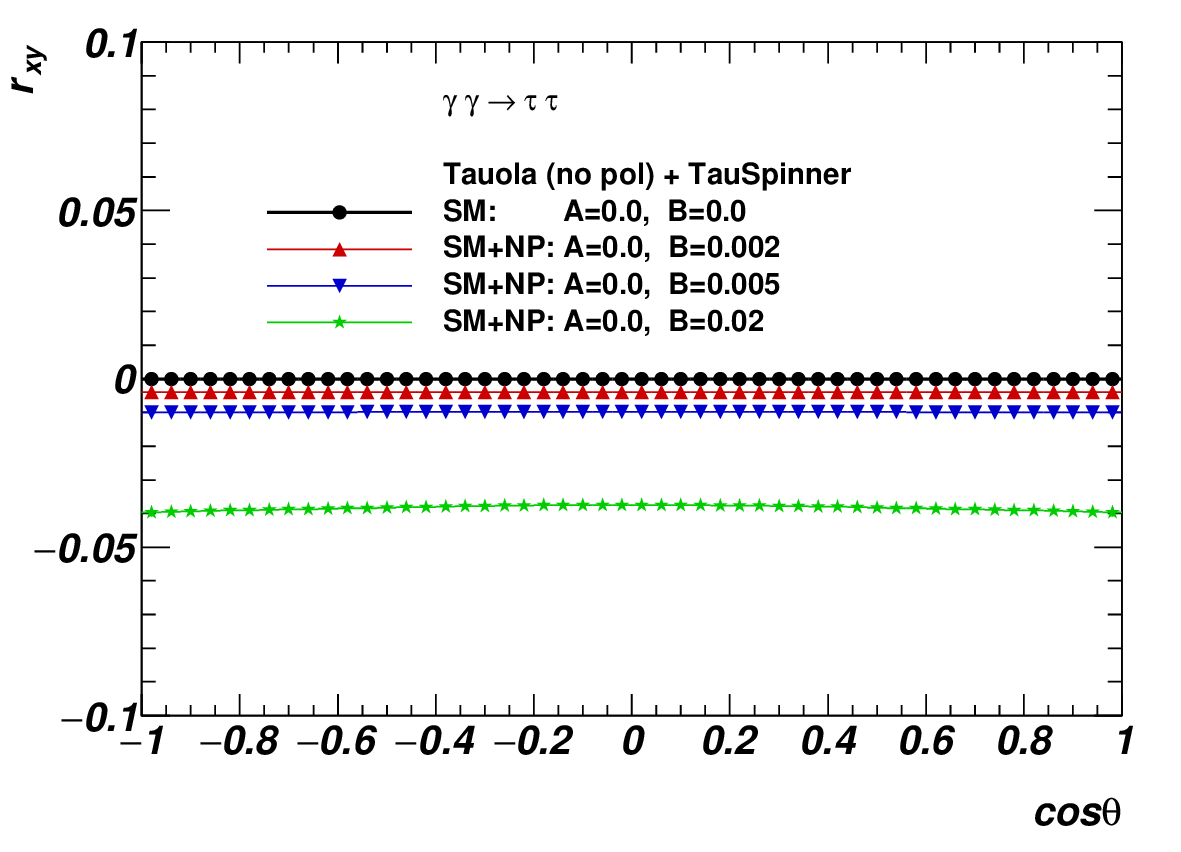}
  \includegraphics[width=7.5cm,angle=0]{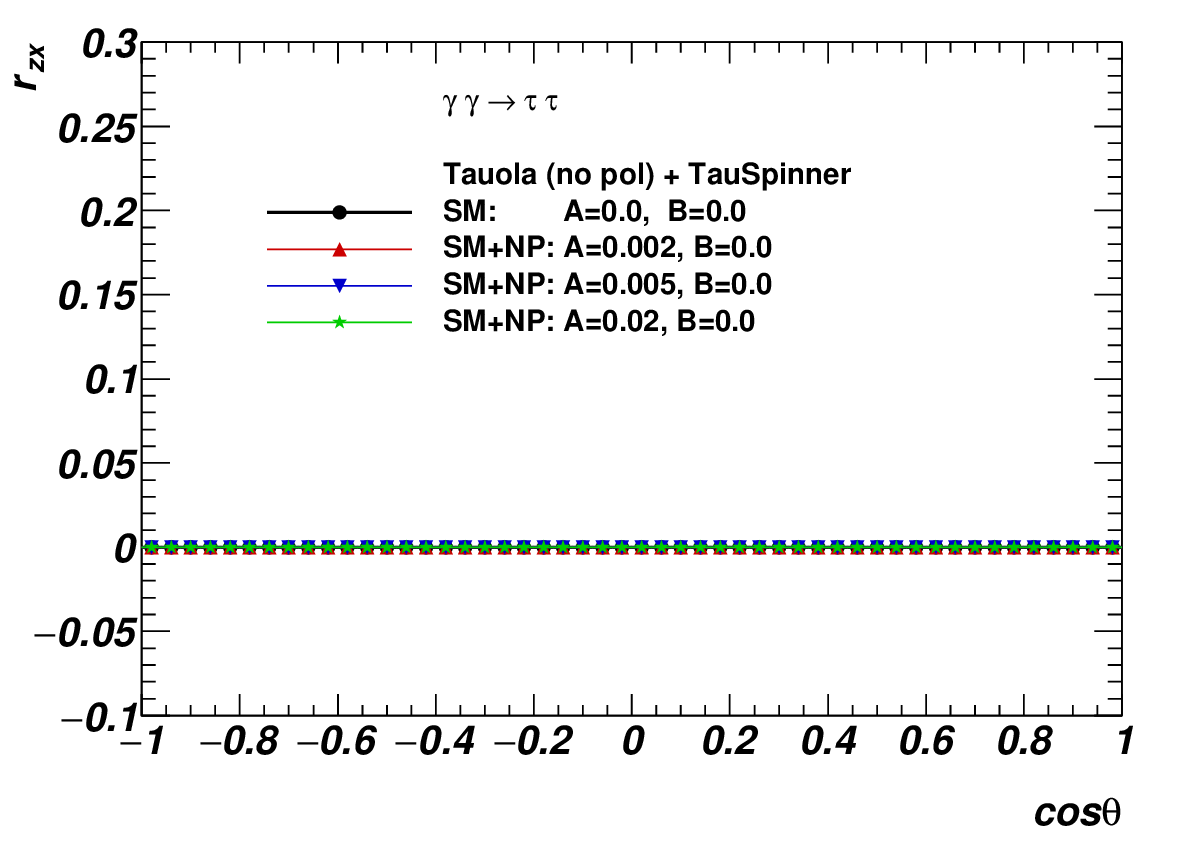}
  \includegraphics[width=7.5cm,angle=0]{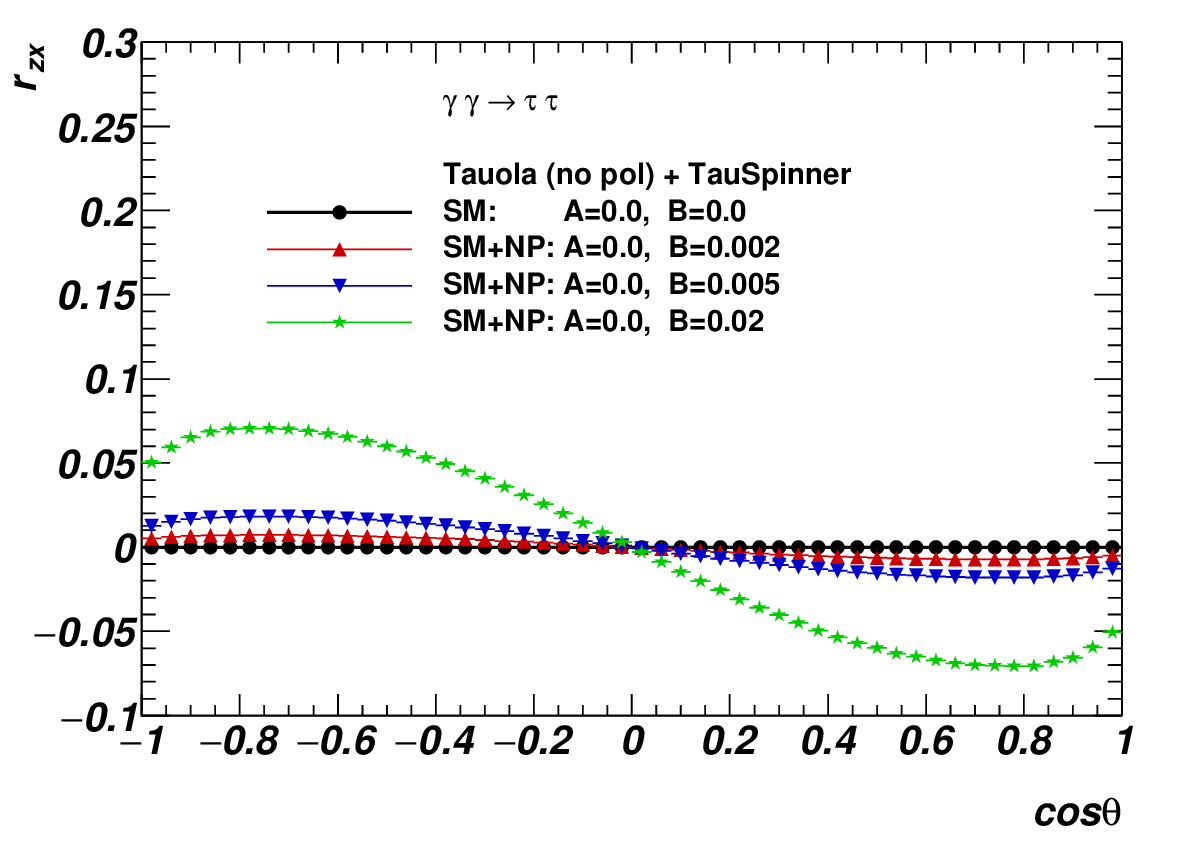}
  \includegraphics[width=7.5cm,angle=0]{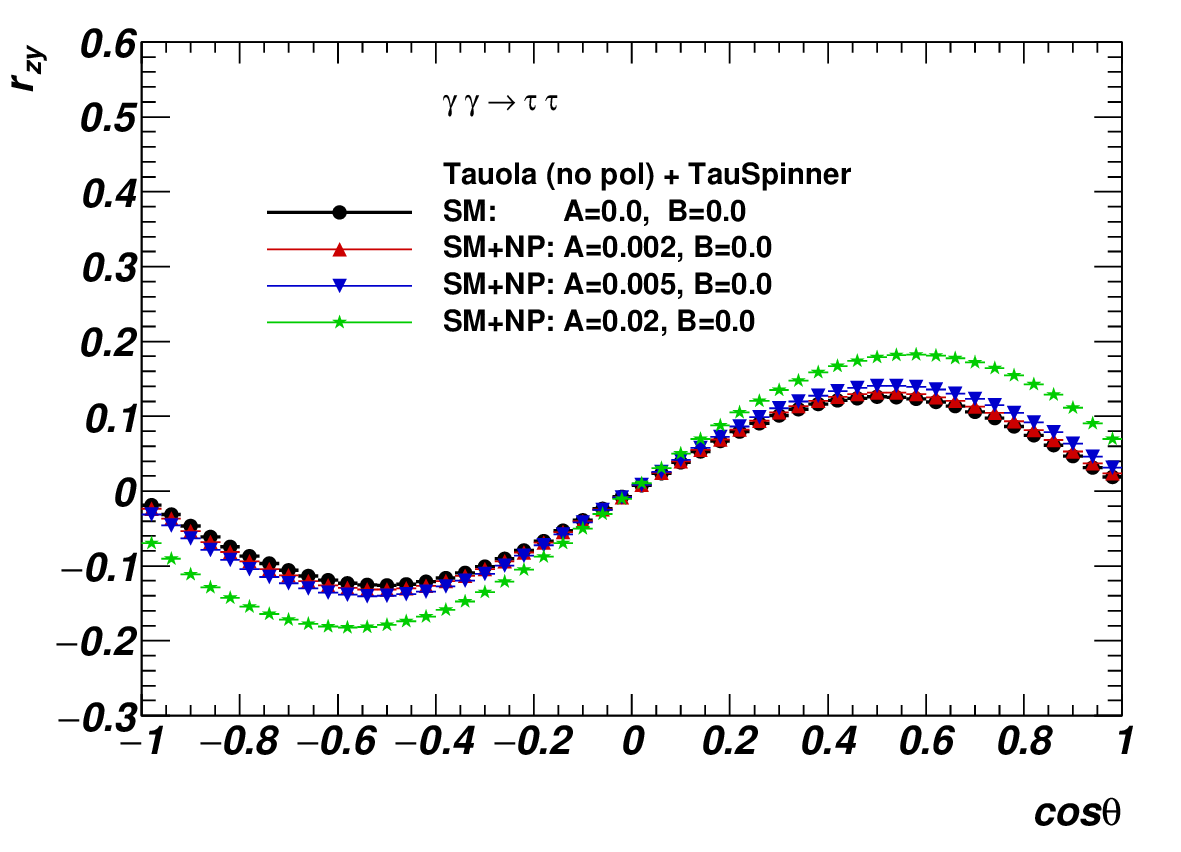}
  \includegraphics[width=7.5cm,angle=0]{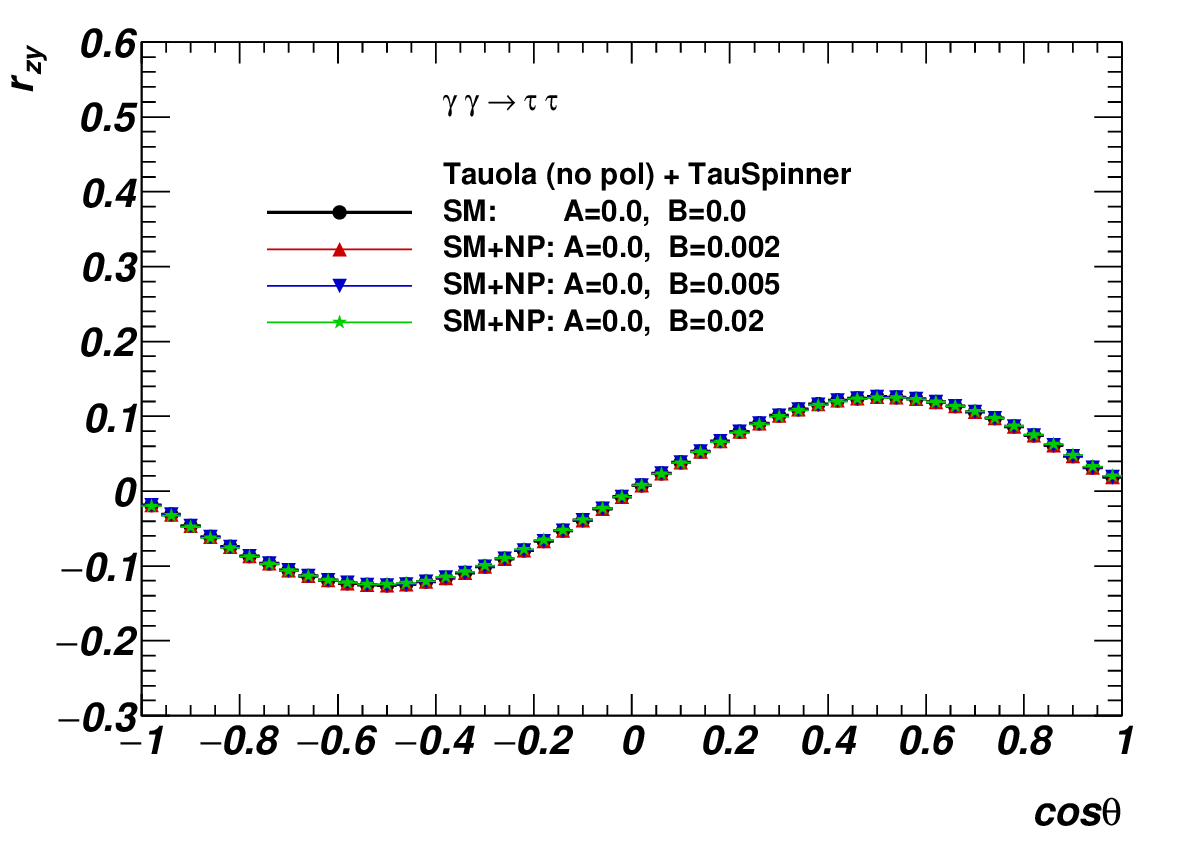}
}
\end{center}
  \caption{Spin-correlation matrix elements $r_{xy}$, $r_{zx}$ and $r_{zy}$ as functions of $\cos\theta$.
    Notation is the same as in Fig.~\ref{Fig:Rtt}.  Elements $r_{xy}$, $r_{zx}$ have negligible SM contributions,
    but non-negligible NP ones, while $r_{zy}$ is sizable already in the SM and attain additional contribution from NP.
 \label{Fig:Rij_b_costheta} }
\end{figure}


\subsection{\texorpdfstring{Spin effects in the $\tau^{\pm} \to \pi^{\pm} \nu_\tau$  decay channels}{}}
\label{sec:pipi}
Let us now turn attention to distributions constructed from the observable $\tau$-decay products. 
In Fig.~\ref{Fig:kinem_SM_pipi} the effect of spin correlations is shown as in the SM ($A=0$, $B=0$)
on few kinematical variables: transverse momenta of the pions, $p_T^{\pi}$, ratio $E_{\pi}/E_{\tau}$,
and ratio of invariant mass of $\pi^+\pi^-$ system to $\tau^+\tau^-$ system, $m_{\pi\pi}/m_{\tau\tau}$.
The $p_T^{\pi}$ and the $E_{\pi}/E_{\tau}$ distributions are rather insensitive 
to the spin correlations in the $\gamma \gamma \to \tau \tau$ process.
However, for the kinematical observable constructed from the four-momenta of both pions,
$m_{\pi\pi}/m_{\tau\tau}$, the effect is apparent. 
The change in the shape for the distribution of $m_{\pi\pi}/m_{\tau\tau}$ is at the level of 10-20\% in 
a wide range around $m_{\pi\pi}/m_{\tau\tau} = 0.5$.

Fig.~\ref{Fig:kinem_BSM_pipi} shows effect of SM+NP extension in the normalisation and spin correlations.
Plots of the ratio (SM+NP)/SM are shown, both including spin correlations.
Once  integrated over full phase space, impact from SM+NP extension is mostly due to change of the cross-section.
However we observe also some change in the shape of the $p_T^{\pi}$ distribution, with the ratio SM+NP to SM 
rising with increasing $p_T^{\pi}$
for $A=0.02$ or $B=0.02$. Some shape effect is also visible in the $m_{\pi\pi}/m_{\tau\tau}$ distribution 
for $A=0.02$.

\begin{figure} 
  \begin{center}                               
{
   \includegraphics[width=7.5cm,angle=0]{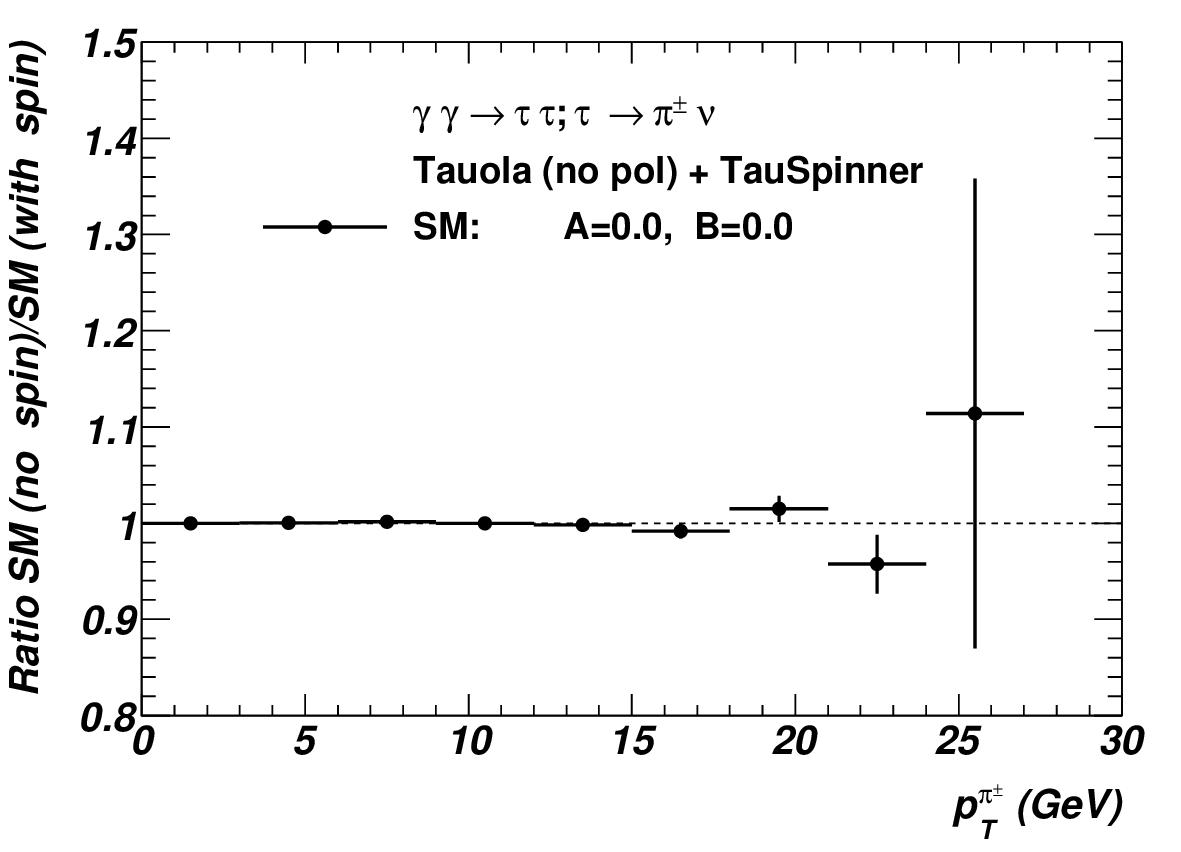}
   \includegraphics[width=7.5cm,angle=0]{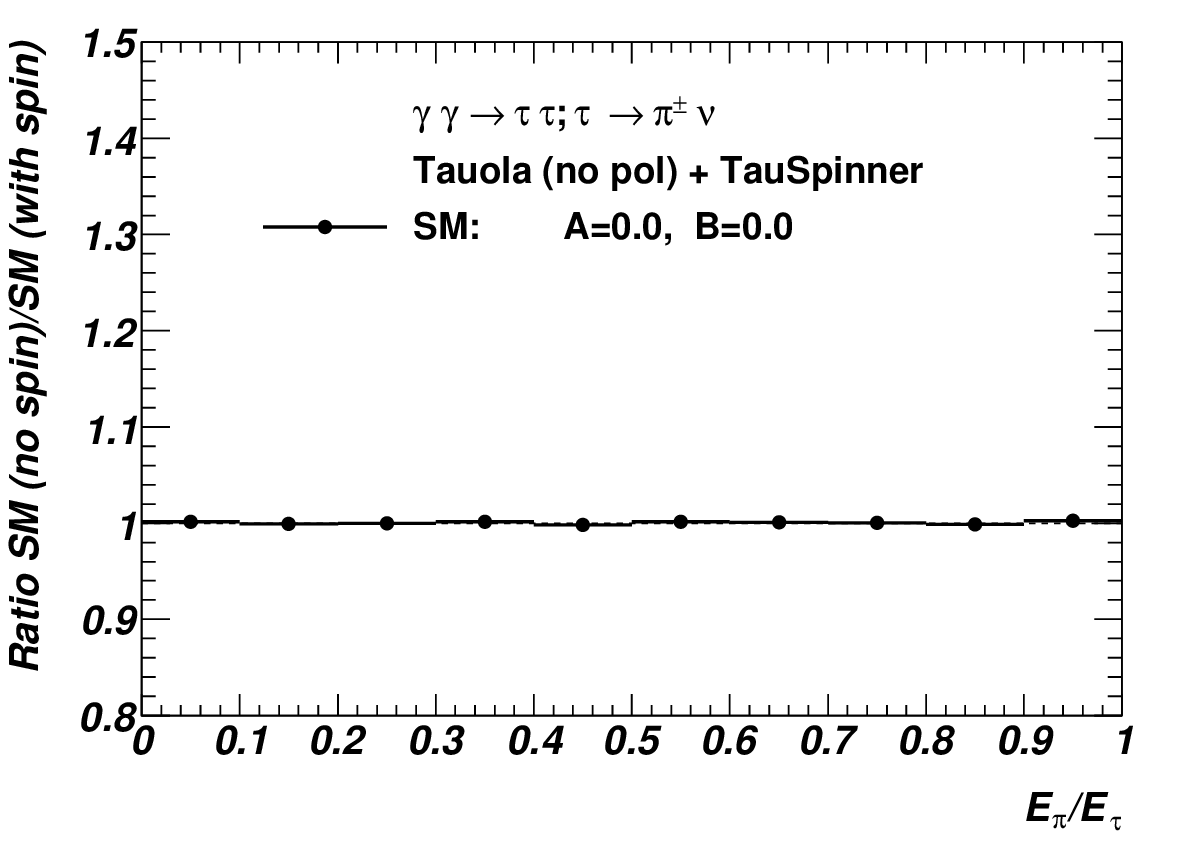}
   \includegraphics[width=7.5cm,angle=0]{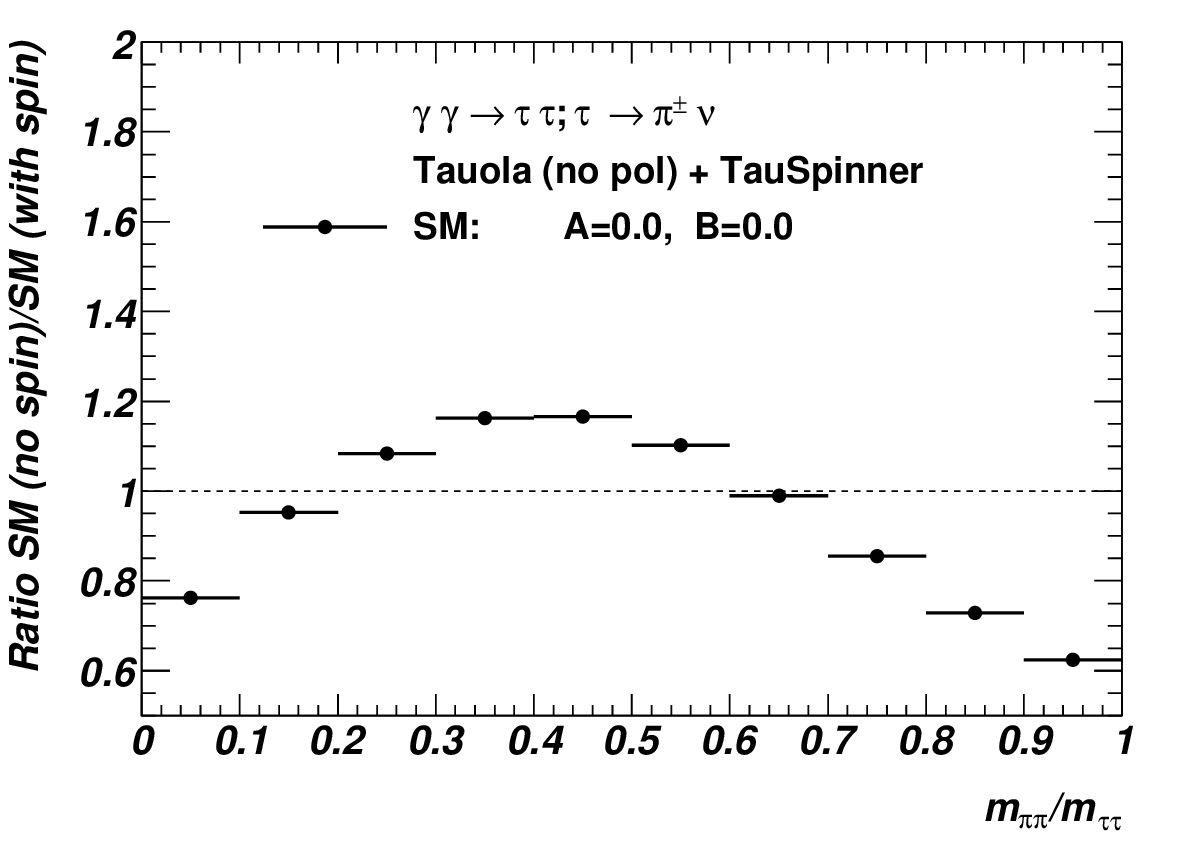}
}
\end{center}
  \caption{Spin-correlation effects for the case of $\tau$ lepton decays: 
	$\tau^+ \to \pi^+ \bar{\nu}_\tau$ and $\tau^- \to \pi^- \nu_\tau$. 
    Shown ratio SM (no spin correlations)/SM (with spin correlations).
 \label{Fig:kinem_SM_pipi} }
\end{figure}
 
\begin{figure} 
  \begin{center}                               
{
   \includegraphics[width=7.5cm,angle=0]{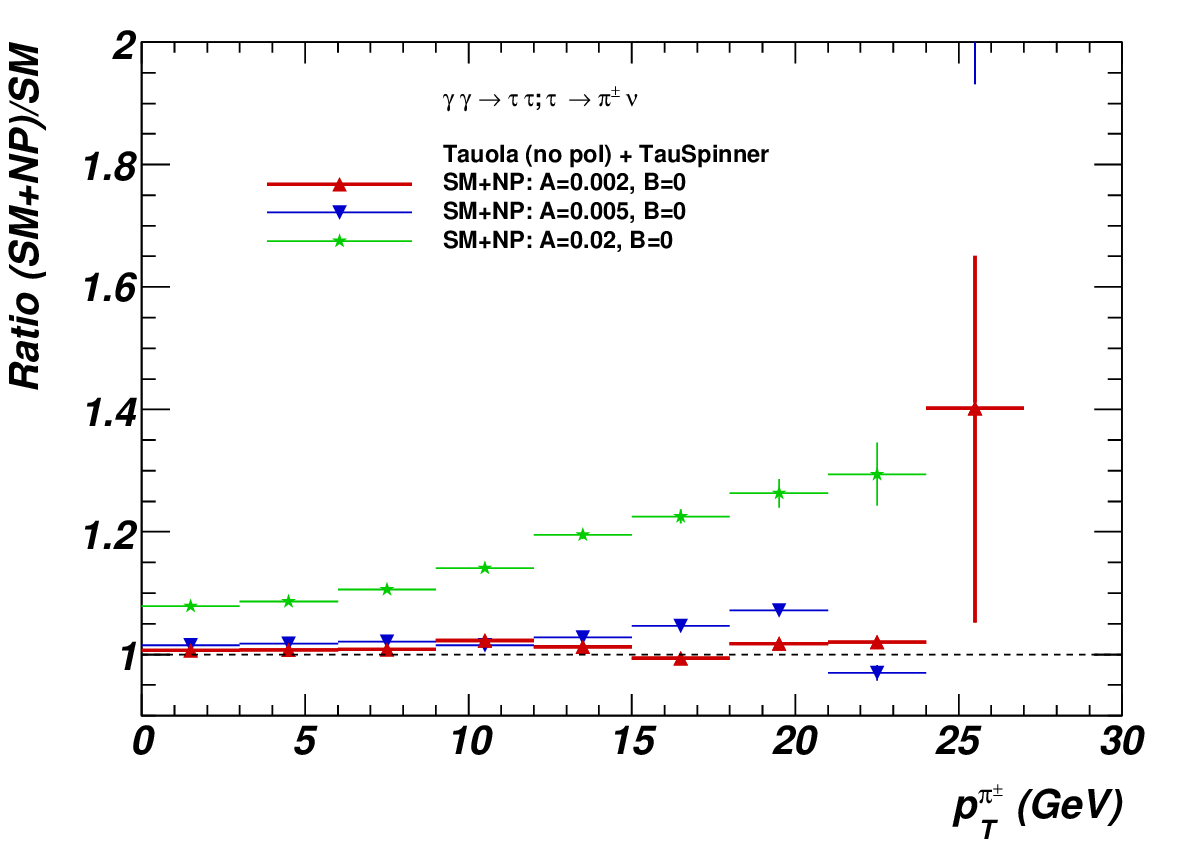}
   \includegraphics[width=7.5cm,angle=0]{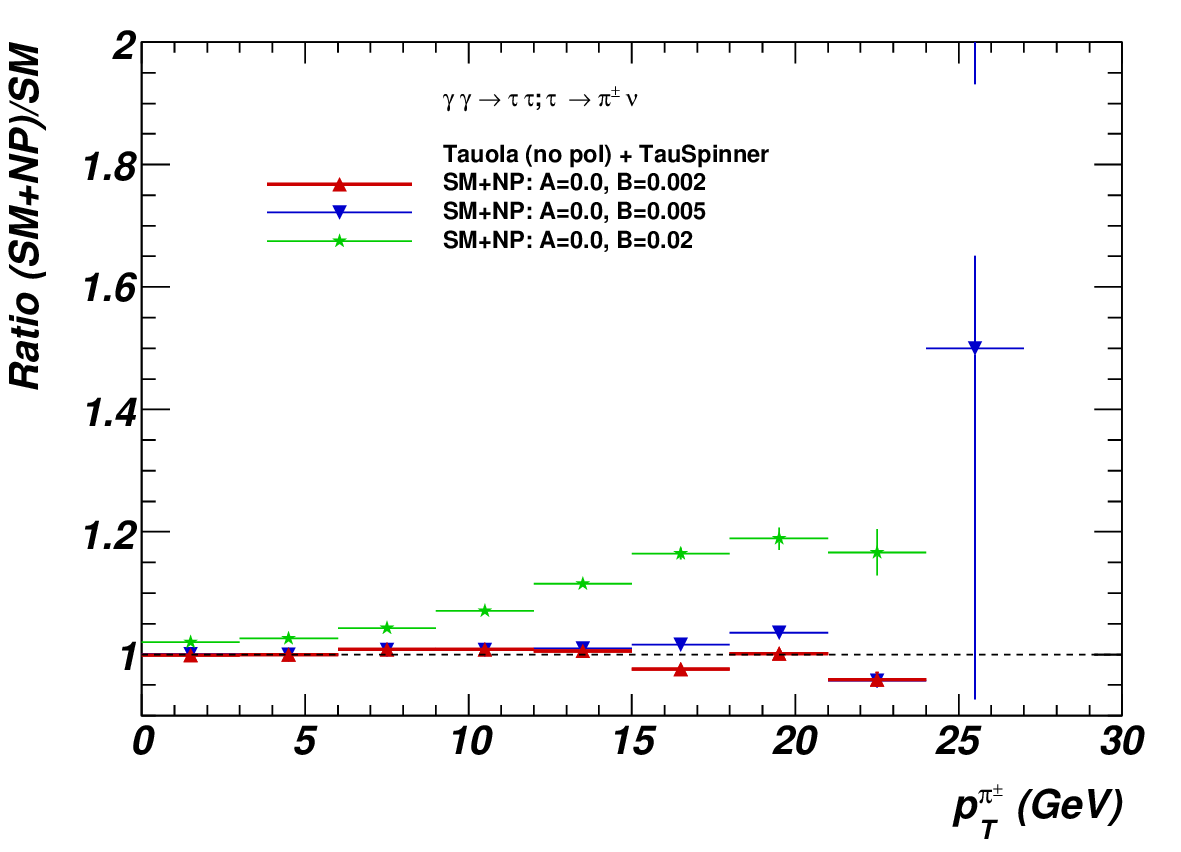}
   \includegraphics[width=7.5cm,angle=0]{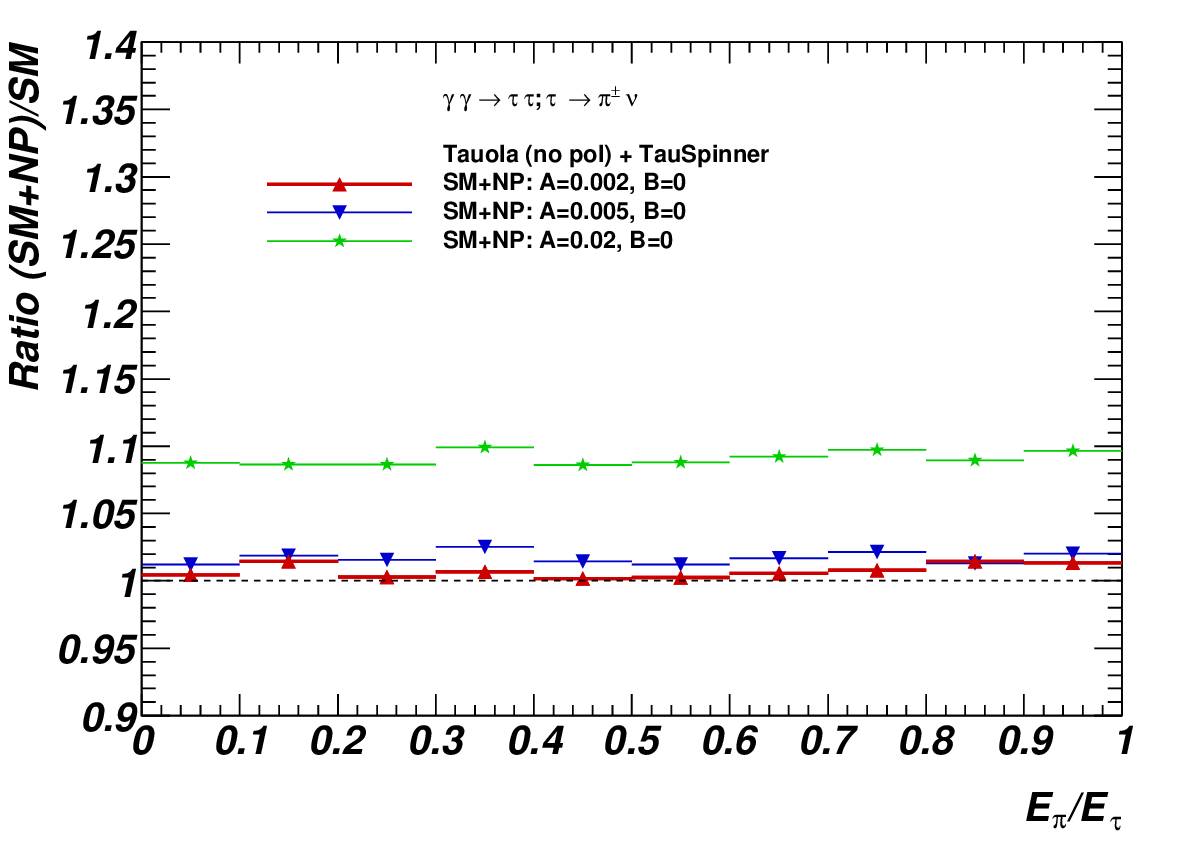}
   \includegraphics[width=7.5cm,angle=0]{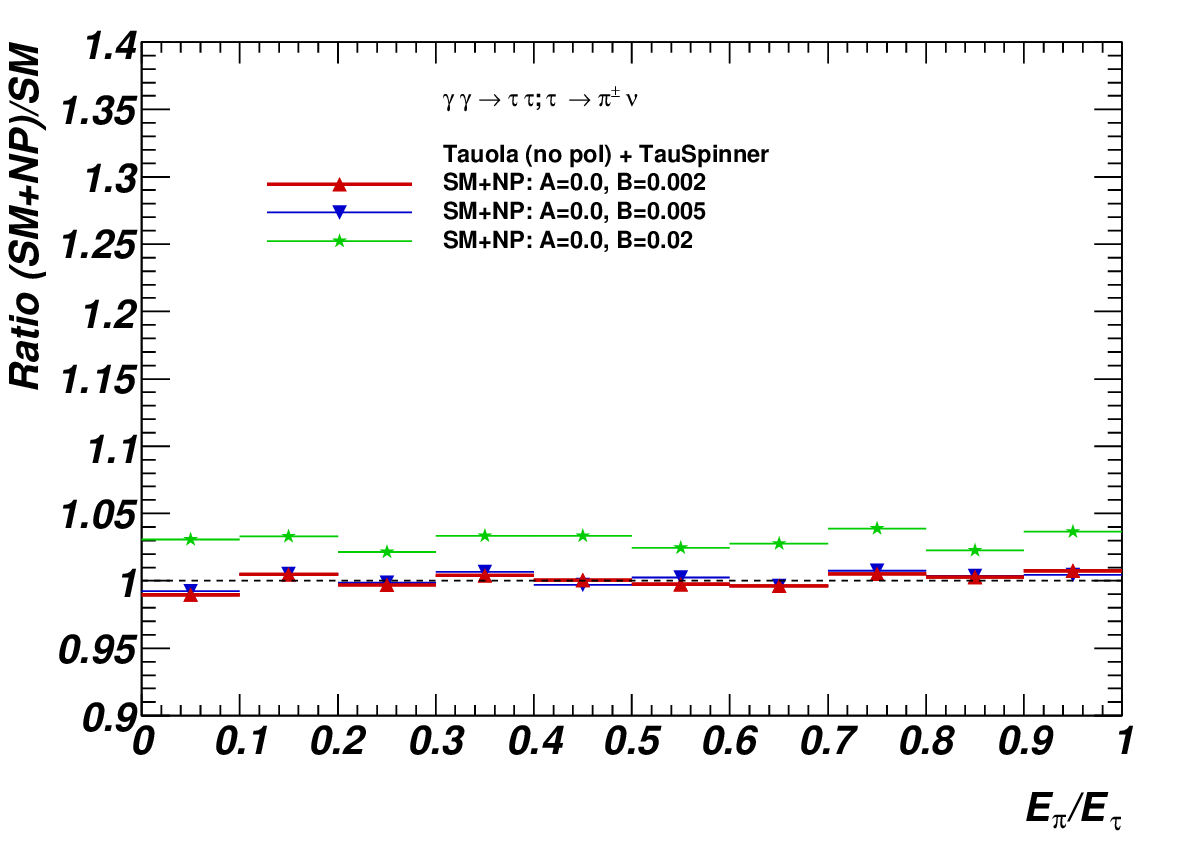}
   \includegraphics[width=7.5cm,angle=0]{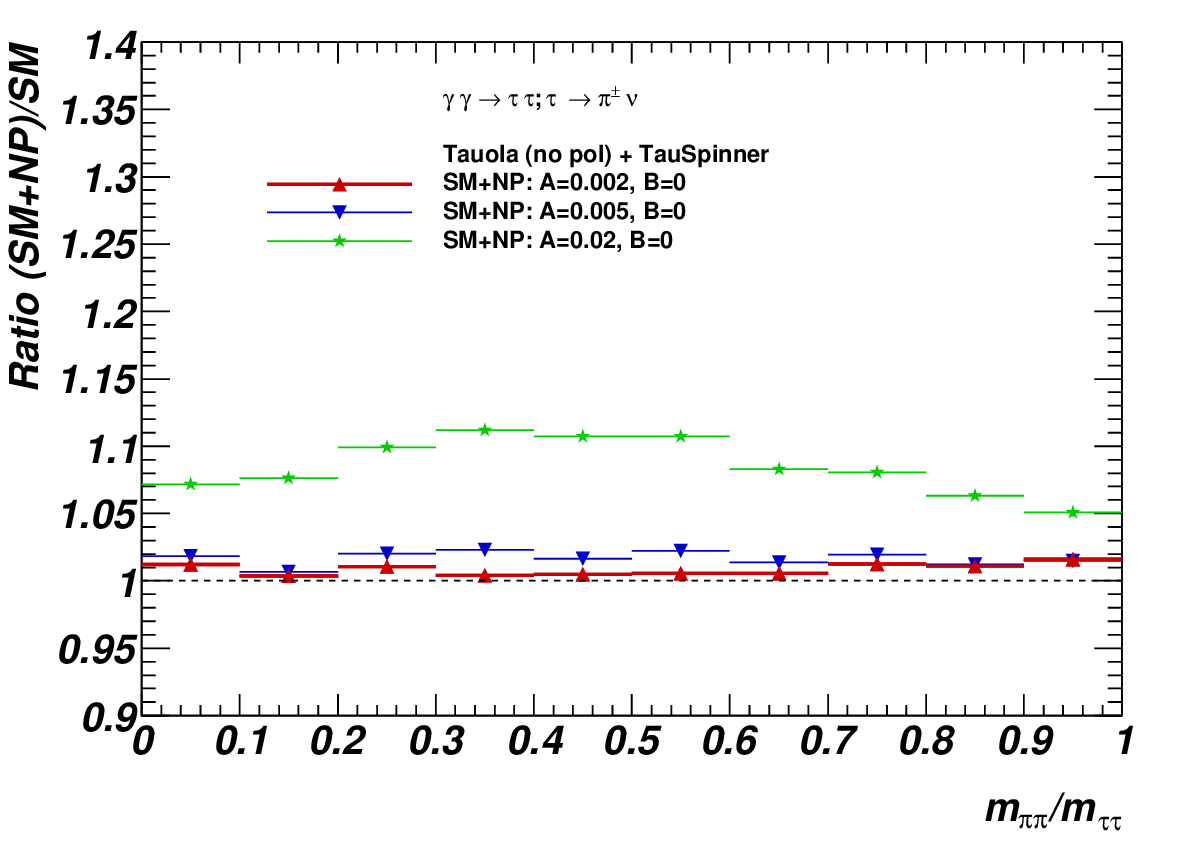}
   \includegraphics[width=7.5cm,angle=0]{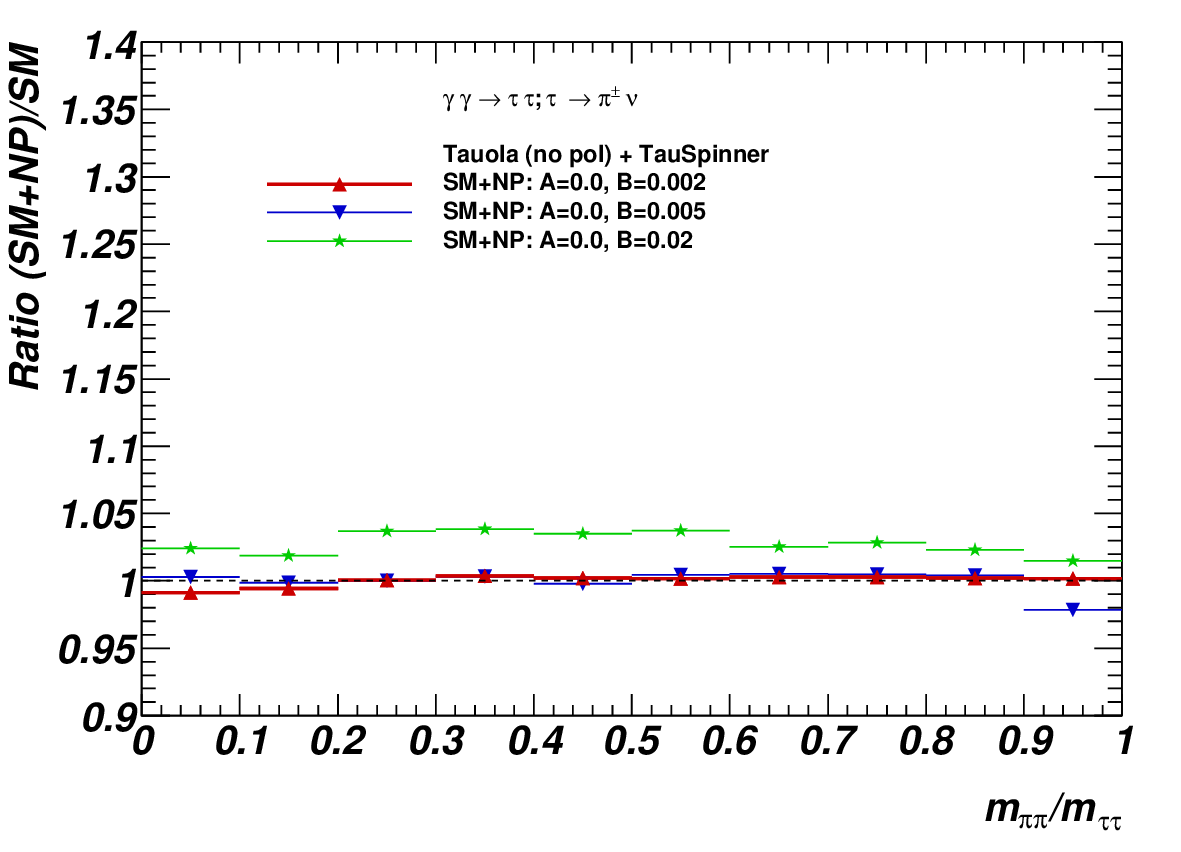}
}
\end{center}
  \caption{The $p_T^\pi$, $E_\pi/E_\tau$ and $m_{\pi\pi}/m_{\tau\tau}$ distributions. Both $\tau$ leptons
decay via $\tau^\pm \to \pi^\pm \nu_\tau$.  Notation is the same as in Fig.~\ref{Fig:Rtt}.
     \label{Fig:kinem_BSM_pipi} }
\end{figure}


\subsection{\texorpdfstring{Spin effects in the $\tau^\pm \to \rho^\pm \nu_\tau$  decay channels}{}}
\label{sec:rhorho}
In Fig.~\ref{Fig:kinem_SM_rhorho} effect of spin correlations is shown as in the SM ($A=0$, $B=0$)
on few kinematical variables: transverse momenta of charged pions $p_T^{\pi}$, 
ratio $E_{\rho}/E_{\tau}$, ratio $Y= E_{\pi^\pm}/E_{\rho}$,
and ratio of invariant mass of $\rho^+\rho^-$
system to $\tau^+\tau^-$ system $m_{\rho\rho}/m_{\tau\tau}$.
Examples of the ratio (SM no spin)/(SM with spin) are shown.
The $p_T^{\pi}$ distribution and distributions of $E_{\rho}/E_{\tau}$,  
$E_{\pi^\pm}/E_{\rho}$ are rather insensitive to the spin correlations in 
the $\gamma \gamma \to \tau \tau$ process.
However, for kinematical observables constructed from the four-momenta of both $\rho$ mesons, like
$m_{\rho\rho}/m_{\tau\tau}$, the effect is apparent.
The change in the shape of the $m_{\rho\rho}/m_{\tau\tau}$ distribution is at the level 
of 10-15\% in a wide range around $m_{\rho\rho}/m_{\tau\tau}$ = 0.5. 

Fig.~\ref{Fig:kinem_BSM_rhorho} shows effect of SM+NP extension in the normalisation and spin correlations.
Shown are the plots of the ratio (SM+NP)/SM, both including spin correlations.
Once integrated over full phase-space, impact from SM+NP extension is mostly due to the change of the cross-section.
However we observe also some change in the shape of $p_T^{\pi}$ distribution, with ratio SM+NP to SM rising 
with increasing  $p_T^{\pi}$
for $A=0.02$ or $B=0.02$. Some shape effect is also visible for the $m_{\rho \rho}/m_{\tau\tau}$ 
distribution with $A=0.02$. But without careful detector study it is not clear how useful the effects can be for the
measurements aiming on sensitivity to NP.

\begin{figure} 
  \begin{center}                               
{
   \includegraphics[width=7.5cm,angle=0]{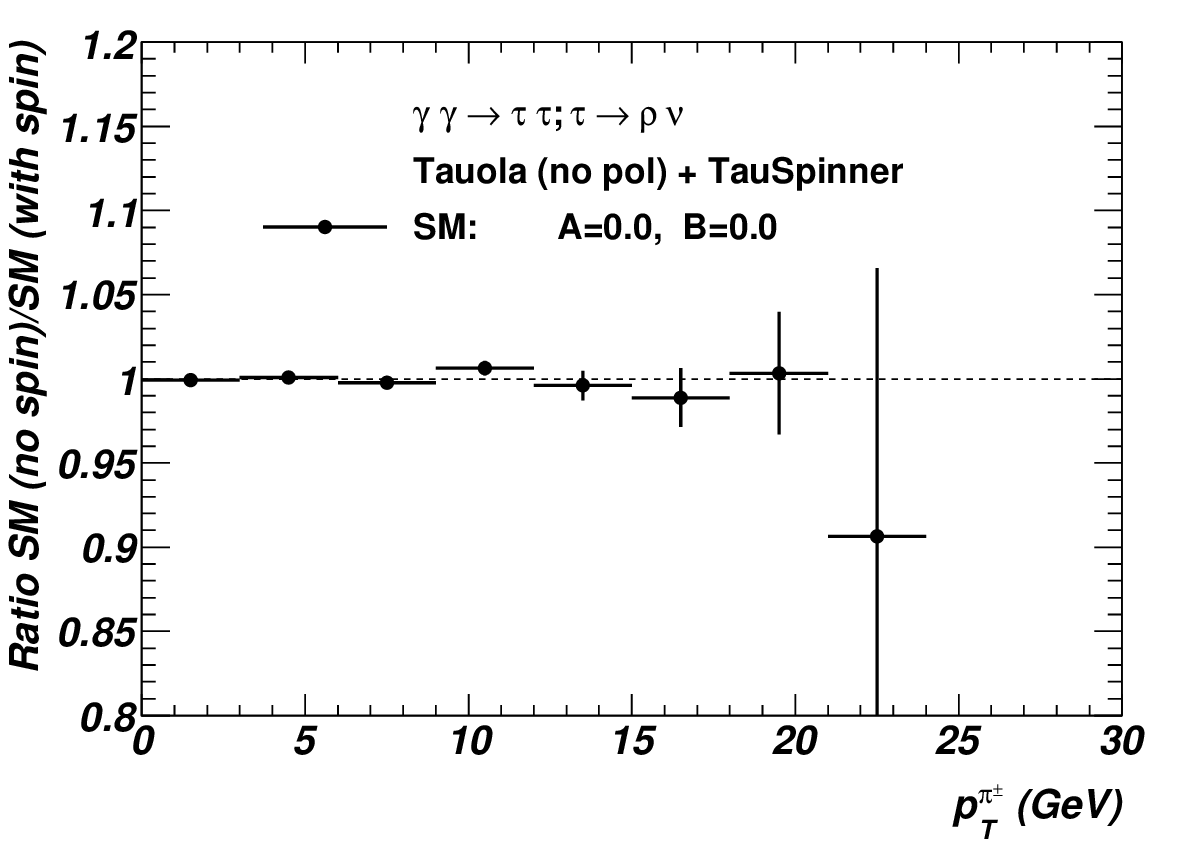}
   \includegraphics[width=7.5cm,angle=0]{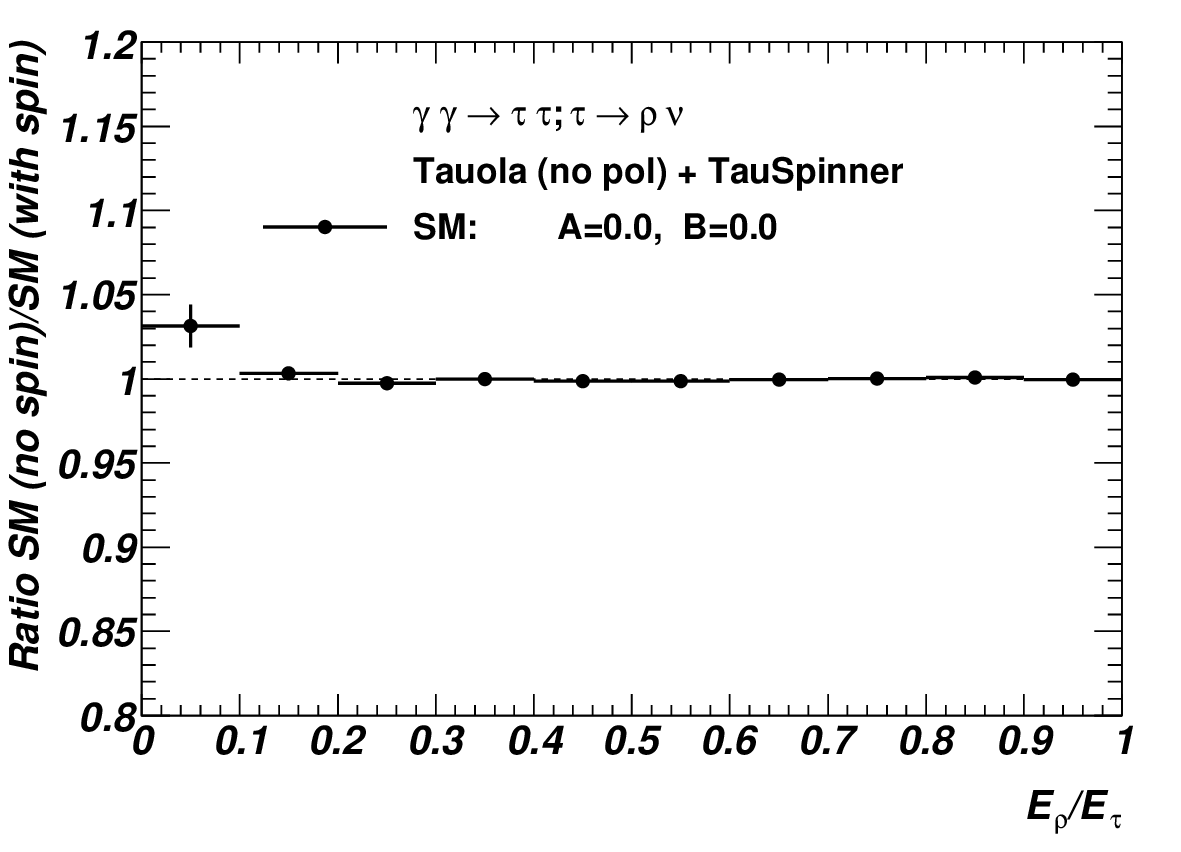}
   \includegraphics[width=7.5cm,angle=0]{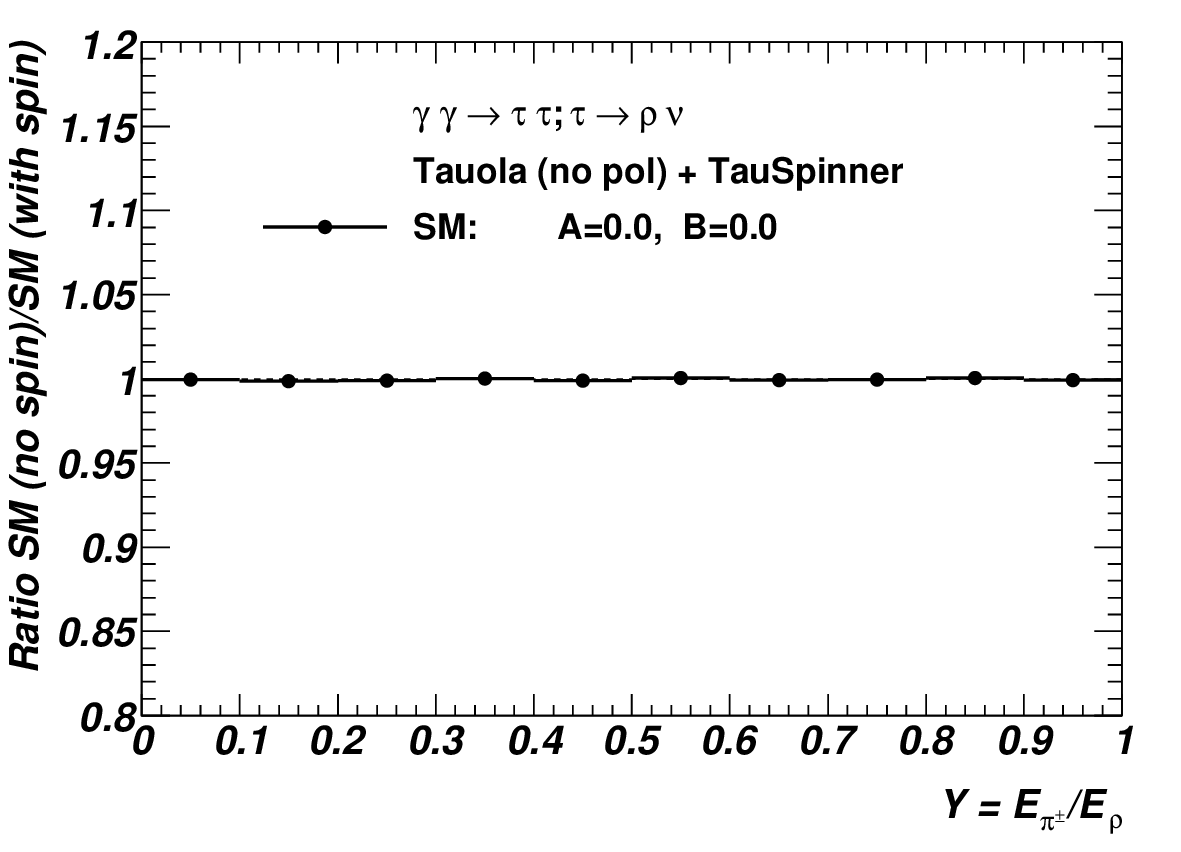}
   \includegraphics[width=7.5cm,angle=0]{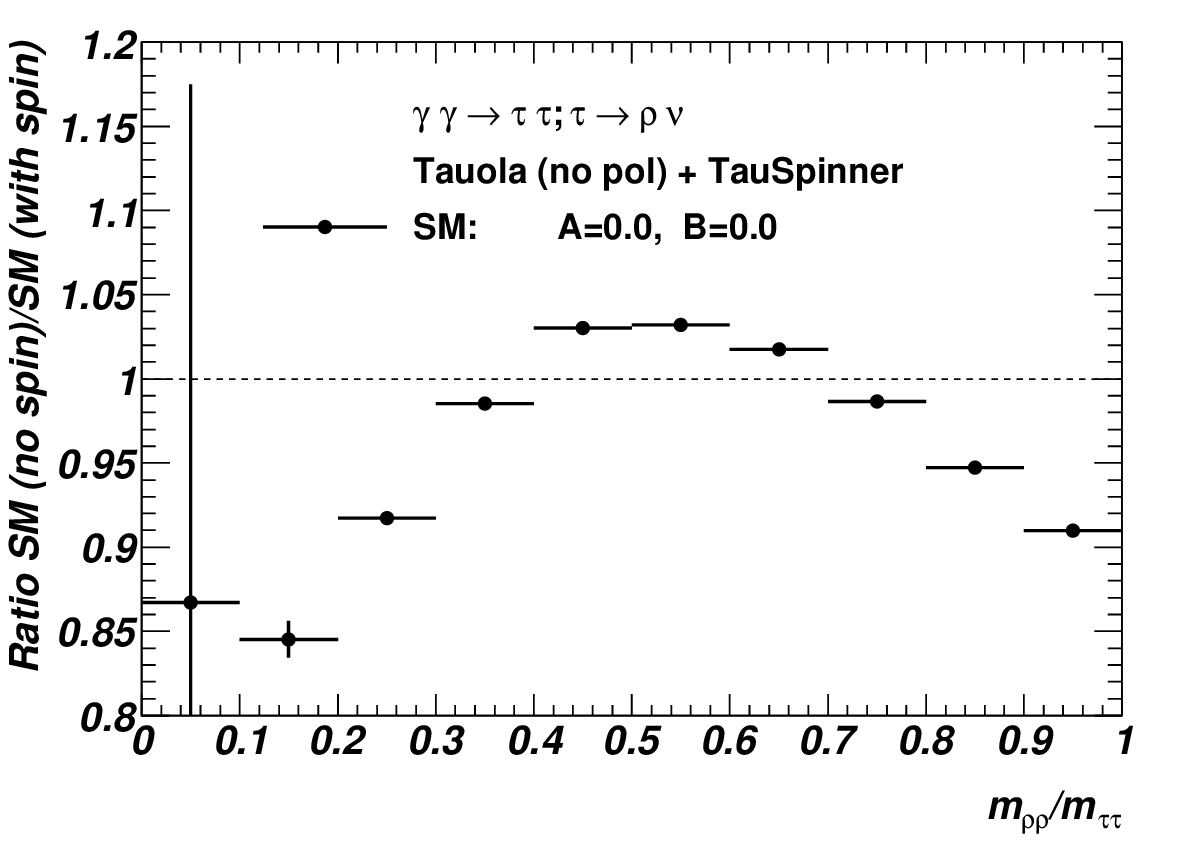}
}
\end{center}
  \caption{Spin correlation effects for the $\tau$-lepton decays: 
	$\tau^+\to \rho^+ \bar{\nu}_\tau$ and $\tau^- \to \rho^- \nu_\tau$.
    Results for the ratio SM (no spin correlations)/SM (with spin correlations) are shown. 
 \label{Fig:kinem_SM_rhorho} }
\end{figure}

\begin{figure}  
\centering																
{
   \includegraphics[width=7.5cm,angle=0]{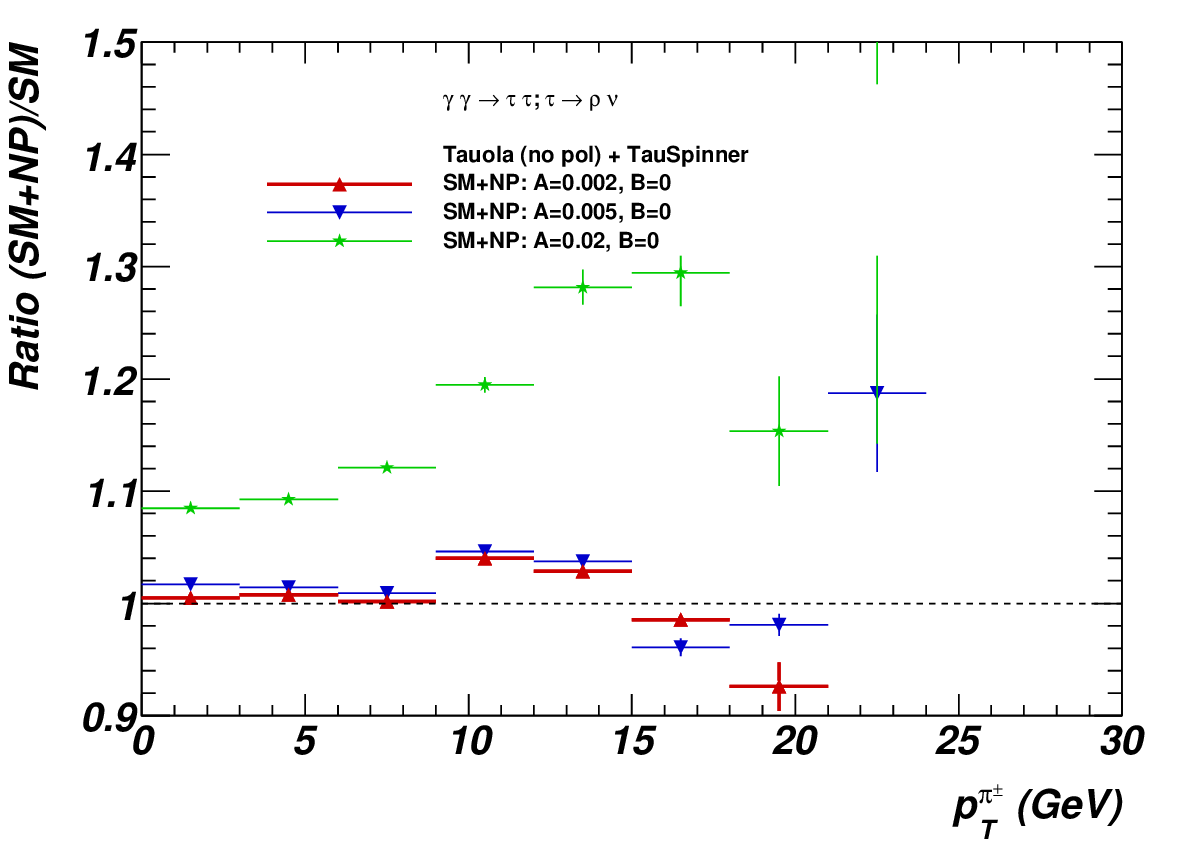}
   \includegraphics[width=7.5cm,angle=0]{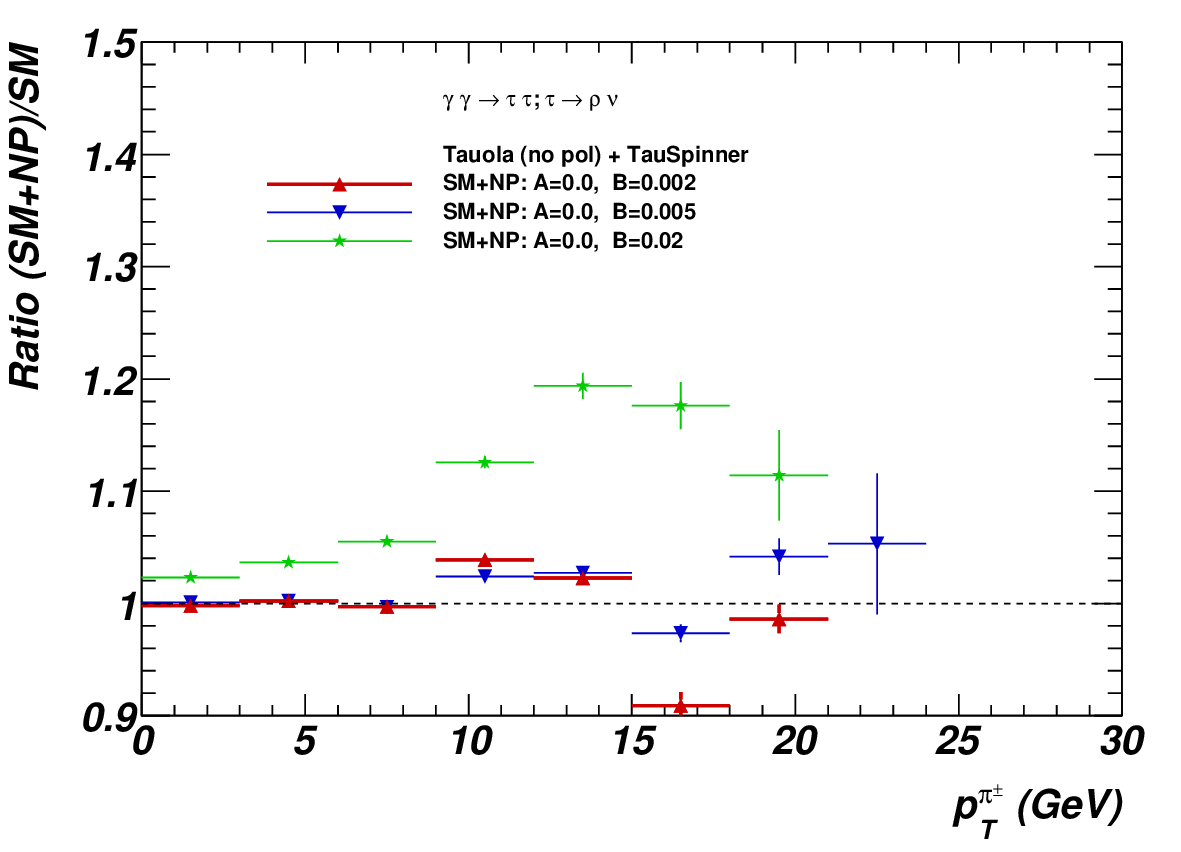}
   \includegraphics[width=7.5cm,angle=0]{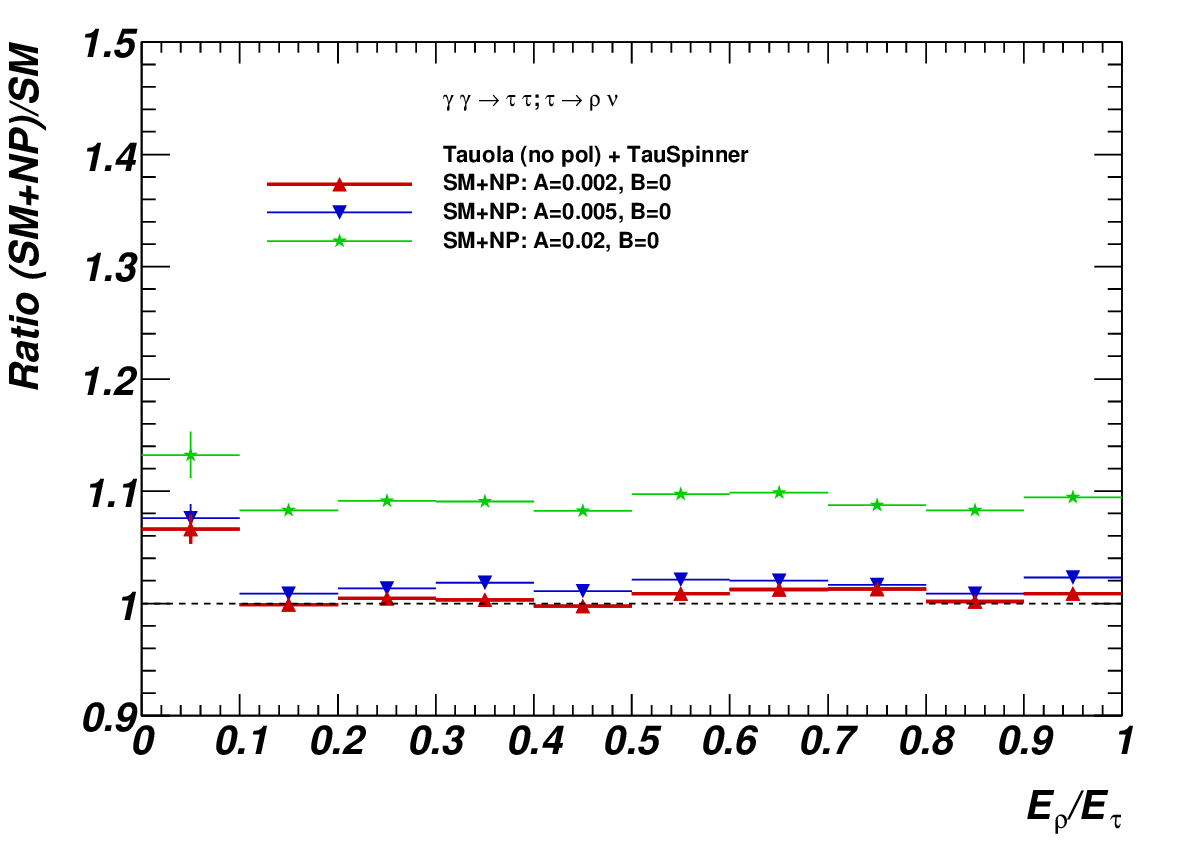}
   \includegraphics[width=7.5cm,angle=0]{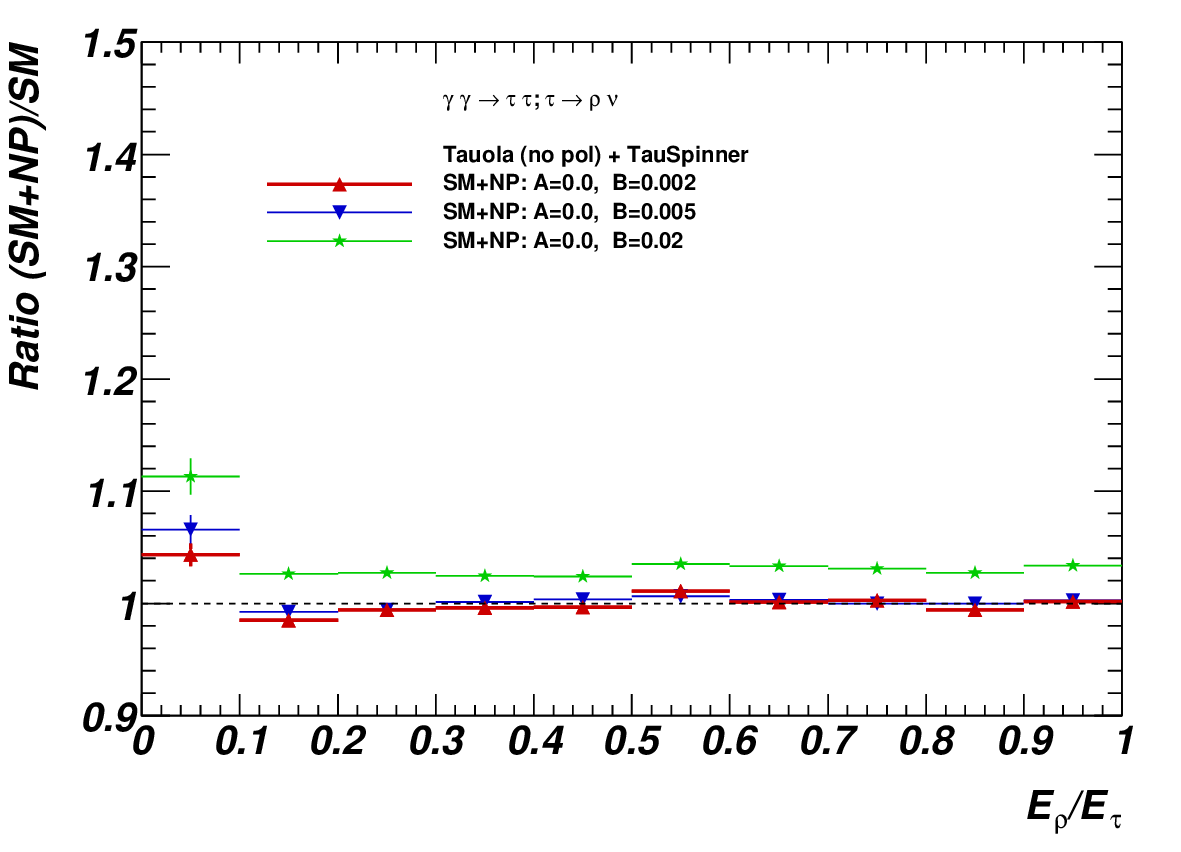}
   \includegraphics[width=7.5cm,angle=0]{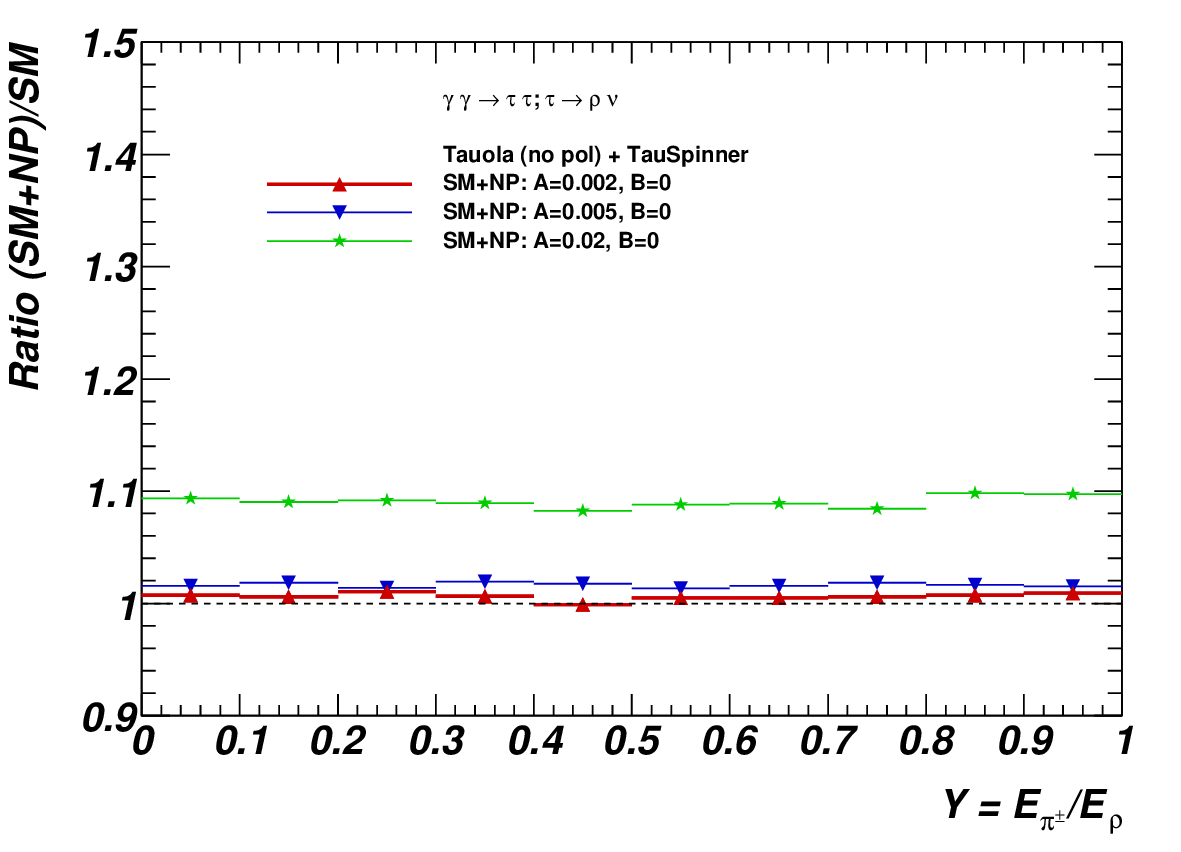}
   \includegraphics[width=7.5cm,angle=0]{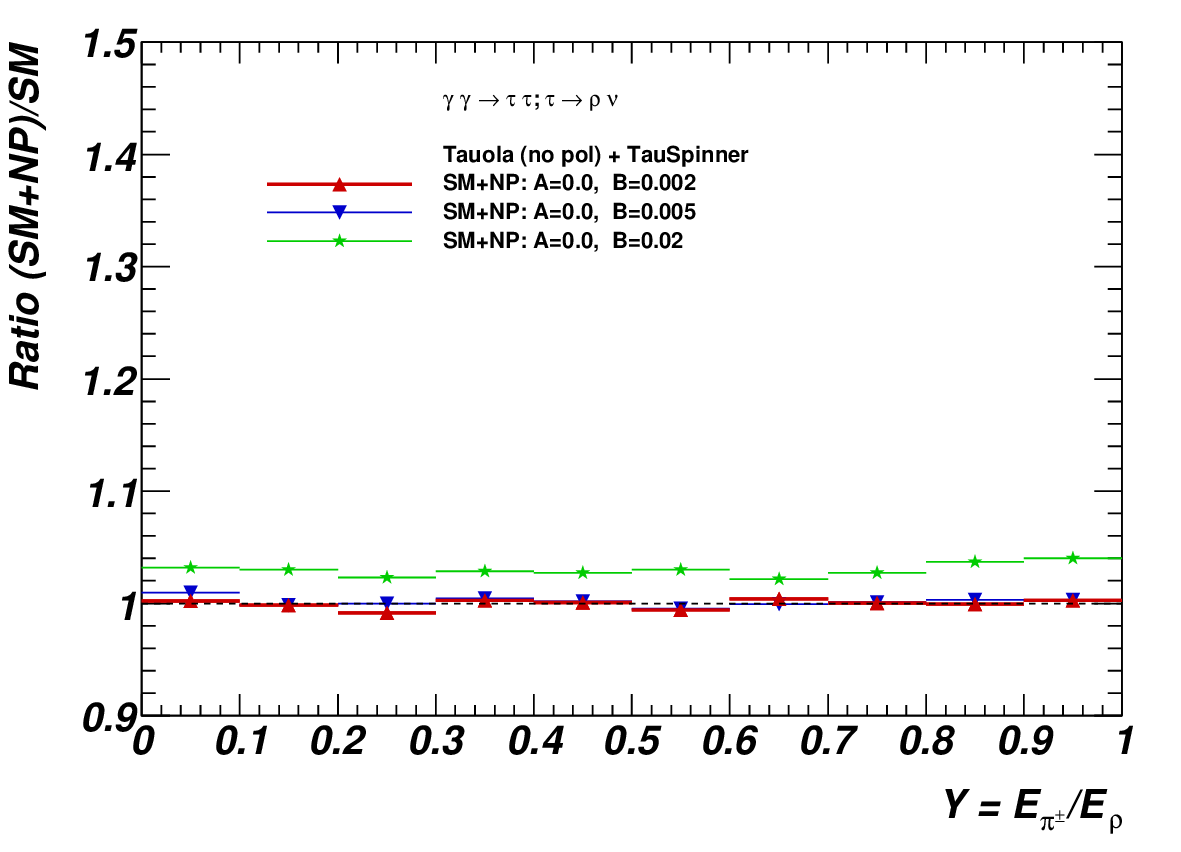}
   \includegraphics[width=7.5cm,angle=0]{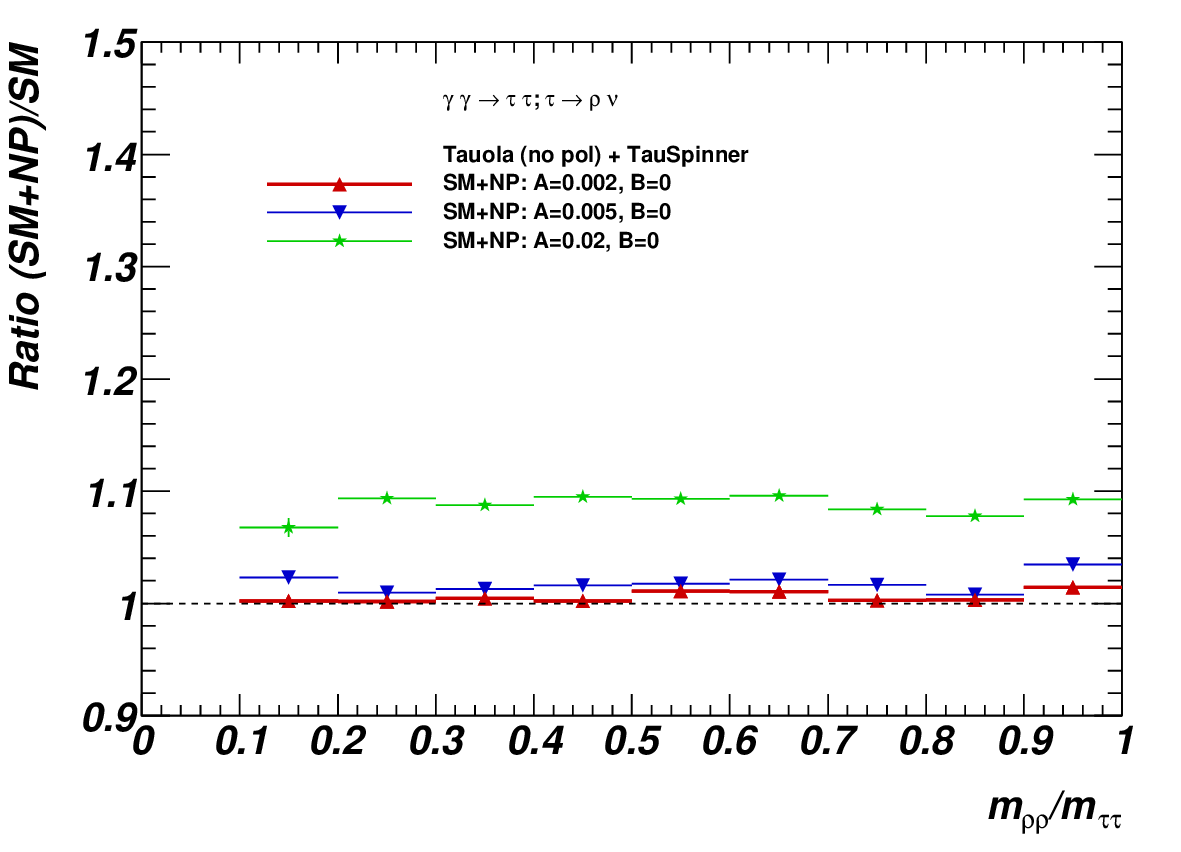}
   \includegraphics[width=7.5cm,angle=0]{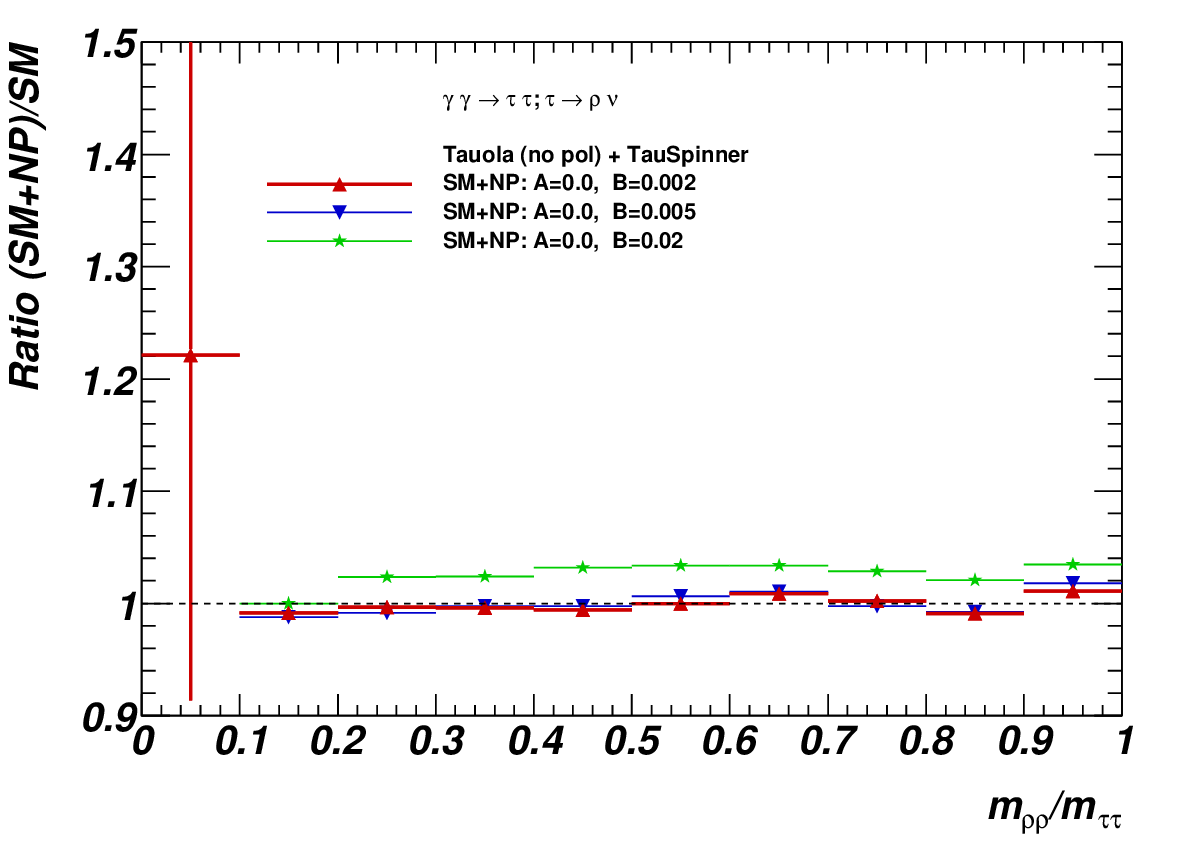}
}
  \caption{Kinematical distributions for both $\tau$ leptons decaying 
	via $\tau^\pm \to \rho^\pm \nu_\tau$.  Notation is the same as in Fig.~\ref{Fig:Rtt}.	
 \label{Fig:kinem_BSM_rhorho} }
\end{figure}


\subsection{\texorpdfstring{Spin effects in the decay channels $\tau^\pm \to \pi^\pm \nu_\tau$,   
$\tau^\mp \to \mu^\mp \nu_\tau \nu_\mu$}{}}
\label{sec:mupi}
 In  the leptonic $\tau$ decays the two neutrinos escape detection. That is why events with the leptonic 
$\tau$-decay channels may be more difficult to interpret. Nonetheless, the leptonic decays represent more 
than 30\% of $\tau$-decay rate and are experimentally easier to trigger on,  thus requiring some attention. 

In Fig.~\ref{Fig:kinem_SM_mupi} effect of spin correlations is shown as in the SM ($A=0$, $B=0$)
on few kinematical variables: transverse momenta, $p_T^{\mu}$ and $p_T^{\pi}$, 
ratios $E_{\mu}/E_{\tau}$ and $E_{\pi}/E_{\tau}$,
transverse momentum of  the $\mu \pi$ system $p_T^{\mu\pi}$ and the 
ratio of invariant mass of $\mu\pi$ system to that of $\tau^+\tau^-$ system,
$m_{\mu\pi}/m_{\tau\tau}$.
Some examples of the ratio (SM no spin)/(SM with spin) are shown.
The $p_T^{\pi}$,  $p_T^{\mu}$ distributions and distributions of the 
ratios $E_{\mu}/E_{\tau}$, $E_{\pi}/E_{\tau}$ are rather insensitive to the spin correlations
in the $\gamma \gamma \to \tau \tau$ process.
However, for the kinematical observables constructed from the four-momenta of both charged particles, 
$\pi$ and $\mu$, like $p_T^{\mu\pi}$ and $m_{\mu\pi}/m_{\tau\tau}$, the effect is apparent.
The change in the shape of the $m_{\mu\pi}/m_{\tau\tau}$ distribution is at the level of 10-15\% in a wide range 
around $m_{\mu\pi}/m_{\tau\tau} = 0.5$.

Figs.~\ref{Fig:kinem_BSM_mupi_a} and \ref{Fig:kinem_BSM_mupi_b} show effect of SM+NP extension in the normalisation and spin correlations.
Shown are the ratio of (SM+NP)/SM, both including spin correlations.
Again, once integrated over full phase space, impact from SM+NP extension is mostly due to change of the cross-section.
However we observe also some change in the shape of $p_T^{\pi}$ distribution, with ratio of SM+NP to SM rising with increasing  $p_T^{\pi}$,
for both $A=0.02$ and $B=0.02$. Some shape effect is also visible in the $m_{\mu\pi}/m_{\tau\tau}$ distribution 
for $A=0.02$.

\begin{figure}  
  \begin{center}                               
    {
   \includegraphics[width=7.5cm,angle=0]{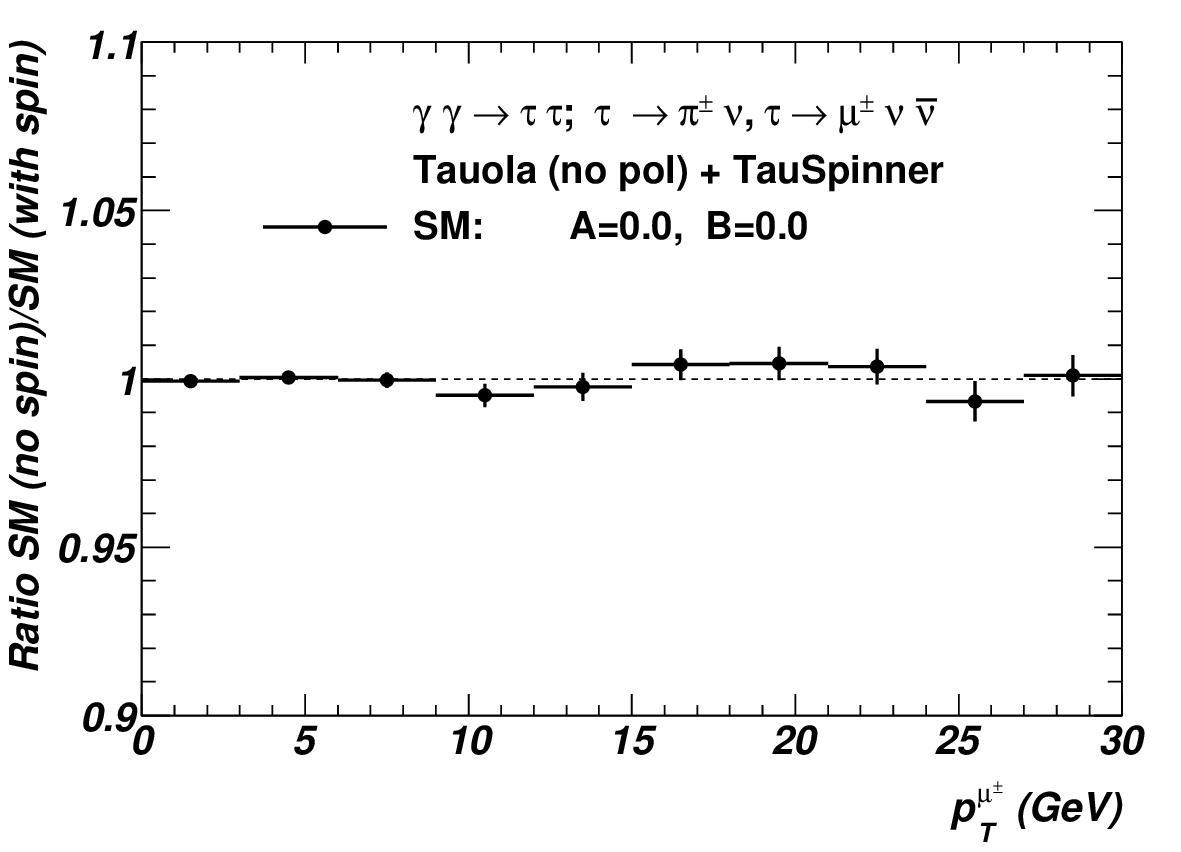}
   \includegraphics[width=7.5cm,angle=0]{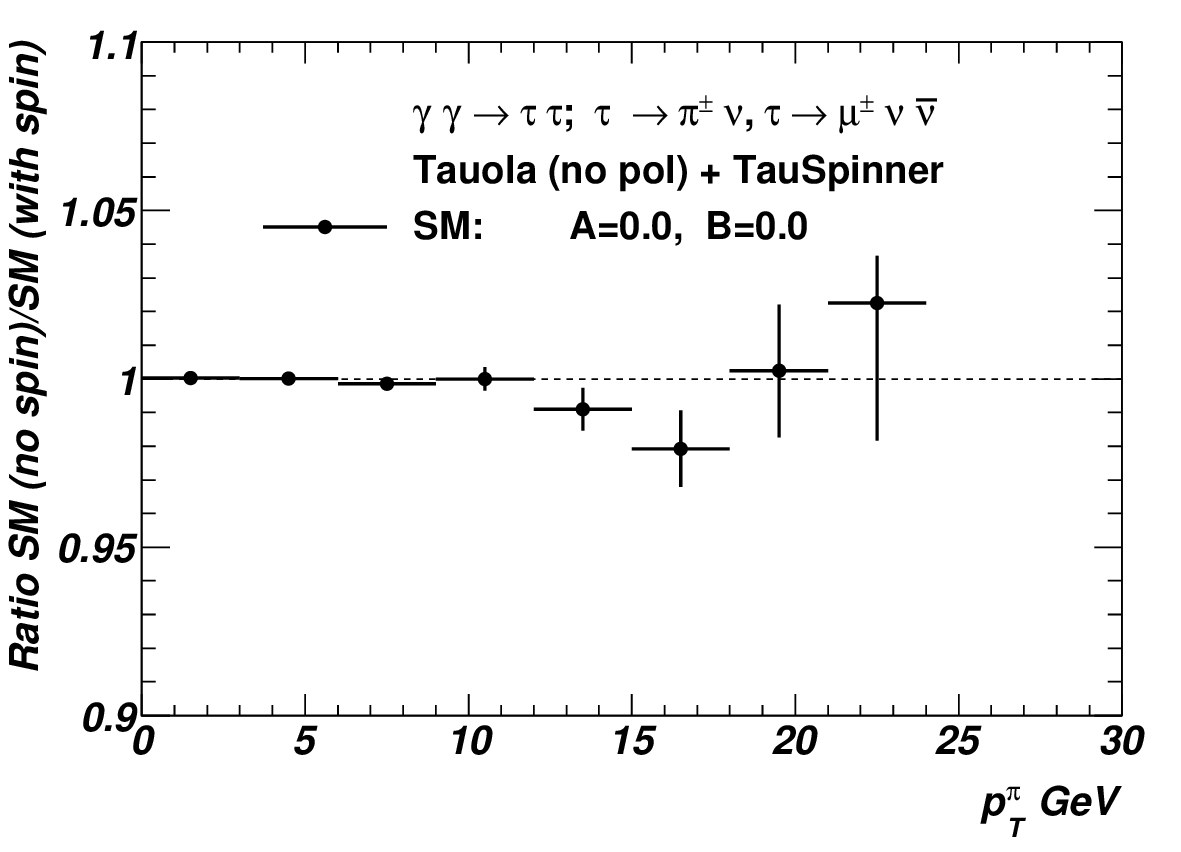}
   \includegraphics[width=7.5cm,angle=0]{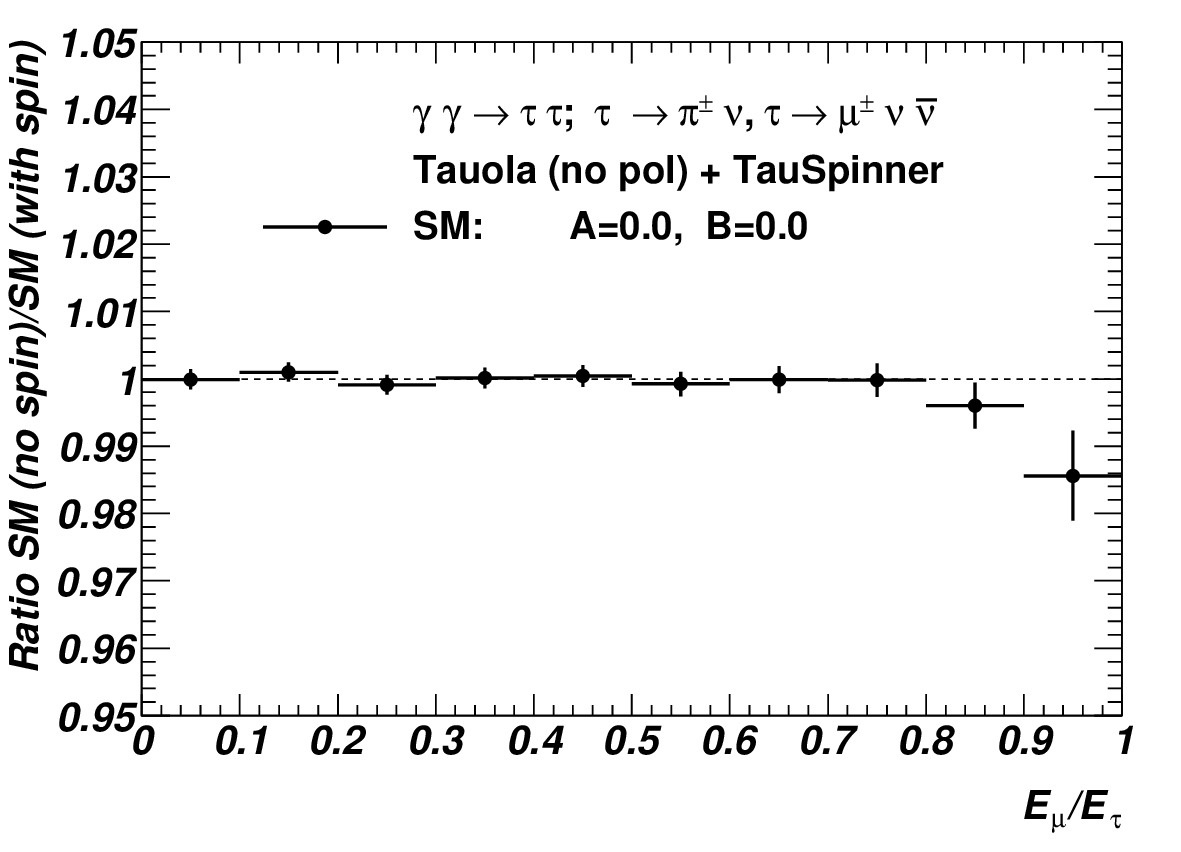}
   \includegraphics[width=7.5cm,angle=0]{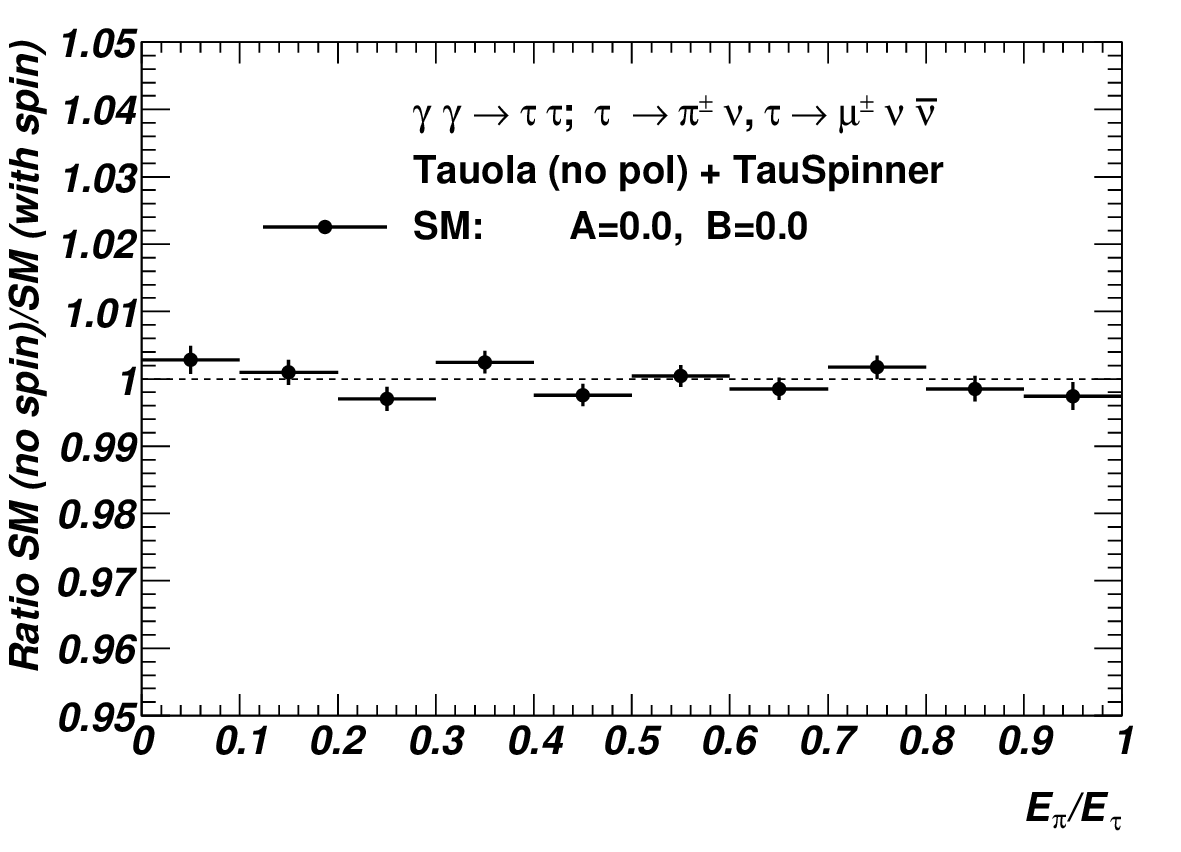}
   \includegraphics[width=7.5cm,angle=0]{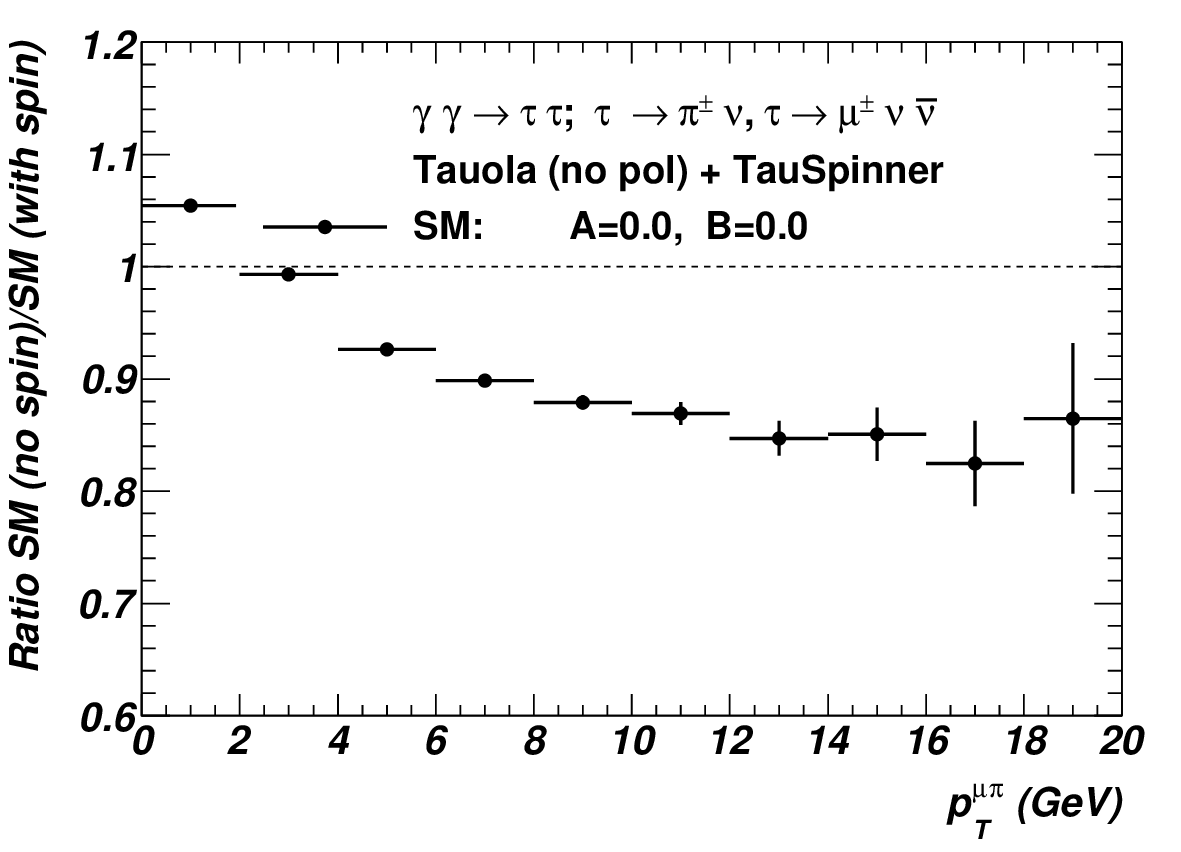}
    \includegraphics[width=6.5cm,angle=0]{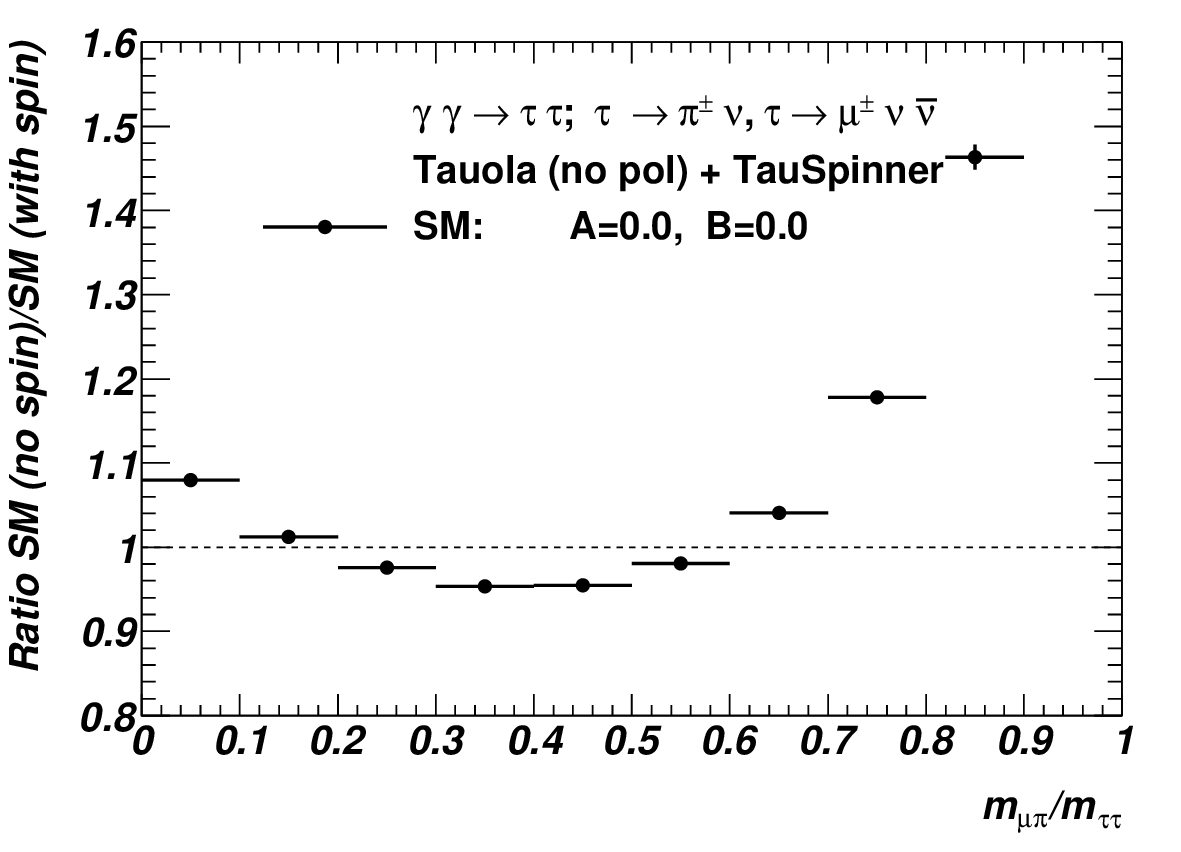}
}
\end{center}
  \caption{Spin correlation effects for the channels: $\tau^\pm \to \pi^\pm \nu_\tau$ and  
	$\tau^\mp \to \mu^\mp \nu_\tau \nu_\mu$.
    Results for the ratios SM (no spin correlations)/SM (with spin correlations) are shown.
 \label{Fig:kinem_SM_mupi} }
\end{figure}

\begin{figure}  
\centering                               
{
  \includegraphics[width=7.5cm,angle=0]{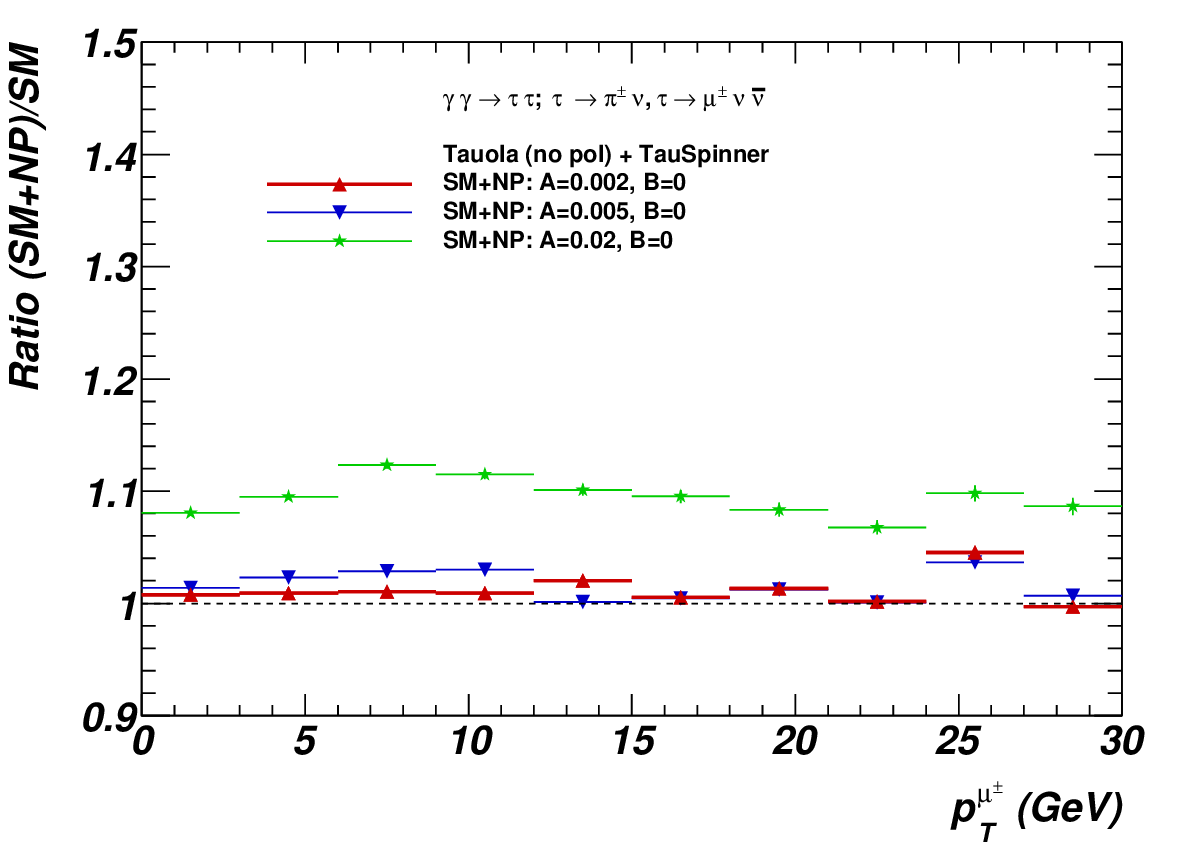}
  \includegraphics[width=7.5cm,angle=0]{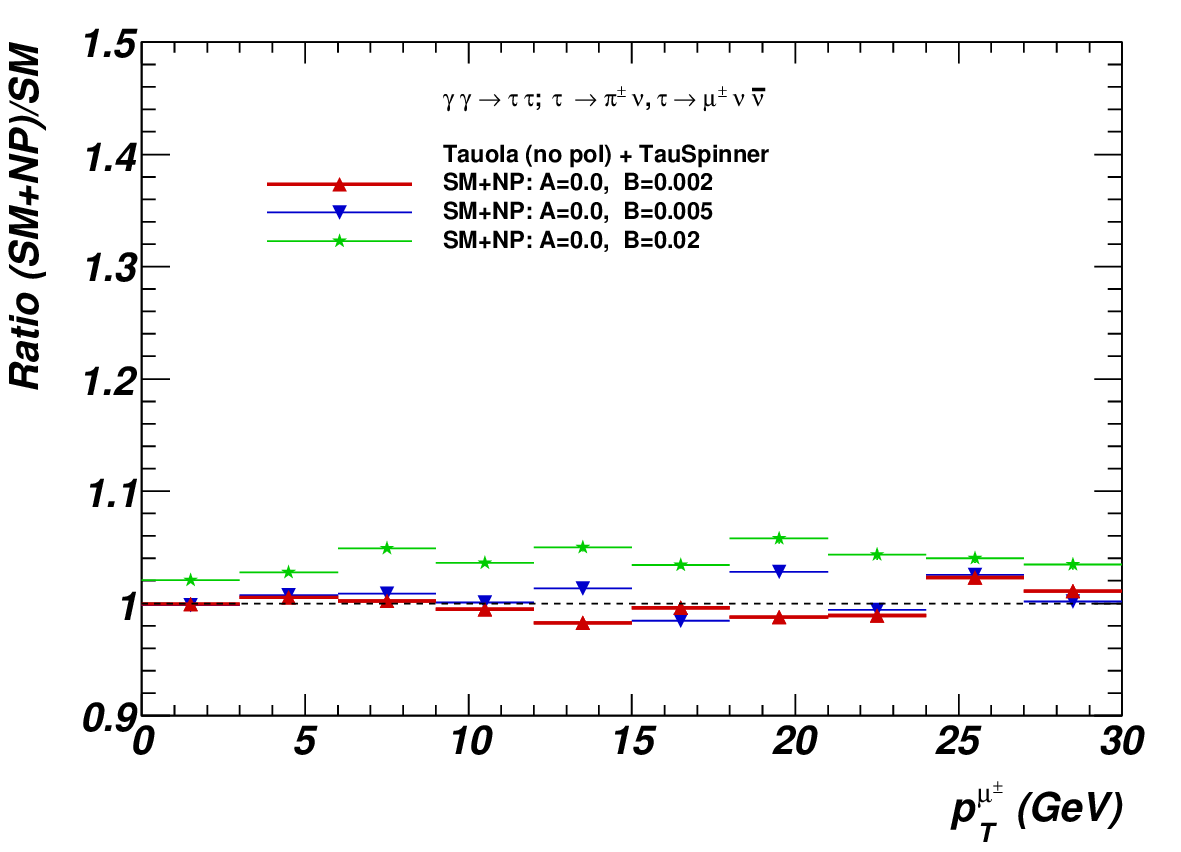}
  \includegraphics[width=7.5cm,angle=0]{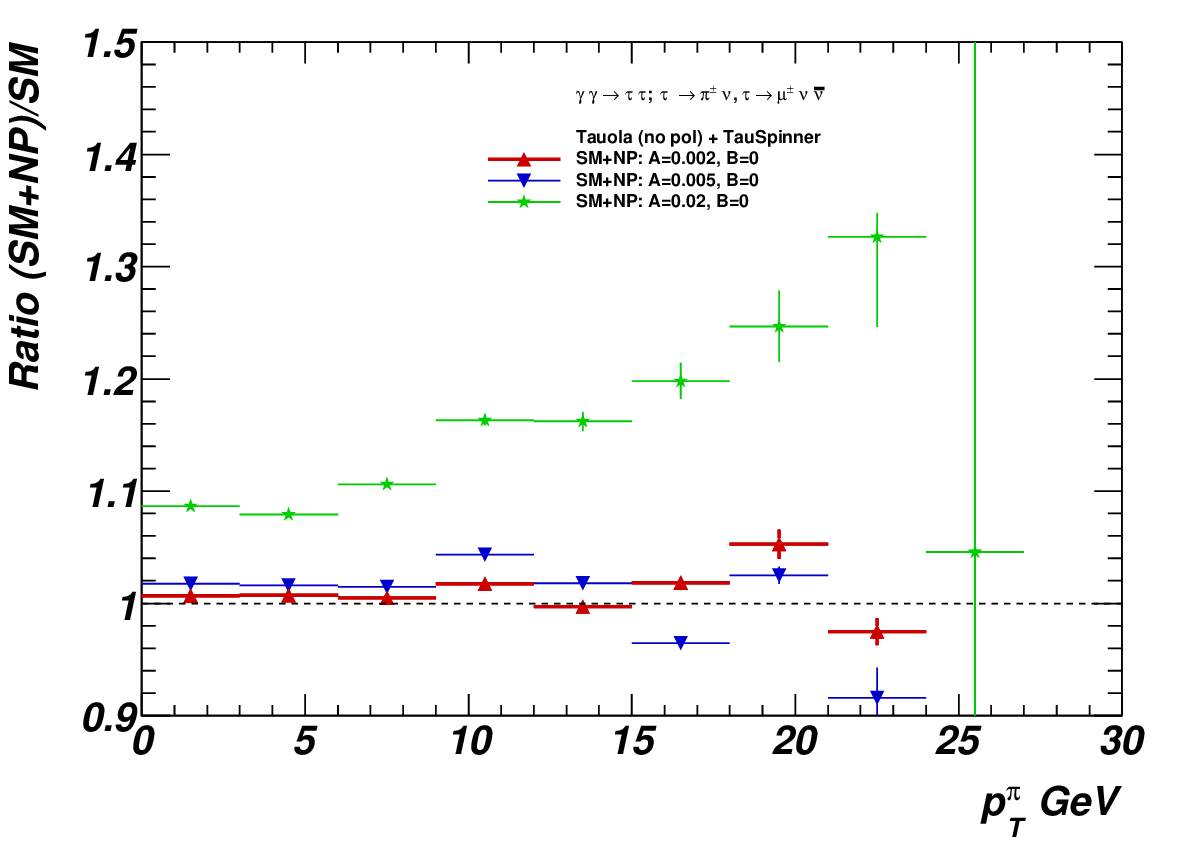}
  \includegraphics[width=7.5cm,angle=0]{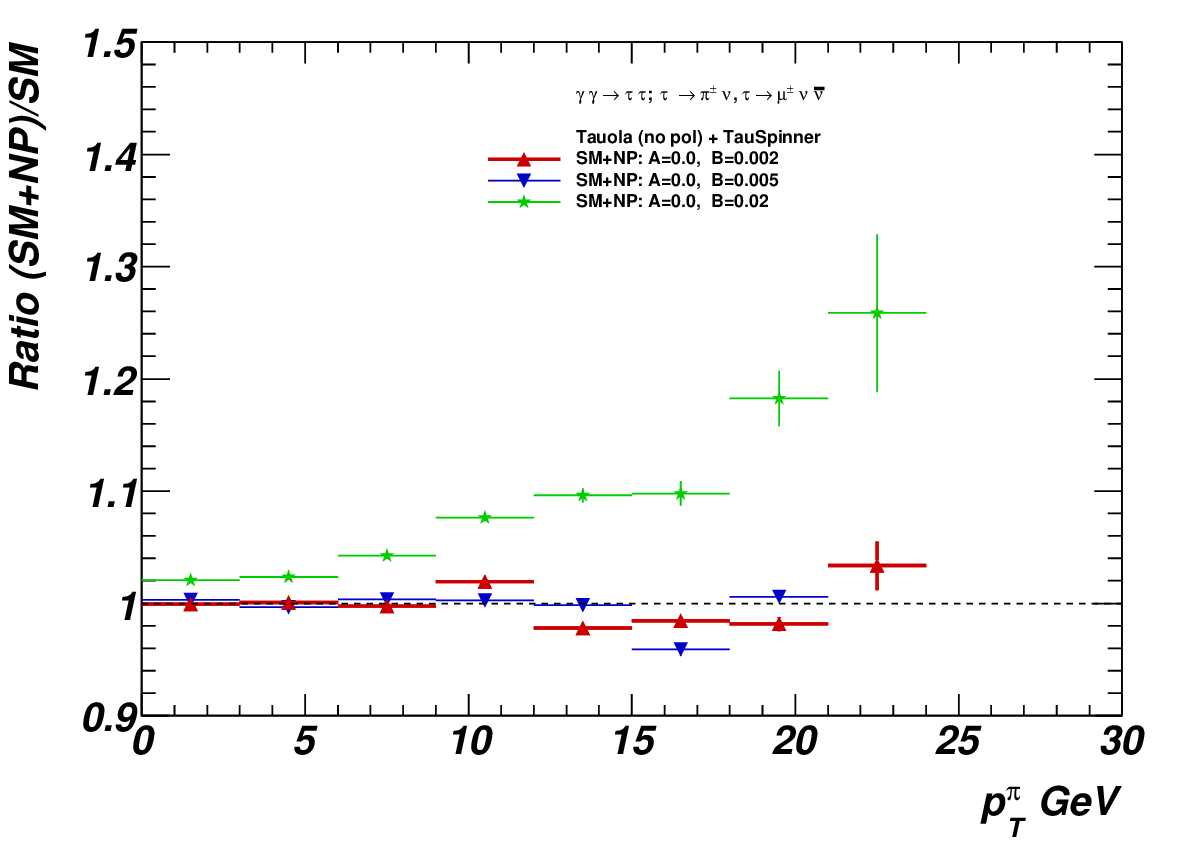}
  \includegraphics[width=7.5cm,angle=0]{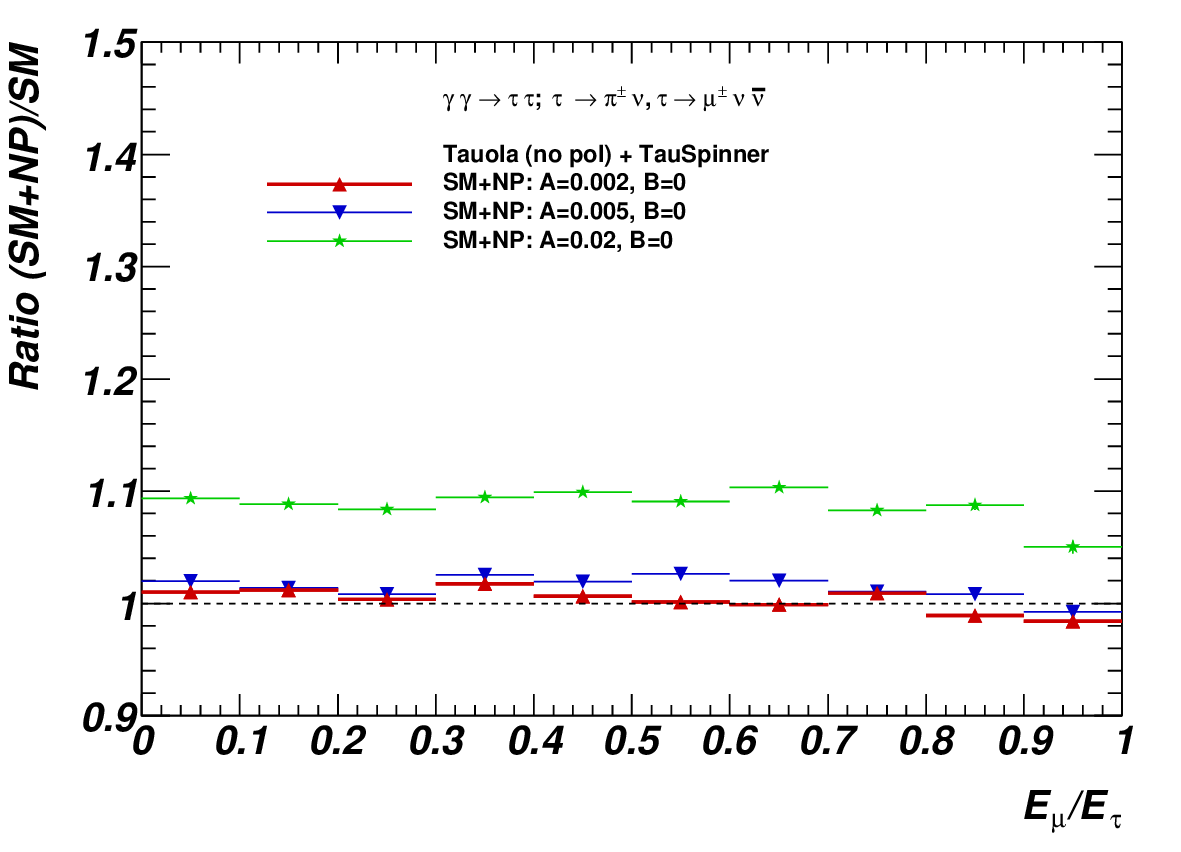}
  \includegraphics[width=7.5cm,angle=0]{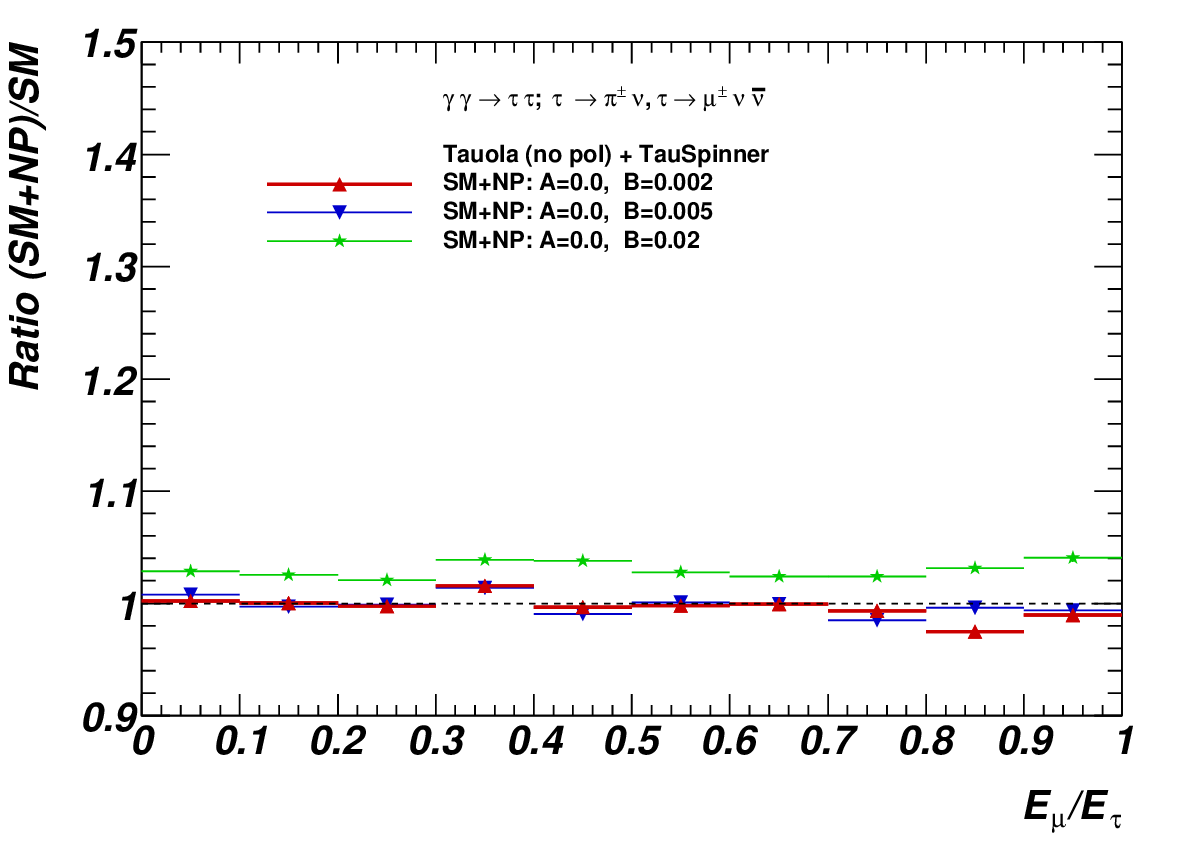}
  \includegraphics[width=7.5cm,angle=0]{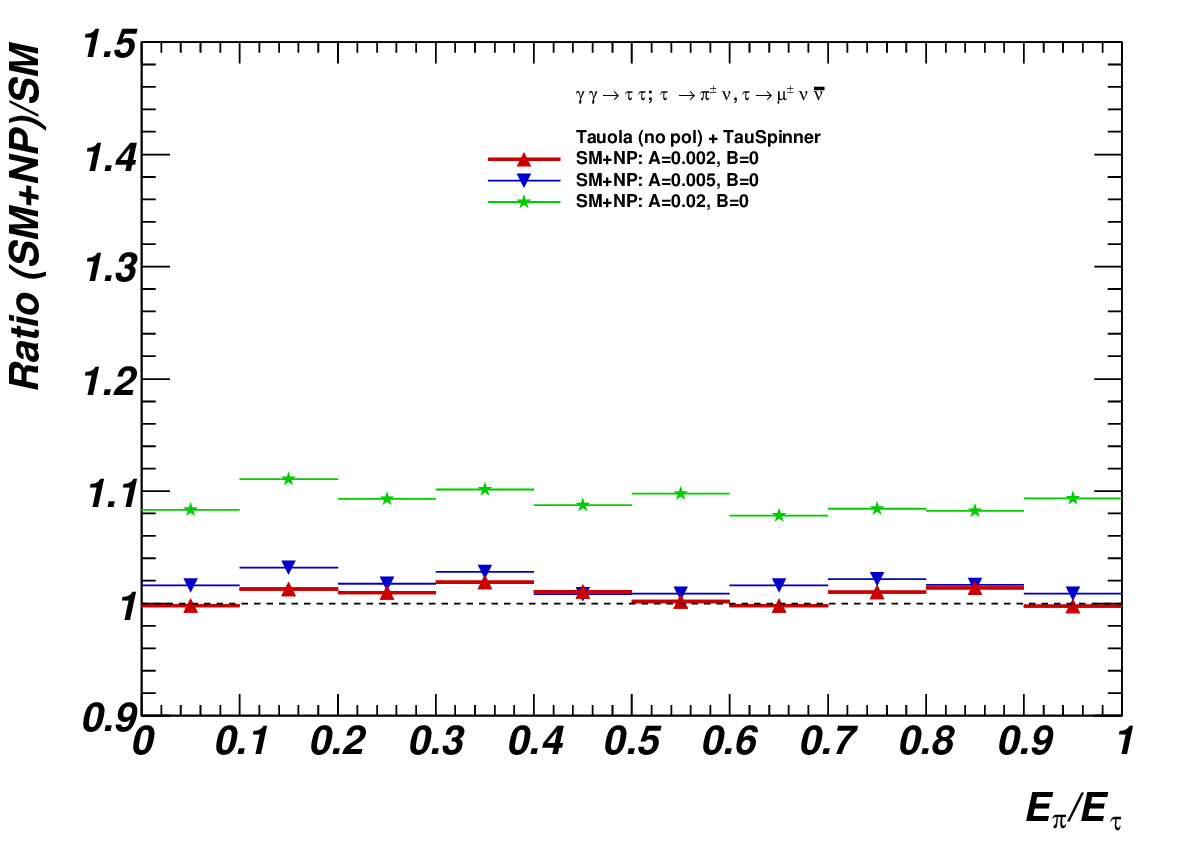}
  \includegraphics[width=7.5cm,angle=0]{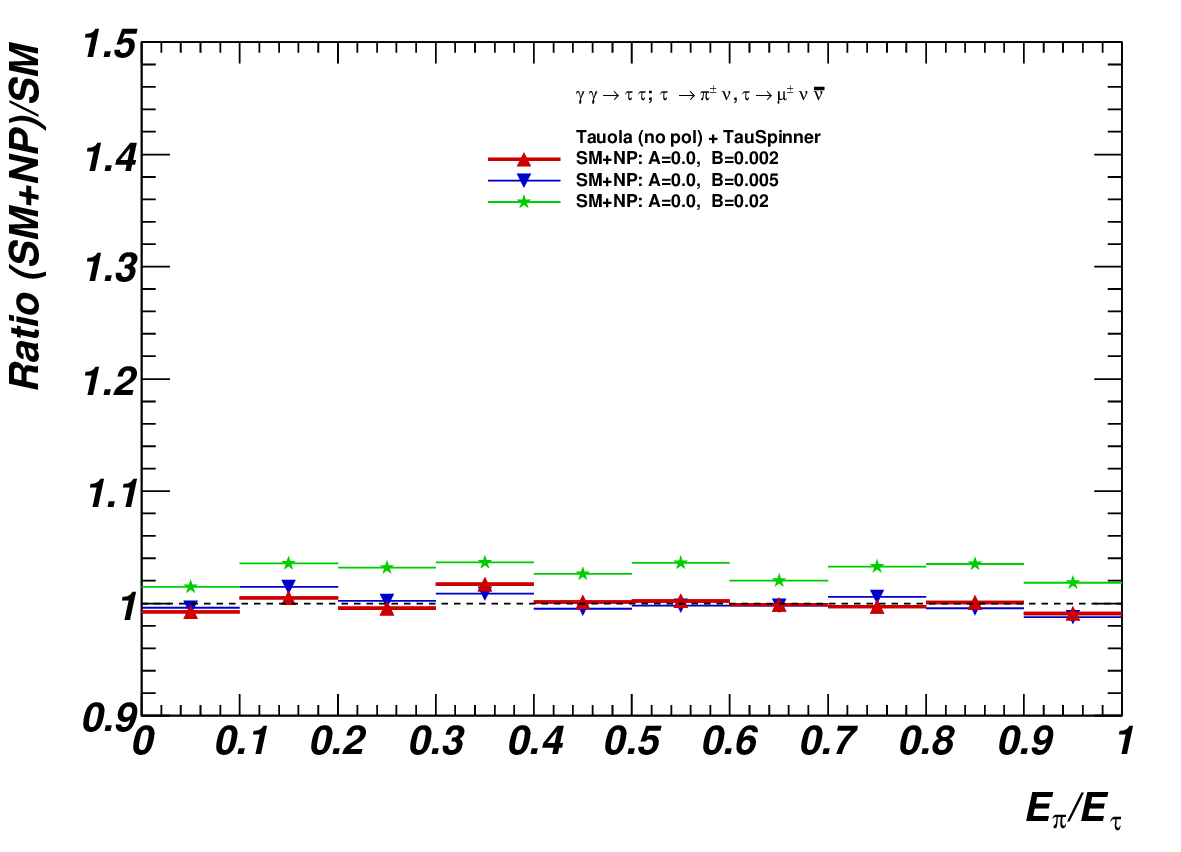}
}
  \caption{Kinematical distributions for one $\tau$ decaying $\tau^\pm \to \pi^\pm \nu_\tau$, and other $\tau$ decaying $\tau^\mp \to \mu^\mp \nu_\tau \nu_\mu$.
Notation is the same as in Fig.~\ref{Fig:Rtt}.	
 \label{Fig:kinem_BSM_mupi_a} }
\end{figure}

\begin{figure}          
  \begin{center}                               
{
  \includegraphics[width=7.5cm,angle=0]{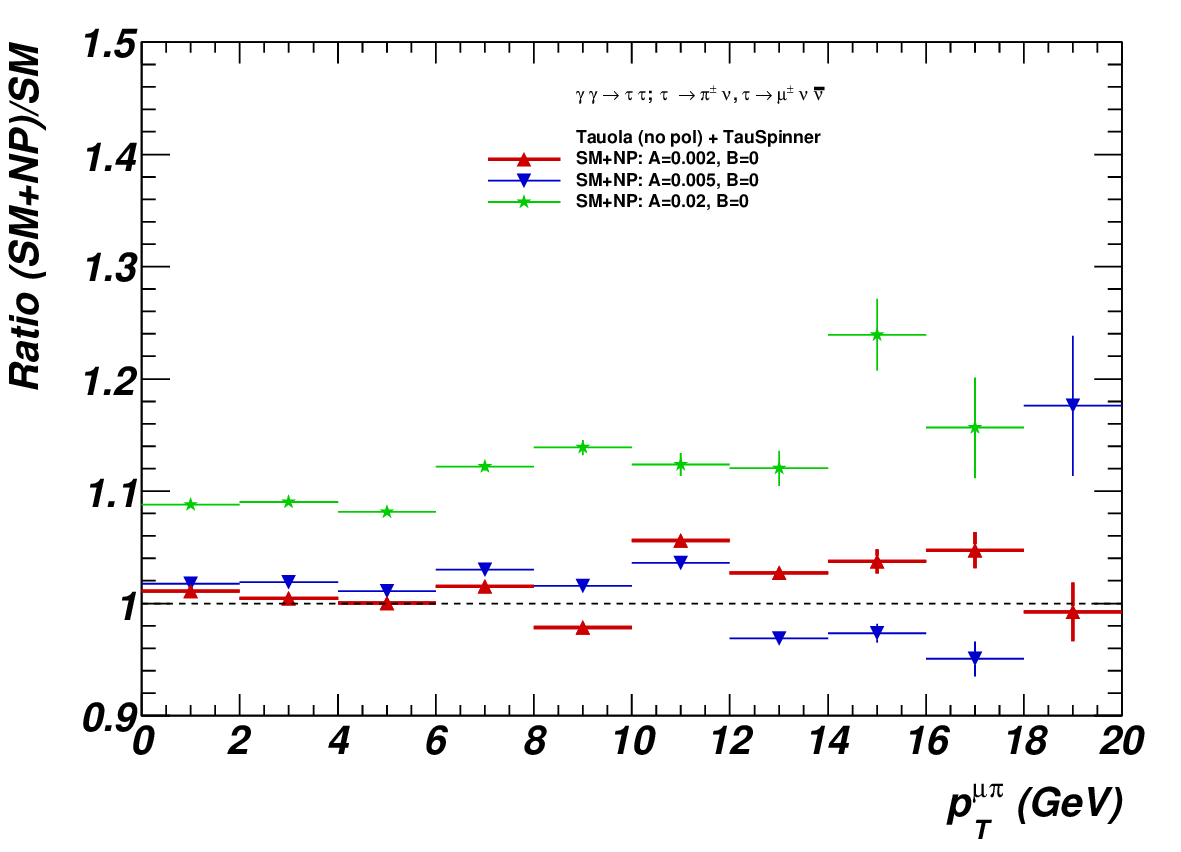}
  \includegraphics[width=7.5cm,angle=0]{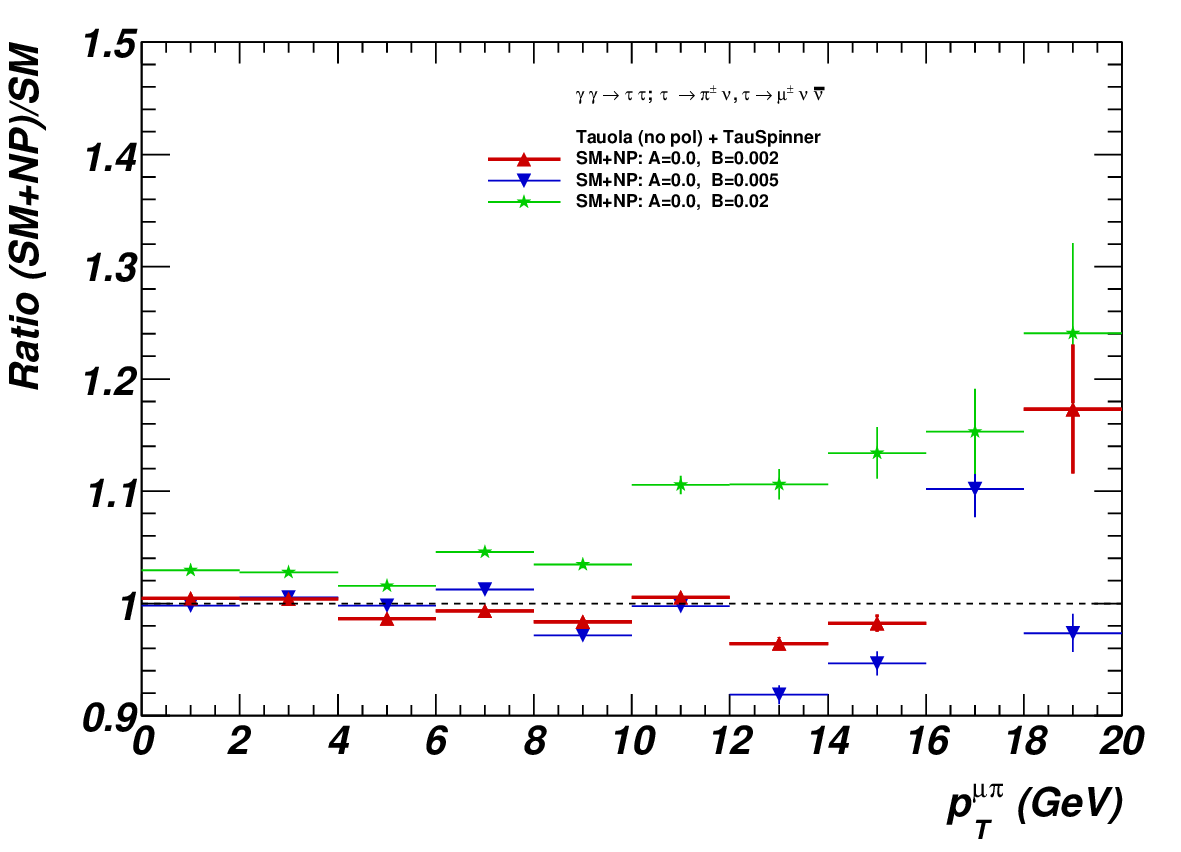}
  \includegraphics[width=7.5cm,angle=0]{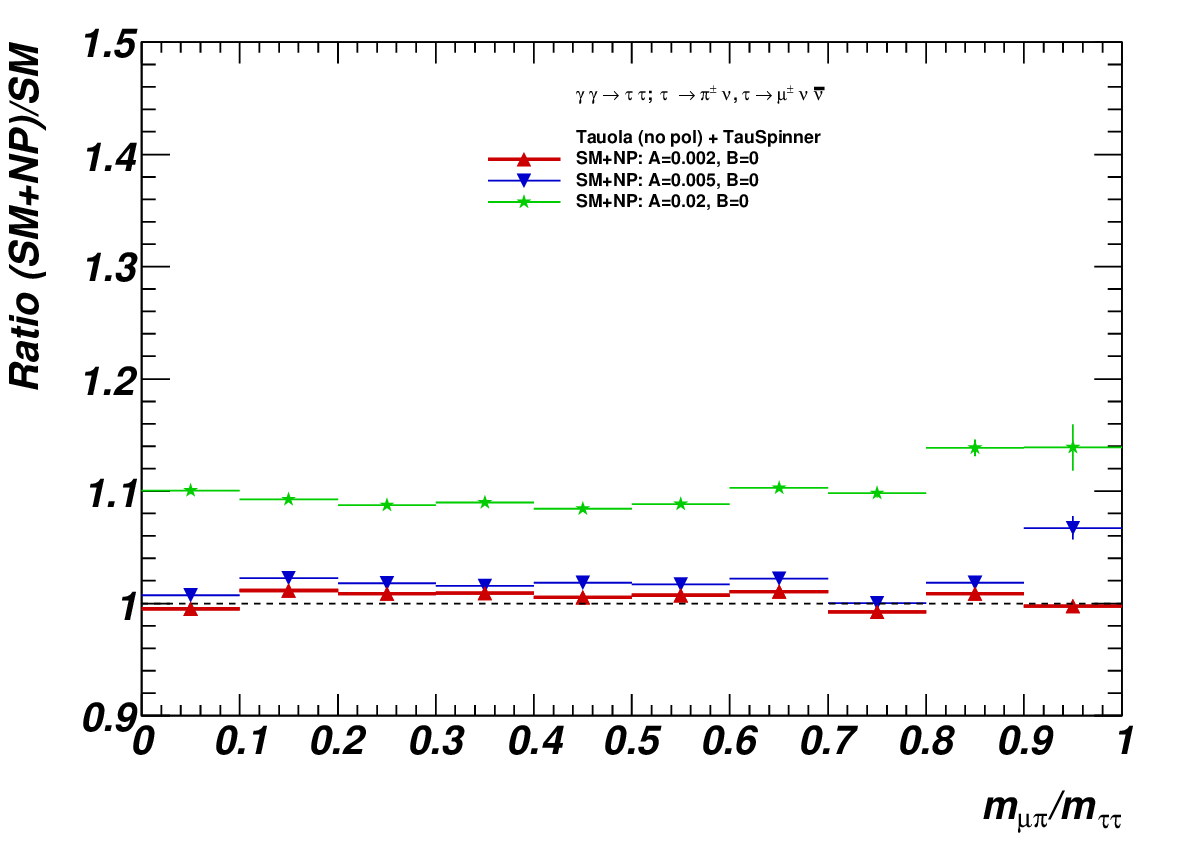}
  \includegraphics[width=7.5cm,angle=0]{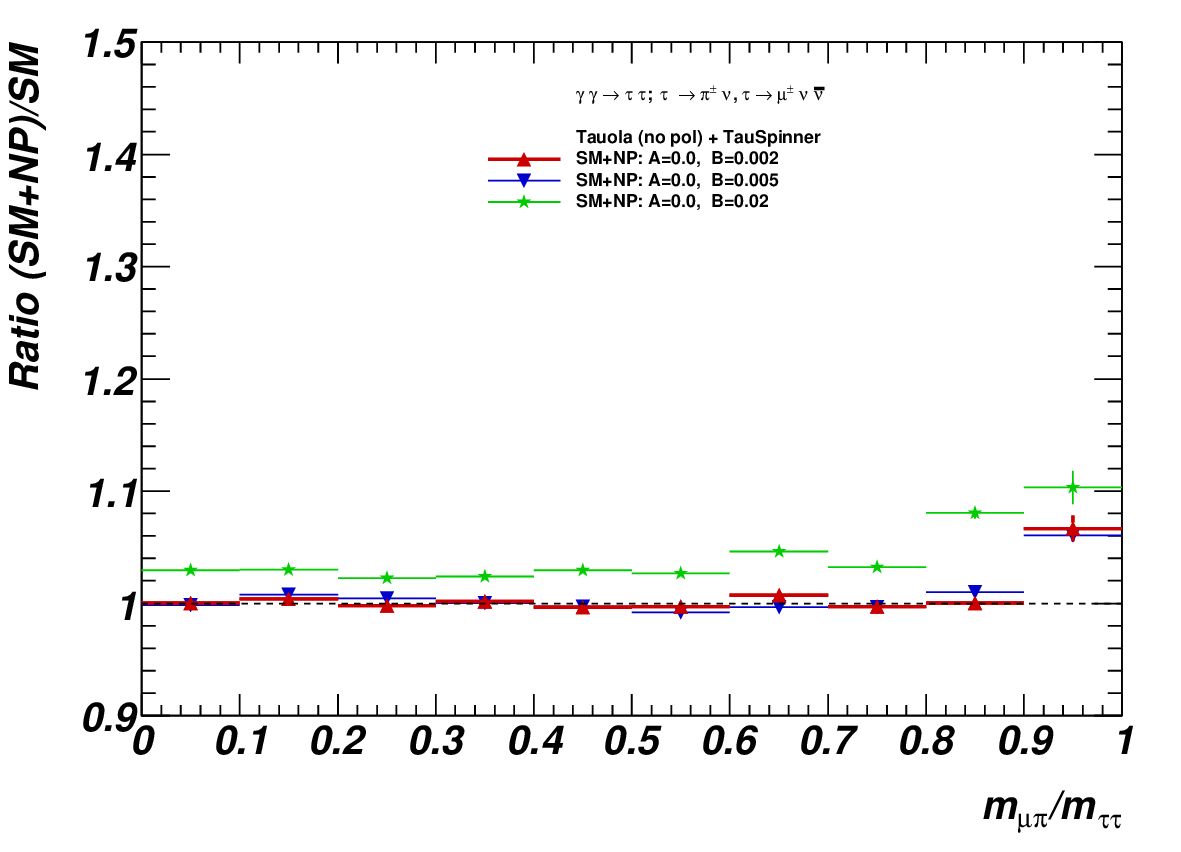}
}
\end{center}
  \caption{Further kinematical distributions for  one $\tau$ decaying $\tau^\pm \to \pi^\pm \nu_\tau$, and other 
	$\tau$ decaying  $\tau^\mp \to \mu^\mp \nu_\tau \nu_\mu$.
Notation is the same as in Fig.~\ref{Fig:Rtt}.
 \label{Fig:kinem_BSM_mupi_b} }
\end{figure}


\section{Summary and outlook}
\label{sec:Outlook}
The effects of $\gamma \gamma$ parton-level contributions to $\tau$-pair production in $pp$ and PbPb collisions
at the LHC are of interest in themselves, but also for evaluation of possible improved sensitivity to constraint anomalous
electric and magnetic dipole moments from the spin effects.

In the present paper, we have addressed the ways to implement effects of such New Physics interactions with the
attributions to the events weights. Physics input is presented with the help of the parton-level
$2\to 2$ processes, where anomalous terms due to electric and anomalous magnetic dipole moments  are included.
Such solution can be applied directly to any other model of NP interactions, provided it can be 
encapsulated into form-factors as of Eq.~(\ref{eq:003}) (in general, rather straightforward  recalculation of functions
used by the algorithm is needed).
Collection of plots to illustrate  sensitivity to dipole moments of the spin-correlation matrix elements is presented.
Examples, sensitive to spin, observable distributions are evaluated,
however without sufficiently detailed evaluation of detector responses they require refinements.
We have first chosen $\tau^\pm \to \pi^\pm \nu_\tau$ decays, as they are the simplest to interpret.
Then we turned to $\tau^\pm \to \rho^\pm \nu_\tau$, because sensitivity to the transverse spin is observable from the visible decay products only. At the end, we have
provided examples, where one of the $\tau$ leptons decays through the leptonic channel.
In this case, interpretation is difficult (due to extra neutrino escaping detection), nonetheless such
configurations represent almost half of all events. 
Such events may also turn useful for the Machine Learning applications. 
Therefore reference distributions may be of interest.
Further exploration of sensitivity requires active participation of physicists involved in the experimental analysis.

Some details on how {\tt TauSpinner} algorithm for event reweighting with the new options 
introduced are provided as well.

\vspace{0.4cm}

\centerline {\bf Acknowledgments}

\vspace{0.4cm}

This project was supported in part from funds of the National Science Centre, Poland,
grant no. 2023/50/A/ST2/00224 and of COPIN-IN2P3 collaboration with LAPP-Annecy.


\appendix

\section{ TauSpinner: technical details}
\label{app:TauSpinner}
Let us recall  steps of the {\tt TauSpinner} initialisation and provide user-guide for configuring  
$\gamma \gamma \to \tau \tau$  component of the spin and production weights.

With the available implementation, weights are calculated assuming that the $\tau\tau$ final state
is a product of either $\bar q q$ or $\gamma \gamma$ scattering.
The calculations of corresponding $R_{ij}$ elements of spin-correlation matrix
are invoked and proportion, with which each of the processes contributed to the sum
in Eq.~(\ref{eq:parton-level}), depends on the parametrisation of the structure functions
$f(x_1,...)$, $f(x_2,...)$. Those represent probabilities of finding in the colliding beam
partons (quark, gluon or photon), carrying momentum fractions $x_1$, $x_2$.

In case of the proton-proton beams, the probabilities $f(x_1,...)$, $f(x_2,...)$ will be taken from
a PDFs library, e.g. parametrisations of~\cite{Klein:2016yzr, Xie:2021equ} include also
the photon structure functions. The parametrisation used should be indicated during initialisation.
For the PbPb collision, the library parametrising photon flux is not available, and user
setup will be required. For example by hand, the $\gamma \gamma$
contribution to $R_{ij}$ can  be taken in a fixed proportion to the summed over flavours $\bar q  q$ one.

\begin{flushleft}
{ \bf Package distribution}
\end{flushleft}

The tarball of the package can be downloaded from the web page \\
{\tt https://tauolapp.web.cern.ch/tauolapp/}

The {\tt TauSpinner} package is distributed in the same tarball as {\tt Tauola} package,
a library for simulating for $\tau$ decay, as they share several components of the code,
interfaces and tests.

The installation script is prepared and is located in main directory\\
{\tt tauola/install-everything.sh}\\
which is installing both  {\tt Tauola/TauSpinner} but also other packages needed for
execution of the code and/or examples:
{\tt HepMC}, {\tt LHAPDF}, {\tt PHOTOS}, {\tt MCTESTER}, {\tt Pythia}.
You can comment it out and provide links with  details of already existing    installations in your own environment.

The examples of use with short README can be found in directory\\
{\tt tauola/TauSpinner/examples} of the distributed tarball.
In particular files
\begin{verbatim}
read_particles_from_TAUOLA.cxx
tau-reweight-test.cxx
\end{verbatim}
provide good starting point.


\begin{flushleft}
{ \bf Initialization}
\end{flushleft}

The user is required to configure both  $\bar q q \to \tau \tau$ and  $\gamma \gamma \to \tau \tau$ processes.
In case the  $\gamma \gamma$ process will  overly contribute, details
of the configuration used for  $\bar q q$ process are not relevant.

For both  $\bar q q$ and $\gamma \gamma$ processes the flag  {\tt ifkorch = 1} is mandatory,
as it invokes consistent flow of the $R_{ij}$ and final weight calculations. Details of implementation
of corresponding matrix elements were given in~\cite{Banerjee:2023qjc} with extension to higher
order terms for electromagnetic dipole moments as discussed in Section~\ref{sec:theory} of the present paper.
The flag {\tt iqed = 0} switches off the SM component $A0$  of the magnetic dipole moment then the
total $A$, $B$ values can be defined by the user.
At the initalisation step, user is required to provide values of the dipole moments used for sample
generation $ A0i, B0i$ and those of the NP model for which weight will be calculated $ Ai, Bi$.
Required is also that user provides proportion of the $\gamma \gamma$ process with respect to the $\bar q q$ one,
at generation {\tt GAMfraci} and for desired NP model  {\tt GAMfrac2i}.
If only $\gamma \gamma$ process is considered, this proportion should be set to be excessively large to make the quark-antiquark contribution negligible,
in the example below it is set to $10^6$.

We expect that in a standard usage the initialisation of parameters will be performed once.
However, re-initialisation of the parameters is possible on event-by-event base.
One can change, e.g. setting of EW corrections or dipole moments and process the same event several times
for weight calculation.
The ratio of the weights calculated, e.g. for the two sets of parameters and the same event, can be then used 
to estimate interesting properties of NP models.

Below shown is snippet of the initialisation, user code should follow,  e.g.
after example  from \\
{\tt tauola/TauSpinner/examples/tau-reweight-test.cxx}
\begin{verbatim}

// initialisation of main flow of TauSpinner 
   double CMSENE = 13000.0; // center of mass system energy
                            // used in PDF calculation. For p p collisions only
   bool Ipp   = true;       // for $pp$ collisions, the only option implemented
                            // but gam gam events from Pb Pb events can be also
                            // processed, detail 
   int Ipol   = 0; // are input samples polarized?
   int nonSM2 = 1; // are we using nonSM calculations? 
   int nonSMN = 0; // If we are using nonSM calculations we may want corrections
                   // to shapes only: y/n  (1/0)
   TauSpinner::initialize_spinner(Ipp, Ipol, nonSM2, nonSMN,  CMSENE);

// initialization for Dizet electroweak tables, here you specify location
// where the tables are present, default: in your run directory
// for more documentation see arXiv:2012.10997, arXiv:1808.08616  
   char* mumu="table.mu";
   char* downdown= "table.down";
   char* upup= "table.up";
   int initResult=initTables(mumu,downdown,upup);

// initialisation for EW parameters
// for more documentation see  arXiv:2012.10997, arXiv:1808.08616 
   double SWeff=0.2315200;
   double DeltSQ=0.;
   double DeltV=0.;
   double Gmu=0.00001166389;
   double alfinv=128.86674175;
   int keyGSW=1;
   double AMZi=91.18870000;
   double GAM=2.49520000;
   ExtraEWparamsSet(AMZi, GAM, SWeff, alfinv,DeltSQ, DeltV, Gmu,keyGSW);
 
// initialisation for the qqbar ->tautau matrix elements
// dipole moments and weak dipole moments are set to 0.0 in the qbarq->tautau ME
// flags relevant here are: 
   int ifkorch = 1;  // global flag to switch on ME calculations of arXiv:2307.03526
   int iqed = 0;     // iqed = 1  (if A0 at SM value to be added) 
   initialize_GSW(0, ifkorch, iqed, 0.0, 0.0, 0.0, 0.0, 0.0, 0.0, 0.0, 0.0);

// initialisation for the gamma gamma ->tautau matrix elements
// works only if global flag ifkorch = 1
// artificial weight for fraction of gamgam vs qbar process, if not calculated
// directly from  structure functions proportions
   double GAMfraci  = 1000000.0;  
   double GAMfrac2i = 1000000.0;
// values for dipole moments used in generated sample (A0i, B0i)
// and in the model for which weight is calculated (Ai, Bi)
   double A0i = 0.0;
   double B0i = 0.0;
   double Ai  = 0.001177;
   double Bi  = 0.0;
   initialize_gamagama(GAMfraci, GAMfrac2i, A0i, B0i, Ai, Bi);
\end{verbatim}
\begin{flushleft}
{\bf Calculation of event weights }
\end{flushleft}

Spin correlations matrix  $R_{ij}$  (coded in {\tt FORTRAN}) is used by weight calculating
method (coded in {\tt C++}).  The function {\tt dipolgamma\_(iqed, E, theta, A0, B0, Rij)} has several input
parameters  {\tt iqed, E, theta, A0, B0} and output matrix {\tt Rij}. The {\tt E, theta} denote respectively the 
energy of the scattering photon in the $2\to 2$ parton level centre-of-mass
frame and scattering angle of the outgoing $\tau$ lepton
with respect to photon beam direction, also in this frame\footnote{%
The elements of the {\tt Rij} matrix are calculated by routines {\tt dipolgamma\_ } and {\tt
 dipolqq\_ } respectively for $\gamma\gamma$ and
antiquark-quark  processes.
They are provided in file {\tt TAUOLA/TauSpinner/src/initwksw.f }.
In addition to calculation,  the frame re-orientation is provided in  these 
{\tt FORTRAN } interfacing routines (Eq.(\ref{eq:framesR})). Also   
the change of index convention from {\tt FORTRAN} 1,2,3,4 to {\tt C++ } 0,1,2,3 is introduced.
Finally, minus sign originating from
$\tau^+$ V+A coupling instead of $\tau^-$ V-A is  introduced in {\tt C++} code.
That explain also 
 sign change in $R_{yy}$ versus publication~\cite{Banerjee:2023qjc}. 
}. The {\tt  A0, B0} denote magnetic and electric moments. The {\tt iqed} flag, switches OFF/ON 
contribution to {\tt A0} from the SM $A(0)_{SM}= 1.17721(5) \cdot 10^{-3}$ of the anomalous magnetic
dipole moment, as calculated in~\cite{Eidelman:2007sb}.

For each event, which is  read from the {\tt input\_file} by:
\begin{verbatim}
int status = readParticlesFromTAUOLA_HepMC(input_file, X, tau, tau2, 
             tau_daughters, tau_daughters2);
\end{verbatim}
calculation of spin weight $WTspin$ and the corresponding relative change to the cross-section $WTprod$,
due to NP values of dipole moments can be invoked.
\begin{verbatim}
double WTspin = calculateWeightFromParticlesH(X, tau, tau2, 
                                              tau_daughters,tau_daughters2);
double WTprod = getWtNonSM();
\end{verbatim}

By definition, always an average of spin weight $<WTspin>$= 1.0. To quantify the impact on particular kinematical
distribution of including spin correlations for a given (SM+NP) model with respect to one used
in generated sample,  one should use product $ WTspin \cdot WTprod$ when filing histograms.
By construction,  for model with A0i, B0i calculated WTprod = 1.0.
It means that only relative change in the cross-section, not the absolute one, can be accessed with
present implementation in {\tt TauSpinner}.


\begin{flushleft}
{\bf Accessing internal variables }
\end{flushleft}

Several functions ({\it getters})  are available to access internal variables  are prepared.
In particular for  each event,  with the methods
\begin{verbatim}
getZgamParametersTR(Rxx, Ryy, Rxy, Ryx);
getZgamParametersL(Rzx, Rzy, Rzz, Rtx, Rty, Rtz);
\end{verbatim}
the set of $R_{ij}$ matrix components can be accessed, as they might be of interest for monitoring purposes.

\section{\texorpdfstring{Elements of the $\gamma \gamma \to \tau \tau $ spin-correlation matrix}{}}
\label{app:Rij_elements}
In this Appendix the elements of the matrix $R_{ij}$ are presented in the form which is actually used in the {\tt FORTRAN} code.
Physics-wise they do not differ from the ones of Eq.~(\ref{eq:R_{ij}}).
\begin{eqnarray} 
R_{11} &=& 
\frac{e^4}{4 \gamma ^4 (1-\beta ^2 \cos^2 \theta)^2}
 [ \beta ^2 \gamma ^2 (-8+2 (4+8 A+10 A^2+4 A^3+A^4+2 (-1+A) (3+A) B^2  \\          \nonumber
&+ & B^4) \gamma ^2-4 (-4 B^2+(A^2+B^2)^2)
\gamma ^4+3 (A^2+B^2)^2 \gamma ^6) \cos^2 \theta+(A^2+B^2)^2 \beta ^6 \gamma ^8 \cos^6 \theta \\ \nonumber
&+& \beta ^4
\gamma ^4 \cos^4 \theta (-4+\gamma ^2 (-8 B^2-(A^2+B^2)^2 (-1+2 \gamma ^2))-(A^2+B^2)^2
\gamma ^2 (-1+\gamma ^2) \cos 2 \theta ) \\ \nonumber
&+& \frac{1}{2} (-16-8 (-1+A (2+A)+4 B^2) \gamma ^2-(4 (A
(2+A) (2+A (4+A)) \\  \nonumber
&+& 2 (-2+A (3+A)) B^2+B^4)+(2+4 A+4 A^2+2 A^3+A^4+2 (1+A+A^2) B^2+B^4) \beta ^2) \gamma ^4 \\ \nonumber
&+& (-8
B^2+2 (A (2+A)+B^2)^2  
+ (-2+(-2+A) A+B^2) (A^2+B^2) \beta ^2) \gamma ^6 \\      \nonumber
&-& (A^2+B^2)^2 \beta ^2 \gamma
^8-2 (-1+\gamma ^2) (2 (1+A) \gamma +(A^2+B^2) \gamma ^3)^2 \cos 2 \theta \\     \nonumber
&+& \beta ^2 \gamma ^4 (2+4 A+4
A^2+2 A^3+A^4+2 B^2+2 A B^2+2 A^2 B^2+B^4  \\    \nonumber
&-& (-2+(-2+A) A+B^2) (A^2+B^2) \gamma ^2+(A^2+B^2)^2 \gamma ^4) \cos 4 \theta ) ].  \nonumber
\end{eqnarray}

\begin{eqnarray} R_{22} &=&
\frac{e^4}{4 \gamma ^4 (1-\beta ^2 \cos^2 \theta)^2}
 [ -8+(8-16 B^2) \gamma ^2-2 (2+4 A^3+A^4-6 B^2+B^4 \\       \nonumber
&+& 4 A (2+B^2)+2 A^2 (5+B^2)) \gamma
^4+2 (A^4-4 B^2+2 A^2 B^2+B^4) \gamma ^6-(A^2+B^2)^2 \gamma ^8 \\ \nonumber
&+& \beta ^2 \gamma ^2 (-8+2 (4+4 A^3+A^4-6 B^2+B^4+4
A (2+B^2)+2 A^2 (5+B^2)) \gamma ^2 \\  \nonumber
&-& 4 (A^4-4 B^2+2 A^2 B^2+B^4) \gamma ^4+3 (A^2+B^2)^2 \gamma ^6)
\cos^2 \theta-\beta ^4 \gamma ^4 (4+8 B^2 \gamma ^2     \\         \nonumber
&+& A^4 \gamma ^2 (-2+3 \gamma ^2)+2 A^2 B^2 \gamma ^2 (-2+3 \gamma
^2)+B^4 \gamma ^2 (-2+3 \gamma ^2)) \cos^4 \theta    \\          \nonumber
&+& (A^2+B^2)^2 \beta ^6 \gamma ^8 \cos^6 \theta ] .
\nonumber 
\end{eqnarray}   

\begin{eqnarray} 
\label{app:R12}
R_{12} &=&
\frac{e^4}{16 \gamma ^2 (1-\beta ^2 \cos^2 \theta)^2}
B  \beta  [ -32+32 A-48 \gamma ^2-192 A \gamma ^2-52 A^2 \gamma ^2-20 B^2 \gamma ^2 \\ \nonumber
&+ &28 \beta ^2 \gamma ^2+120 A \beta ^2 \gamma ^2+27
A^2 \beta ^2 \gamma ^2+3 B^2 \beta ^2 \gamma ^2+112 A \gamma ^4+44 A^2 \gamma ^4+12 B^2 \gamma ^4 \\
\nonumber
&-& 172 A \beta ^2 \gamma ^4-65 A^2 \beta ^2 \gamma
^4-9 B^2 \beta ^2 \gamma ^4+72 A \beta ^4 \gamma ^4+27 A^2 \beta ^4 \gamma ^4+3 B^2 \beta ^4 \gamma ^4
\\ \nonumber
&+& 4 (8+(-4+8 \beta ^2+B^2 (3+\beta
^2)) \gamma ^2+B^2 (-1-2 \beta ^2+\beta ^4) \gamma ^4 \\        \nonumber
&+& 4 A (2+(-6+9 \beta ^2) \gamma ^2+(5-12 \beta ^2+6
\beta ^4) \gamma ^4)+A^2 \gamma ^2 (-5+7 \gamma ^2+9 \beta ^4 \gamma ^2 \\       \nonumber
&+& \beta ^2 (9-18 \gamma ^2))) \cos 2 \theta +\beta ^2 \gamma ^2 (4+B^2 (1+(1+\beta ^2) \gamma ^2)+4 A (6+(-5+6 \beta ^2) \gamma ^2)  \\       \nonumber
&+& A^2 (9+(-7+9 \beta ^2) \gamma ^2)) \cos 4 \theta  ]  .        \nonumber
\end{eqnarray}
                                                         
\begin{eqnarray} 
\label{app:R33}
R_{33} &=&
-\frac{e^4}{4 \gamma ^4 (1-\beta ^2 \cos^2 \theta)^2}
 [  -8-8 (-3+2 B^2) \gamma ^2-2 (10+4 A^3+A^4+B^4-4 \beta ^2 \\      \nonumber
&+& 4 A (2+B^2+2 \beta ^2)+2 A^2 (5+B^2+2
\beta ^2)+B^2 (-22+8 \beta ^2)) \gamma ^4+2 (4+3 A^4 \\  \nonumber
&+& 3 B^4-4 \beta ^2+2 A^2 (10+3 B^2-8 \beta ^2)-4 A^3 (-2+\beta
^2)+4 B^2 (-4+3 \beta ^2)     \\    \nonumber
&-& 4 A (B^2 (-2+\beta ^2)+4 (-1+\beta ^2))) \gamma ^6+(A^4 (-5+2
\beta ^2)+2 A^2 B^2 (-5+2 \beta ^2) \\    \nonumber
&+& B^2 (-16 (-1+\beta ^2)+B^2 (-5+2 \beta ^2))) \gamma ^8-2
(A^2+B^2)^2 (-1+\beta ^2) \gamma ^{10}   \\ \nonumber
&+& \gamma ^2 (-2 \gamma ^2 (-1+\gamma ^2) (2+2 A+A^2 \gamma ^2+B^2
\gamma ^2)^2+\beta ^2 (-8+2 (4+4 A^3       \\ \nonumber
&+& A^4-14 B^2+B^4+4 A (2+B^2)+2 A^2 (5+B^2)) \gamma ^2-16 (A+3
A^3  \\ \nonumber
&+& A^4+3 A B^2+B^2 (-1+B^2)+2 A^2 (2+B^2)) \gamma ^4+(16 A^3+11 A^4+16 A B^2 \\ \nonumber
&+& B^2 (-16+11 B^2)+2 A^2
(8+11 B^2)) \gamma ^6-2 (A^2+B^2)^2 \gamma ^8)  \\            
\nonumber
&+& 4 \beta ^4 \gamma ^4 (2-2 A^3 (-3+\gamma ^2)+B^2
(-2+6 \gamma ^2)+A^4 (1-\gamma ^2+\gamma ^4)    \\ 
\nonumber
&+& B^4 (1-\gamma ^2+\gamma ^4)+A (8-2 B^2 (-3+\gamma ^2))+2
A^2 (6-\gamma ^2    \\ \nonumber
&+& B^2 (1-\gamma ^2+\gamma ^4)))) \cos^2 \theta-\beta ^2 \gamma ^4 (\beta ^2 (4-2
(-4-8 A+A^4-8 B^2        \\ 
\nonumber
&+& B^4+2 A^2 (-2+B^2)) \gamma ^2+(16 A^3+7 A^4+16 A B^2+7 B^4+2 A^2 (8+7 B^2)) \gamma
^4     \\ 
\nonumber
&+& 2 (A^2+B^2)^2 \gamma ^6)+2 \beta ^4 \gamma ^4 (-4 A^3-4 A B^2+A^4 (-1+\gamma ^2) \\ \nonumber
&+& 2 A^2 (-2+B^2 (-1+\gamma
^2))+B^2 (4+B^2 (-1+\gamma ^2)))-4 \gamma ^2 (2+2 A^3 (1+\gamma ^2) \\    
\nonumber
&+& 2 B^2 (1+\gamma
^2)+A^4 (1-\gamma ^2+\gamma ^4)+B^4 (1-\gamma ^2+\gamma ^4) \\ 
\nonumber
&+& 2 A (2+B^2 (1+\gamma ^2))+2 A^2 (2+\gamma
^2+B^2 (1-\gamma ^2+\gamma ^4)))) \cos^4 \theta \\ 
\nonumber
&+& (A^2+B^2)^2 \beta ^4 \gamma ^8 (2-2 \gamma ^2+\beta
^2 (1+2 \gamma ^2)) \cos^6 \theta ] .    
\nonumber
\end{eqnarray}                                                     

\begin{eqnarray} R_{13} &=&
-\frac{e^4}{2 \gamma  (1-\beta ^2 \cos^2 \theta)^2}
 \cos\theta [ 4-4 \beta ^2-4 B^2 \beta ^2-4 \gamma ^2+4 B^2 \gamma ^2+4 \beta ^2 \gamma ^2-6 B^2 \beta ^2 \gamma ^2 \\           \nonumber
&-& 2 B^4
\beta ^2 \gamma ^2-4 B^2 \gamma ^4+B^4 \gamma ^4+4 B^2 \beta ^2 \gamma ^4-B^4 \gamma ^6+B^4 \beta ^2 \gamma ^6          \\ \nonumber
&+& 4 A (2+(-2+\beta ^2+B^2
(1-2 \beta ^2)) \gamma ^2+B^2 (-1+\beta ^2) \gamma ^4)+4 A^3 \gamma ^2 (1-\gamma ^2   \\ \nonumber
&+& \beta ^2 (-2+\gamma
^2))+A^4 (\gamma ^4-\gamma ^6+\beta ^2 \gamma ^2 (-2+\gamma ^4))         \\ \nonumber
&+& 2 A^2 (-(-1+\gamma ^2) (2+2
\gamma ^2+B^2 \gamma ^4)+\beta ^2 \gamma ^2 (-3+2 \gamma ^2+B^2 (-2+\gamma ^4))) \\ \nonumber
&-& 2 \beta ^2 \gamma ^2 (2 (-1+\beta
^2)+2 A^3 (-1+(-1+\beta ^2) \gamma ^2)+A^4 (-1+\gamma ^2+(-1+\beta ^2) \gamma ^4) \\ \nonumber
&+&B^4 (-1+\gamma
^2+(-1+\beta ^2) \gamma ^4)+B^2 (-2 (1+\gamma ^2)+\beta ^2 (1+2 \gamma ^2)) \\ \nonumber
&+& 2 A (-2+\beta
^2+B^2 (-1+(-1+\beta ^2) \gamma ^2))+A^2 (-4+\beta ^2-2 \gamma ^2+2 \beta ^2 \gamma ^2 \\ \nonumber
&+& 2 B^2 (-1+\gamma ^2+(-1+\beta
^2) \gamma ^4))) \cos^2 \theta+(A^2+B^2)^2 \beta ^4 \gamma ^4 (1+(-1+\beta ^2) \gamma
^2) \cos^4 \theta ]  \sin\theta  .       \nonumber
\end{eqnarray}                 

\begin{eqnarray} R_{23} &=&
- \frac{e^4}{8 \gamma  (1-\beta ^2 \cos^2 \theta)^2}
B  \beta  [ -16+8 \gamma ^2-6 B^2 \gamma ^2-4 \beta ^2 \gamma ^2+7 B^2 \beta ^2 \gamma ^2+2 B^2 \gamma ^4 \\      \nonumber
&-& 3 B^2 \beta ^2 \gamma ^4+B^2
\beta ^4 \gamma ^4-A^2 \gamma ^2 (-10+\beta ^2+14 \gamma ^2-21 \beta ^2 \gamma ^2+7 \beta ^4 \gamma ^2)
\\   \nonumber
&-& 4 A (4+6 (-2+\beta ^2)
\gamma ^2+5 (2-3 \beta ^2+\beta ^4) \gamma ^4)-\beta ^2 \gamma ^2 (4+B^2 (1+\gamma ^2-\beta ^2 \gamma ^2) \\ \nonumber
&+& 4 A (6+5
(-1+\beta ^2) \gamma ^2)+A^2 (9+7 (-1+\beta ^2) \gamma ^2)) \cos 2 \theta  ]  \sin 2 \theta  .  \nonumber
\end{eqnarray}                                           

\begin{eqnarray} 
R_{44}&=& 
-\frac{e^4}{4 \gamma ^4 (1-\beta ^2 \cos^2 \theta)^2}
 [ 8-8 \gamma ^2+2 (-2+4 A^3+A^4+2 B^2+B^4+4 A (-2+B^2) \\          \nonumber
&+& 2 A^2 (-1+B^2)) \gamma ^4-2 (8 A^3+A^4+8
A B^2+2 A^2 (4+B^2)+B^2 (8+B^2)) \gamma ^6       \\ \nonumber
&-& (A^2+B^2)^2 \gamma ^8+\beta ^2 \gamma ^2 (8+8 A^3 \gamma ^2
(-1+4 \gamma ^2)+4 B^2 \gamma ^2 (-1+8 \gamma ^2)    \\ \nonumber
&+& A^4 \gamma ^2 (-2+4 \gamma ^2+3 \gamma ^4)+B^4 \gamma ^2 (-2+4
\gamma ^2+3 \gamma ^4)+8 A \gamma ^2 (2+B^2 (-1+4 \gamma ^2)) \\ \nonumber
&+&  2 A^2 \gamma ^2 (2+16 \gamma ^2+B^2 (-2+4 \gamma
^2+3 \gamma ^4))) \cos^2 \theta-\beta ^4 \gamma ^4 (-4+16 A^3 \gamma ^2 \\ \nonumber
&+& 16 B^2 \gamma ^2+16 A B^2 \gamma ^2+A^4 \gamma
^2 (2+3 \gamma ^2)+B^4 \gamma ^2 (2+3 \gamma ^2)        \\ \nonumber
&+& 2 A^2 \gamma ^2 (8+B^2 (2+3 \gamma ^2))) 
\cos^4 \theta +(A^2+B^2)^2 \beta ^6 \gamma ^8 \cos^6 \theta ] .             \nonumber
\end{eqnarray}                                  

\vspace{1cm}
\bibliographystyle{utphys_spires}
\bibliography{Gamma-gamma_appb}

\end{document}